\documentclass[journal=jctcce,manuscript=article,chaptertitle=true]{achemso}

\usepackage{chemformula,bm,caption,subcaption} 
\usepackage[T1]{fontenc} 

\author{R\'obert Izs\'ak}
\affiliation{Riverlane, St Andrews House, 59 St Andrews Street, Cambridge, CB2 3BZ,  UK}
\email{robert.izsak@riverlane.com}
\author{Aleksei V. Ivanov}
\author{Nick S. Blunt}
\author{Nicole Holzmann}
\affiliation{Riverlane, St Andrews House, 59 St Andrews Street, Cambridge, CB2 3BZ,  UK}
\author{Frank Neese}
\affiliation{Max-Planck Institut für Kohlenforschung 
Kaiser-Wilhelm-Platz 1
D-45470 Mülheim an der Ruhr 
Germany}

\title[An \textsf{achemso} demo]
  {Measuring Electron Correlation. The Impact of Symmetry and Orbital Transformations}

\abbreviations{RHF, UHF, FCI, strong correlation}
\keywords{strong correlation, hydrogen}

\begin{document}


\begin{abstract}
In this perspective, the various measures of electron correlation used in wavefunction theory, density functional theory and quantum information theory are briefly reviewed. We then focus on a more traditional metric based on dominant weights in the full configuration solution and discuss its behaviour with respect to the choice of the $N$-electron and the one-electron basis. The impact of symmetry is discussed and we emphasize that the distinction between determinants, configuration state functions and configurations as reference functions is useful because the latter incorporate spin-coupling into the reference and should thus reduce the complexity of the wavefunction expansion. The corresponding notions of single determinant, single spin-coupling and single configuration wavefunctions are discussed and the effect of orbital rotations on the multireference character is reviewed by analysing a simple model system. In molecular systems, the extent of correlation effects should be limited by finite system size and in most cases the appropriate choices of one-electron and $N$-electron bases should be able to incorporate these into a low-complexity reference function, often a single configurational one.
\end{abstract}


\section{Electron Correlation}

Electron correlation has often been called the ``chemical glue'' of nature due to its ubiquitous influence in molecules and solids\cite{kurth2000role,martin2022electron}. The problem was studied in the smallest two-electron systems, such as He and H$_2$ already in the 1920s and these early efforts are discussed in detail elsewhere\cite{lowdin1958correlation,kotani1961quantum,slater1963quantum,kutzelnigg2002einfuhrung}. The correlation concept usually presupposes an independent particle model, typically Hartree--Fock (HF) mean-field theory, that serves as a reference compared to which the exact solution is correlated. The associated idea of correlation energy goes back to Wigner's work in the 1930s on the uniform electronic gas and metals\cite{wigner1934interaction,wigner1938effects}, and it is nowadays commonly defined, following L\"owdin, as the difference between the exact and the HF energy\cite{lowdin1958correlation}. There is a vast literature on the subject and here we will confine ourselves to some of the most pertinent reviews. Early work on the correlation problem is reviewed by L\"owdin in the 1950s\cite{lowdin1958correlation} and more recently\cite{lowdin1995historical}, as well as by Sinano\u{g}lu\cite{sinanoglu1961many,sinanoglu1964many} within the context of his many-electron theory. The early perspective of McWeeny is a short but useful discussion of the problem in terms of density matrices\cite{mcweeny1967nature}. Kutzelnigg, Del Re and Berthier penned an overview on the statistical measures of correlation\cite{kutzelnigg1968correlation}, and, as P. v. Herigonte, they also reviewed the situation in the seventies\cite{herigonte1972electron}. This was later followed up several times by Kutzelnigg\cite{kutzelnigg2000electron,kutzelnigg2003theory}. Bartlett and Stanton not only discuss correlation effects in molecules, but also offer a tutorial on post-HF wavefunction methods\cite{bartlett1994applications}. Similarly educational is the chapter written by Knowles, Sch\"utz and Werner\cite{knowles2000ab} and the feature articles of Tew, Klopper, Helgaker\cite{tew2007electron} and Martin\cite{martin2022electron}. More recent work on the homogeneous electron gas is reviewed by Senatore and March\cite{senatore1994recent} as well as Loos and Gill\cite{loos2016uniform}, the former authors also discuss orbital-based and quantum Monte Carlo methods.  For a review taking an entanglement-based approach to the correlation problem, we refer the reader to Chan\cite{chan2012low}. There are also several book-length monographs dedicated to the subject by Fulde\cite{fulde1995electron}, March\cite{march1996electron} and Wilson\cite{wilson2014electron}, as well as a number of collections\cite{advchemphys014,march1999electron}. In the remainder of this section, we will consider some selected topics about correlation effects before stating the scope of this perspective.

An important aspect of electron correlation is its strength and the various measures proposed to quantify it. For the homogeneous electron gas, where some analytical energy expressions can be obtained at least under some circumstances, the discussion is often concerned with the relative magnitudes of kinetic and potential energies in the limiting cases of high and low electron densities. While Wigner was able to obtain an expression for the correlation energy at the low-density limit\cite{wigner1934interaction,wigner1938effects}, in general, perturbation theory diverges already at second order\cite{raimes1972many}, a problem that can only be remedied partly at the high-density limit following Gell-Mann and Brueckner\cite{gell-mann1957correlation}. In the high-density limit, the kinetic energy dominates over the repulsive potential energy, the electrons are delocalized and behave like a gas, thus, an independent particle model should provide a good description. In the low-density limit, it is the Coulomb interactions responsible for correlation effects that dominate, forcing the electrons to localize on site (form a lattice).  Interestingly, the low- and high-density limits have their analogues in molecular systems\cite{senatore1994recent}: in the hydrogen molecule, the high-density case would correspond to short bond lengths and a delocalized molecular orbital\cite{hund1927deutung,mulliken1928assignment,lennard1929electronic,slater1963quantum} (MO) description, while the low-density limit would correspond to large internuclear separation where the electrons are localized around the nuclei, better described by a local valence bond\cite{heitler1927wechselwirkung,pauling1931nature,coulson1949notes,slater1963quantum} (VB) approach. While this is a useful conceptual link, it must be remembered that the homogeneous electron gas is a better description of metallic solids than it is of more inhomogeneous systems like molecules. In the next section, the contrast between solids and molecules with regard to strong correlation will be discussed before we turn our attention fully to molecular systems and examine the various measures used in wavefunction theory, density functional theory and quantum information theory.

From a wavefunction point of view, correlation strength is often associated with the nature of the reference function which is usually the HF state, as mentioned above. An interesting point that draws attention to the significance of the chosen reference is the observation that from a statistical point of view HF already contains some correlation introduced by antisymmetrization and the exclusion principle\cite{kutzelnigg1968correlation}. Following Kutzelnigg, Del Re and Berthier\cite{kutzelnigg1968correlation}, two variables may be called uncorrelated if
\begin{equation}
\langle \mathbf{r}_1 \cdot \mathbf{r}_2\rangle_{ij} =
\langle \mathbf{r}_1 \rangle_{i} \langle \mathbf{r}_2\rangle_{j},
\label{uncorr}
\end{equation}
where the quantities corresponding to the $i$th and $j$th particles in an $N$-electron system can be simply related to the usual expectation values as
\begin{equation}
\langle \mathbf{r}_1 \rangle = N\langle \mathbf{r}_1 \rangle_{i}, \quad
\langle \mathbf{r}_1 \cdot \mathbf{r}_2\rangle = \frac{N(N-1)}{2}\langle \mathbf{r}_1 \cdot \mathbf{r}_2\rangle_{ij}.
\end{equation}
Applying this to position vectors, the implication is that the positions of two electrons are uncorrelated. This is a weaker notion than that of independence, which is often taken to mean lack of correlation. Statistically, independence (or, correlation in the sense of dependence) could be defined\cite{kutzelnigg1999cumulant} by requiring that the two-body cumulant $\lambda_2$
\begin{equation}
\lambda_2 (1,1';2,2') = \gamma_2(1,1';2,2') - \gamma_1(1,1')\gamma_1(2,2')
\end{equation}
obtained from the one- and two-body reduced density matrices (1- and 2-RDM), $\gamma_1$ and $\gamma_2$, should vanish. The fulfilment of this condition would imply a relationship like Eq.~\eqref{uncorr} for any observable. This is the sense in which Wigner and Seitz\cite{wigner1933constitution,wigner1934constitution} understood correlation, and this would correspond to a simple Hartree product of orbitals as a reference function, or, more generally, to a direct product of subsystem wavefunctions. Antisymmetrization then implies switching from this product reference function to an antisymmetric Slater determinant constructed from spin-orbitals. More generally, it is also possible to use generalized (antisymmetrized/wedge) product functions\cite{mcweeny1959density} to obtain the total wavefunction from antisymmetric group functions of subsystems. Using an antisymmetrized reference, it is customary to talk about Fermi and Coulomb correlation, the former being excluded from the conventional definition of the correlation energy. To make this explicit, the cumulant can be rewritten as
\begin{equation}
\lambda_2(1,1';2,2') = \gamma_2(1,1';2,2') - \gamma_1(1,1')\gamma_1(2,2') + \gamma_1(1,2')\gamma_1(2,1'),
\label{cumulant}
\end{equation}
Here, the last term accounts for the exchange contribution, and thus, $\lambda_2$ only contains the parts of $\gamma_2$ that cannot be factorized in terms of $\gamma_1$ in any way. In this sense, $\lambda_2$ is the most general descriptor of correlation effects. It should also be noted that by using partial trace relations\cite{kutzelnigg1999cumulant}, it can be shown that $\text{Tr} (\lambda_2) = \text{Tr}(\gamma_1^2-\gamma_1)$, and hence correlation effects are also related to the idempotency of the 1-RDM. This notion goes beyond that of Wigner and Seitz and is closer to L\"owdin's idea of correlation. The significance of Fermi correlation in particular was discussed by several authors in the past\cite{mcweeny1967nature,kutzelnigg1968correlation,tew2007electron}, and more recently in a pedagogical manner by Malrieu, Angeli and co-workers\cite{giner2016fermi}. Nevertheless, as pointed out by Kutzelnigg and Mukherjee\cite{kutzelnigg1999cumulant}, Eq.~\eqref{cumulant} is not identical to L\"owdin's definition either since it contains no reference to the Hartree--Fock state, apart from that fact that the closed-shell HF determinant, for example, would be enough to make $\lambda_2$ vanish and would thus be uncorrelated in this sense. The definition in Eq.~\eqref{cumulant} is an intrinsic one in the sense that $\gamma_1$ and $\gamma_2$ can be calculated for any (reference) state. More generally, any second-quantized operator string can also be evaluated with respect to a general reference (vacuum) state using generalized normal order and the associated general form of Wick's theorem that itself relies on n-body cumulants\cite{kutzelnigg1997normal,evangelista2022automatic}.

The foregoing discussion indicates the role of the reference state in the definition of correlation effects and highlights the fact that reference states should conveniently incorporate certain correlation effects to make subsequent treatment easier. As antisymmetrization dictates using Slater determinants rather than simple orbital products, spin symmetry may be enforced by demanding that the reference be constructed as a fixed linear combination of determinants to account for spin symmetry\cite{tew2007electron}, and there might be other criteria\cite{tsuchimochi2009strong,chan2012low,kutzelnigg2012separation}. In such cases, one may distinguish between correlation recovered by the reference and residual effects. From such an operational point of view, the question is whether the strongest correlation effects can already be captured at the reference level to ensure a qualitatively correct starting point. In practice, one may choose to build reference states and represent the wavefunction using the following $N$-electron basis states:
\begin{itemize}
\item{
\textbf{Determinant (DET):} An antisymmetrized product of orbitals. They are eigenfunctions of $\hat{S}_z$ but not necessarily of $\hat{S}^2$.
}
\item{
\textbf{Configuration State Function (CSF):} A function that is an eigenstate of both $\hat{S}_z$ and $\hat{S}^2$. CSFs can be obtained from linear combinations of determinants, but there are at most as many CSFs as DETs for a given number of unpaired electrons and total spin.
}
\item{
\textbf{Configuration (CFG):} At an abstract level, a configuration is just a string of spatial orbital occupation numbers (0,1,2) for each orbital in a given $N$-electron wavefunction. By extension, a configuration is also the set of determinants or CSFs that share the same spatial occupation numbers, but differ in the spin-orbital occupation numbers. 
}
\end{itemize}
We will discuss the role of this choice and offer a classification of wavefunction expansions and potential reference states based on the properties of the exact solution, which leads us to define what we call the multireference character of the wavefunction. We will also examine the role of the orbital basis and ask the question how orbital rotations influence this apparent multireference character in a simple model system. On the other hand, we are not concerned here with the practical construction of good reference states, even though the topics discussed undoubtedly have consequences in that regard. At this stage, we simply ask what the difference is between a canonical and a local orbital representation in terms of the apparent multireference character and what the chances are of incorporating the strongest correlation effects into the reference function, be that via enforcing spin symmetry or generating a new many-electron basis by rotating orbitals.

\section{Solids versus Molecules}

Strong correlation is one of those important but elusive concepts that is often used in discussions of the properties of molecular and solid-state systems.  A brief survey of the literature that follows below will reveal that it is often equated with various other notions, such as the multireference character of the wavefunction, the failure of density functional theory (DFT) and quantum entanglement. It is not immediately clear how these various notions are related to one another or to strong correlation, or whether any one should be preferred over the other. On the other hand, it seems relatively clear that a large part of the discussion on strong correlation focuses on extended solid-state systems, where density functional and dynamical mean-field\cite{georges_1992, georges_1996} approaches are commonly used\cite{anisimov2010electronic}, although many-body methods\cite{kuramoto2020quantum} and quantum Monte Carlo techniques\cite{foulkes2001quantum, shuai_2018} are also available as well as a number of exactly solvable model problems\cite{korepin1994exactly}. The theoretical tools used to describe periodic solids of infinite extension and finite-size molecular systems can be quite different in ways that reflect the underlying physics. Complexity and locality may have different manifestations for these systems in terms of the number of parameters or the type of basis sets required to describe them. In terms of symmetry, the obvious difference is the presence of translational symmetry in solids. Furthermore, symmetry for an infinite number of degrees of freedom also makes spontaneous symmetry breaking possible in ground-state solutions\cite{gross1996role}. For all these reasons, it seems reasonable to separate the discussion of electron correlation in solids from that in molecules.

As mentioned before, in the homogeneous electron gas model strong correlation effects are related to the low density limit. Wigner argued that at this limit the electrons are localized at lattice points that minimize the potential energy\cite{wigner1934interaction} and vibrate around the lattice points\cite{wigner1938effects} with just the zero point energy demanded by the uncertainty principle. Leaving the vibrational part aside, the total energy of the system at very low densities has the form\cite{wigner1938effects,raimes1963wave,march1996electron,fulde1999solids}
\begin{equation}
E = \langle\hat{T}\rangle + \langle\hat{V}\rangle,\quad
\langle\hat{T}\rangle = \frac{\alpha}{r^2},\quad
\langle\hat{V}\rangle = \frac{\beta}{r},
\label{EHEG}
\end{equation}
where $\alpha$ and $\beta$ are constants and $r$ is the Fermi radius, which in this context is the radius of a sphere available for a single electron. If the density is low, $r$ is large, and that means that the mean kinetic energy $\langle\hat{T}\rangle$ is negligible compared to the mean potential energy $\langle\hat{V}\rangle$, which is sometimes taken to be a criterion for strong correlation. Wigner was able to obtain a correlation energy formula by assuming a vanishingly small kinetic energy at the low-density limit\cite{wigner1934interaction} and interpolating between the low- and high-density limits, and considered the zero point kinetic energy, which would add a term proportional to $r^{-\frac{3}{2}}$ in Eq.~\eqref{EHEG}, only in a later study\cite{wigner1938effects,coldwell-horsfall1960zero,loos2016uniform}. While the result works reasonably at density values actually present in some solids, some criticisms merit attention especially when application to molecules are concerned. While admitting the good agreement with measurements in metallic systems and similar results derived from plasma theory by Bohm and Pines\cite{bohm1951collective,pines1956electron}, Slater criticized\cite{slater1956electronic} the assumption of uniform positive background, which should be an even more serious problem when considering more inhomogeneous systems, such as the dissociating H$_2$ molecule\cite{lowdin1958correlation}. We note in passing that it is possible to account for inhomogeneities in the electron gas\cite{brueckner1969correlation}, but perhaps the conceptual simplicity of the homogeneous electron gas model is more relevant for the current discussion. Concerns were also raised by L\"owdin\cite{lowdin1958correlation} about the model not obeying the virial theorem\cite{slater1933virial} in the sense expected in molecular systems. It has since been shown that the uniform electron gas obeys the more general form of the virial theorem\cite{march1958kinetic,argyres1967virial},
\begin{equation}
2\langle\hat{T}\rangle + \langle\hat{V}\rangle = -r\frac{\mathrm{d} E}{\mathrm{d} r},
\end{equation}
which means that in the low-density limit, the total energy $E$ is not stationary (although the potential energy may be in the sense needed to find an optimal lattice). In molecular systems, on the other hand, stationary points of the total electronic energy are of special interest as in the case of H$_2$ at equilibrium and infinite separations. In these cases, the virial theorem takes the form $2\langle\hat{T}\rangle + \langle\hat{V}\rangle = 0$ which does not allow the kinetic energy to be vanishingly small\cite{lowdin1958correlation}. In this regard, see also the work of Ruedenberg\cite{ruedenberg1962physical} for an analysis of the formation of the chemical bond that does respect the virial theorem. For the electron gas, the main use of the virial theorem is actually the separation of kinetic and potential energy contributions in the correlation energy\cite{march1958kinetic}. Thus, it may be concluded that the criterion of the relative magnitude of the mean kinetic and potential energy is appropriate within the homogeneous electron gas model, which is a better approximation of metallic solids than molecules. Nevertheless, extensions of the same basic idea are possible. Within the Hubbard model\cite{hubbard1963electron}, for example, correlation effects are commonly considered as being strong if the on-site electronic repulsion is larger than the bandwidth (resonance energy) or the hopping term, which favours antiferromagnetic coupling between the various sites\cite{fulde1995electron}. Repulsion is a potential energy term, while the hopping integral has to do with the kinetic energy of electrons. We will discuss the appropriate extension to the molecular Hamiltonian in later sections.

As in this perspective we are more concerned with molecular systems,  it is especially interesting that solids containing $d$ or $f$ electrons are considered strongly correlated\cite{fulde1995electron}, as this would prompt one to look for strong correlation in molecular systems containing transition metals or lanthanides. Before leaving the discussion of solid state behind, it should also be recalled that sometimes the nature of the problem at hand calls for the combination of solid-state and molecular approaches, as in studying strongly correlated (bond-breaking) problems in surface chemistry via embedding approaches, which pose challenges of their own\cite{janesko2017strong}. The foregoing discussion also underlines the problem of finding quantitative measures of strong correlation, or, perhaps more precisely, of the various other properties associated with it. After reviewing the various attempts made along these lines, we will expound our own perspective on strongly correlated systems in chemistry as viewed through the lens of the multireference character of the wavefunction expansion. We will use a simple metric that can be obtained from the exact solution and examine its properties using the hydrogen molecule as a model system. As mentioned before, this problem has been studied since the early days of quantum chemistry\cite{lowdin1958correlation,kotani1961quantum,slater1963quantum,kutzelnigg2002einfuhrung} and also has some practical relevance in that it can be taken as a model for the antiferromagnetic coupling in transition metal compounds such as a typical Cu(II) dimer. We note in passing that the hydrogen chain is also a common model problem in the study of extended systems\cite{motta2017towards,motta2020ground}.

\section{Wavefunction Theory}

Within the framework of wavefunction-based post-Hartree--Fock approaches, it is customary to distinguish between single and multireference methods to which the concepts of dynamic and static or non-dynamic correlation also correspond\cite{sinanouglu1963many,sinanoglu1964many,knowles2000ab,bartlett1994applications}. Dynamic correlation is also called short-range correlation as it has to do with the electrostatic repulsion of electrons and the difficulty of describing the resulting Coulomb cusp (see this study\cite{prendergast2001impact} for a different perspective). When it comes to other correlation effects,  some authors use static and non-dynamic correlation interchangeably and sometimes refer to it as long-range correlation\cite{knowles2000ab,mok1996dynamical}, others\cite{bartlett1994applications} prefer to reserve static correlation for effects induced by enforcing spin symmetry in order to construct a correct zeroth order description, and call those associated with other sources, such as bond breaking, non-dynamical.  Tew, Klopper and Helgaker follow a similar distinction except that they include spin symmetry in Fermi correlation and identify static correlation as the correlation associated with near degeneracy and thus needed for a correct zeroth order description\cite{tew2007electron}. Yet others\cite{hollett2011two} distinguish within static/non-dynamical effects between those that can be captured by spin-unrestricted HF and those that cannot. A summary of more formal attempts to define static and dynamic correlations is also available\cite{benavides2017towards}. The important conclusion from this discussion seems to be a point already raised by Wigner and Seitz\cite{wigner1934constitution} that the main difference is between correlation effects that arise from symmetry and those arising from interelectronic repulsion, regardless of how we choose to classify these further. Thus, within symmetry effects one might further distinguish between antisymmetry and spin symmetry, within repulsion effects between those associated with spatial (near) degeneracy and the rest.

Ignoring the question of spin symmetry for the moment, non-dynamic/static correlation may be identified with strong correlation in the present context, and thus, it is also associated with multireference methods, where `reference' is a generic term for some function that serves as an initial state on which a more sophisticated Ansatz can be built. This also suggests that there are two ways of dealing with static correlation, either improving the reference or improving the wavefunction Ansatz built on it. In principle, single reference methods will yield the exact solution if all possible excitation classes are included in the Ansatz, and it was furthermore demonstrated that they can also be adapted to strong correlation by removing some of the terms from the working equations\cite{bulik2015can}. More pragmatically, one might ask at what excitation levels an Ansatz, such as coupled cluster (CC), needs to be truncated to yield acceptable results for some properties\cite{karton2006w4,hait2019levels}, but this is costly on the one hand and does not offer a simple picture of strong correlation either. One way to improve the reference is to find the exact solution within at least a complete active space (CAS)\cite{roos1980complete}. Unfortunately, this requires the construction of many-electron basis states the number of which scales exponentially with the size of the active space. However, as observed by Chan and Sharma\cite{chan2011density}, this apparent exponential complexity does not always reflect the true complexity of the system since the principle of locality often forces a special structure on the wavefunction expansion. This entails that the amount of information needed to construct the wavefunction should be proportional to the system size, which agrees well with the observation that static correlation in molecules can typically be tackled with only a handful of terms in the wavefunction expansion\cite{knowles2000ab}. As an illustration, the hydrogen molecule around the equilibrium bond length is well-described by a single-determinant built from canonical orbitals and associated with the configuration $\sigma^2_g$, while at larger separations another contribution from $\sigma^2_u$ becomes equally important\cite{chan2011density,knowles2000ab}. In the latter limit, the wavefunction takes the following bi-determinant form
\begin{align}
\Psi_0 
&= \frac{1}{\sqrt{2}}(\lvert i\bar{i}\rvert - \lvert a\bar{a}\rvert) \nonumber \\
&= \frac{1}{\sqrt{2}}(\lvert\mu\bar{\nu}\rvert - \lvert\bar{\mu}\nu\rvert),
\label{SState}
\end{align}
where $i$ and $a$ are the occupied and unoccupied molecular orbitals in the minimal basis HF solution and the shorthand notation $|i\bar{i}|$ is used to denote determinants, with the overbar signifying a $\beta$ spin. While these canonical orbitals are delocalized, $\mu$ and $\nu$ denote atomic orbitals, each localized on one of the hydrogens and orthonormal at large distances. As is well-known\cite{kotani1961quantum,kutzelnigg2002einfuhrung,mouesca2014density}, the molecular orbital (MO) and valence bond (VB) approaches offer equivalent descriptions of this wavefunction, but the locality of the VB picture leads to an interpretation in terms of ``left-right'' correlation, i.e., the phenomenon that at large internuclear separation an electron preferably stays close to the left nucleus if the other one is close to the right nucleus, and \emph{vice versa}. This form of static correlation can be accounted for in a spin-unrestricted single-determinant framework leading to different orbitals for different spins beyond a certain bond distance, but, for molecular systems, approximations preserving the symmetries of the exact wavefunction are usually preferable. It is all the more interesting that the constrained-pairing mean-field theory of Scuseria and co-workers is designed to capture strong correlation\cite{tsuchimochi2009strong} by breaking and restoring various symmetries\cite{scuseria2011projected}, while Kutzelnigg took a different route towards obtaining a similarly economic reference function by separating bond-breaking correlation effects via a generalized valence bond (GVB)/antisymmetrized product of strongly orthonormal geminals (APSG) approach\cite{kutzelnigg2012separation}. 

It is also apparent from these considerations that most traditional methods follow the ``bottom-up'' approach in which a systematic Ansatz is built on a simple reference function and then truncated using some simple criterion (e.g., the excitation level). The typical arsenal of this approach includes many body perturbation theory, CI and CC approaches in both the single and multireference cases\cite{bartlett1994applications,knowles2000ab,helgaker2014molecular}. The most popular of them is CC theory\cite{shavitt2009many}, which was introduced into chemistry by \v{C}i\v{z}ek\cite{cizek1966correlation}, Paldus and Shavitt\cite{paldus1972correlation}, was reformulated in terms of a variational principle by Arponen\cite{arponen1983variational} and whose mathematical properties are still an active area of research\cite{faulstich2019analysis,csirik2021coupled}. Even more actively studied are its local variants\cite{schutz2001low,riplinger2013efficient} and its explicitly correlated\cite{shiozaki2009explicitly}, excited state\cite{izsak2020single} and multi-reference\cite{lyakh2012multireference} extensions. Beyond these methods, there is also room for a ``top-down'' approach in which one aims directly at the exact solution (full CI, full CC) and applies some strategy to neglect contributions that are less important while trying to save as much of the accuracy as possible. The relative sparsity illustrated above with the H$_2$ example can be exploited in selected CI methods of the CIPSI (configuration interaction with perturbative selection through iterations) type\cite{huron1973iterative} for much larger systems. The CIPSI method is the first of many selected CI (SCI) approaches that attempt to solve the problem of selecting relevant contributions in a deterministic fashion, but it is also possible to apply stochastic sampling processes, including QMC approaches\cite{austin2012quantum} such as the FCIQMC method\cite{booth2009fermion,petruzielo2012}. FCIQMC can be performed without approximation, in which case the method is exact but suffers from a sign problem in general\cite{spencer2012}. Instead, the initiator approximation is typically applied, which involves truncating contributions between certain determinants with low weight\cite{cleland2010, cleland2011}. These and other methods have been reviewed in more detail recently by Eriksen\cite{eriksen2020shape}, including Eriksen and Gauss’ own approach based on many body expansions\cite{eriksen2018many,eriksen2019many}. Another approximate FCI method particularly suitable for strongly correlated systems, the density-matrix renormalization group\cite{white1992density,chan2011density} (DMRG) approach will be discussed later among entanglement-based approaches. We also note that some of these methods have been analysed in terms of scaling and the gains due to the reduction in the number of important contributions in the wavefunction expansion\cite{chilkuri2021comparison,li2020compression,manni2021modeling,li2021resolution} (see also some remarks on orbital rotations below).

It has been observed that multireference character, interpreted here as the fact that the wavefunction expansion contains more than one determinant of significant weight for some system, does not necessarily imply strong correlation, and the additional criteria of the breakdown of single reference perturbation theory and the appearance of degeneracies were suggested\cite{shee2021revealing}. More explicitly, strong correlation could be said to occur whenever the perturbation (the difference between the full Hamiltonian and the Fockian) is larger than the gaps between the Hartree--Fock orbital energies (spectrum of the Fockian)\cite{chan2012low}. This could be regarded as an extension of the similar definition within the Hubbard model. For non-metallic states, strong correlation would also be implied if the correlation energy is larger than the excitation energy to the first excited state of the same symmetry as the ground state\cite{benavides2017relating}. Nevertheless, the multireference character can be used to characterize the exact wavefunction, even if its usefulness for approximate approaches may be questioned. This leads us towards a definition of multireference character via the full configuration interaction expansion,
\begin{equation}
\Psi = \sum_I C_I\Phi_I,
\label{FCI}
\end{equation}
where $\Phi_I$ are some basis states, typically determinants. If any one coefficient $C_I$ dominates (see more later) the expansion, then the associated $\Phi_I$ could be taken as the reference in a single reference (determinant) treatment. Thus, multireference character is associated with wavefunction coefficients $C_I$ which for orthonormal basis states are equivalent to the overlap between the exact state $\Psi$ and the basis state $\Phi_I$. Unfortunately, this criterion is impractical to decide \emph{in advance} whether a multireference treatment of a system is necessary, and other diagnostics\cite{jiang2012multireference}, such as the one relying on the norm of the singles vector ($T_1$) in a CC calculation\cite{lee1989theoretical} truncated at the level of single and double excitations, were also found insufficient\cite{jiang2012multireference} or even misleading\cite{liakos2011interplay}. Metrics relying on natural orbital populations are more promising as they give information about the number of electrons uncoupled by correlation driven processes\cite{head2003characterizing,ramos-cordoba2017local,hait2019levels}. The deviation of the one-body reduced density matrix from idempotency has also been used as a measure\cite{lowdin1955quantum,ramos2016separation} and well as measures associated directly with the two-body cumulant\cite{juhasz2006cumulant,kutzelnigg2012separation,ramos2016separation}. More pragmatically, the static correlation energy has also been defined as the exact correlation energy obtained from a minimal basis calculation involving valence electrons, the idea being that in such a calculation the most important configurations would be represented, although this approach also needs a hierarchy of approximations to be useful\cite{crittenden2013hierarchy}. A few other schemes have also been suggested to decompose the correlation energy into dynamical and non-dynamical parts using simple formulae\cite{mok1996dynamical,ramos2016separation}, and more recently, using cumulants\cite{ramos2016separation}.

For the purposes of this perspective, the FCI-based criterion in Eq.~\eqref{FCI} will be sufficient as it enables us to base our discussion on the properties of the exact wavefunction. While it is common to use determinants as basis states, this is not the only possibility. Using CSFs or CFGs for example would yield basis states that can be written as sums of determinants or constructed in some other way\cite{pauncz2012spin,chilkuri2021comparison}. We will investigate their effects on the multireference character in a later section. Already in Eq.~\eqref{SState}, while both the MO and VB descriptions are multi-determinant expressions, multiple configurations only enter into the (canonical) MO picture. We will also examine the role of orbital transformations in defining the multireference character of the wavefunction. Again, in Eq.~\eqref{SState}, the concept of left-right correlation seems to be connected more to the local VB picture than to the MO one. Our primary goal here is to comment on multireference character, rather than other notions about strong correlation, but before doing that, we briefly examine other approaches to the problem.

\section{Density Functional Theory}

In their recent paper, Perdew et al. remark that strong correlation is sometimes taken to mean ``everything that density functional theory gets wrong''\cite{perdew2021interpretations}. Most practical DFT approaches rely on the Kohn--Sham procedure which in most cases corresponds to a single-determinantal implementation. Unlike the Hartree--Fock solution however, the Kohn--Sham determinant is not an approximation to the exact wavefunction, rather, it is a tool to obtain the ground-state density or spin densities. There seem to be two opinions in the literature on what this signifies. Perdew et al. argue that as a consequence, the Kohn--Sham determinant need not exhibit the symmetries of the Hamiltonian as those will be reflected in the density in a more indirect fashion. This also implies that despite the appearance of a mean-field approach, Kohn--Sham DFT is in principle able to account for correlation effects\cite{perdew2009some}. Others argue\cite{filatov1998spin,gagliardi2017multiconfiguration,cremer2001density} that spin-adapted reference functions should be used at least in some open-shell cases which may be multi-determinantal, as it is done in the spin-adapted Hartree--Fock case in wavefunction theory. In any case, a single determinant may not always suffice and the question remains to what extent DFT can treat strongly correlated systems. In trying to answer this question, Perdew et al.\cite{perdew2009some} distinguish between strong correlation arising in condensed matter models and static correlation originating in near or exact degeneracies, such as the closing HOMO-LUMO gap in the H$_2$ bond dissociation process. On problems of the latter type, they note the typical failure of common semilocal approximations. 

While multi-determinental techniques have been introduced in DFT\cite{filatov1998spin,gagliardi2017multiconfiguration,cremer2001density}, it is more common to rely on broken symmetry DFT\cite{ziegler1977calculation,noodleman1981valence,daul1994density,mouesca2014density} in which multiplet energies are obtained from a combination of single-determinant energies. This scheme may or may not hold exactly, depending on the specific case\cite{daul1994density,filatov1998spin}. Perhaps the simplest example of spin-symmetry breaking is the fact that the spin-unrestricted approach yields only one of the determinants $\lvert\mu\bar{\nu}\rvert$ or $\lvert\bar{\mu}\nu\rvert$ in Eq.~\eqref{SState} which can be interpreted as a resonance structure in the valence bond sense. Although such determinants are not pure spin states, they are degenerate. From such broken symmetry solutions and mono-determinantal pure spin states, it is often possible to calculate the energies of multi-determinantal spin states: in H$_2$, the appropriate combination of the energy of the broken symmetry solution and the triplet energy yields the energy of the excited-state singlet\cite{mouesca2014density}. The use of this simple two-level model in the description of magnetic coupling is analysed in detail elsewhere\cite{neese2004definition}. The more general broken symmetry (BS) DFT procedure starts with a mixture of canonical orbitals obtained from a closed-shell calculation and optimizes the mixing angles during the iterative solution to obtain a mixture of multiple configurations\cite{cremer2001density}. A similar idea underlies the permuted orbital (PO) approach where the broken symmetry states are generated by permuting orbitals, typically the HOMO and the LUMO\cite{cremer2001density}, and can be used to construct multi-determinant spin eigenfunctions. Based on the capabilities of the BS and PO approaches, Cremer proposes a DFT-oriented multireference typology\cite{cremer2001density}: Type 0 wavefunctions need a single configuration or determinant and the spin-unrestricted approach can handle them; Type I consists of a single configuration with multiple determinants for which the PO approach is assigned; Type II is the case of multiple configurations each with a single determinant to be treated by BS-DFT; Type III systems with multiple configurations each with multiple determinants are the most difficult and can be handled by explicitly multireference methods, although possibly the PO approach might work. While we find the distinction between determinant and configuration/CSF references commendable, in the cases studied by Cremer there was no need to distinguish between configurations and CSFs, a feature that we will come back to in our own classification. As for interpreting broken symmetry solutions, it should be recalled that while the exact functional for finite systems is not symmetry broken, approximations are. Strong correlation in the exact wavefunction describes ``fluctuations'' in which the spins on the different nuclei are interchanged ($\lvert\mu\bar{\nu}\rvert$ vs. $\lvert\bar{\mu}\nu\rvert$)\cite{gunnarsson1976exchange}. Symmetry breaking in approximate methods is in fact taken to reveal strong correlations\cite{perdew2021interpretations} with a mechanism associated with it: at large distances, when H$_2$ becomes a pair of H atoms, an infinitesimal perturbation can reduce the wavefunction and the spin densities to a symmetry broken form, in which an electron with a given spin is localized around one H atom\cite{perdew2021interpretations,gunnarsson1976exchange}. All this is in line with Anderson's famed essay\cite{anderson1972more} on emergent phenomena on different scales involving increasingly larger numbers of particles, though it also implies that symmetry breaking may be more relevant for solids than for molecules\cite{gross1996role,perdew2021interpretations}.

It may be worth pointing out that the physical contents of broken symmetry wavefunctions can be conveniently ``read'' by means of the corresponding orbital transformation (COT)\cite{amos1961single}. This transformation orders the spin-up and spin-down orbitals into pairs of maximum similarity. This results in a wavefunction that can be read analogously to a generalized valence bond (GVB) wavefunction\cite{ladner1969improved,bobrowicz1977self} consisting of doubly occupied orbitals, singlet coupled electron pairs in two different orbitals and unpaired (spin-up) electrons. Obviously BS-DFT can properly represent the singlet coupled pairs, but the analysis shows that the spatially non-orthogonal pseudo-singlet pairs obtained from the COT serve a similar purpose, namely to variationally adjust the ionic- and neutral components of the wavefunction\cite{neese2004definition}. This way of reading BS wavefunctions has found widespread application in the field of metal-radical coupling (e.g. examples\cite{ghosh2003noninnocence,herebian2003analysis,chlopek2007electronic}). A more thorough review is available\cite{neese2009prediction}.

The theorems due to Hohenberg and Kohn\cite{hohenberg1964inhomogeneous} form the basis of the variational formulation of DFT, although questions about their precise meaning\cite{kutzelnigg2006dft} remain pertinent enough to motivate further research\cite{lieb2002density,penz2022structure}, not to mention the fact that these theorems do not generalize easily to excited states\cite{casida2012progress}. The favorable linear scaling property of DFT approaches is often derived from Kohn's principle of nearsightedness\cite{kohn1996density,prodan2005nearsightedness}, i.e., that under certain conditions the density at some point is mostly determined by the external potential in a local region around that point. Practical approaches to density functional theory often consist of proposing new functional forms\cite{li2020recent} in lieu of the unknown exact density functional of the energy. In contrast, the energy is a known functional of the 2-RDM\cite{mayer1955electron,lowdin1955quantum,bopp1959ableitung}, although enforcing constraints\cite{mazziotti2012structure} to actually optimize this functional is not trivial. Fortunately, the energy can also be written as a functional of the 1-RDM\cite{gilbert1975hohenberg}, and different variants of density matrix functional theory are usually parametrized in terms of the 1-RDM and the 2-body cumulant\cite{muller1984explicit,kutzelnigg2006density,piris2014perspective,pernal2016reduced}. Like DFT, this approach also relies on approximate functionals, but it manages to overcome some of the failings of the former. The appropriate constrains for the 1-RDM itself\cite{coleman1963structure} are relatively easy to enforce, although the reconstructed 2-RDM should then still satisfy its own $N$-representability conditions\cite{piris2014perspective}. One might then construct functionals\cite{cioslowski2003approximate} inspired by wavefunction approaches, such as geminal theories, or use auxiliary quantities to parametrize the cumulant\cite{piris2013natural}. In density cumulant functional theory\cite{kutzelnigg2006density,simmonett2010density} (DCFT), the functional is given in terms of the best idempotent estimate to the 1-RDM and the 2-body cumulant, and the critical $N$-representability conditions affect only the relatively small part due to the latter. Connections with traditional DFT can also be established, e.g., in the DCFT case, an analogue of the DFT exchange-correlation functional can be obtained by neglecting the Coulomb terms in which the non-idempotent part of the 1-RDM occur\cite{kutzelnigg2006density}. The nearsightedness principle applies to the 1-RDM as well\cite{kohn1996density}, and density matrix functionals have the advantage over DFT that the kinetic, Coulomb and exchange energies are all known functionals of the 1-RDM, only the correlation energy functional remains unknown\cite{kutzelnigg2006dft-aiqc}. Excited state applications of RDM functionals have also been considered\cite{pernal2016reduced,copan2018linear}. A collection of studies on the various aspects of RDM-based methods is available\cite{cioslowski2000many}, while the relationship between wavefunction based methods, DFT and RDM functionals is discussed in Kutzelnigg’s review\cite{kutzelnigg2006dft-aiqc}. In the current context, it is relevant to mention that some density matrix functionals have been applied with success to strongly correlated problems\cite{gidofalvi2008active,piris2013natural,pernal2016reduced,mullinax2015can,lathiotakis2014local,piris2017global,gibney2022density}.

So far we have only considered functional approaches in which the variable is the density or the RDMs, but it is also possible to set up a variational formulation\cite{martin1959theory,luttinger1960ground,baym1961conservation} for functionals of Green's functions, which themselves can be regarded as generalizations of RDMs with their own cumulant expansion\cite{ziesche2000cumulant}. All three of these quantities form the basis for embedding theories enabling the treatment of large systems\cite{sun2016quantum}. Dynamical mean-field theory\cite{georges_1992, georges_1996,kotliar2006electronic} is a much applied method in solid state physics that can be derived from such a variational approach and has also been investigated from the perspective of molecular quantum chemistry\cite{lin2011dynamical,zgid2011dynamical}. Another method that falls into this category is the popular GW approximation of Hedin\cite{hedin1965new,aryasetiawan1998gw,reining2018gw}. The properties of these and other similar methods have also been analysed from the perspective of strong correlation\cite{kotliar2006electronic,zgid2011dynamical,phillips2014communication,pokhilko2021interpretation}, excited states\cite{aryasetiawan1998gw,blase2018bethe} and symmetry breaking\cite{pokhilko2021interpretation,pokhilko2021evaluation}. Although both have some success in multi-reference systems, GW approaches are typically applied to weakly-correlated systems while dynamical mean-field approaches have succeeded where DFT or other Green's function approaches failed\cite{kotliar2006electronic,zgid2011dynamical,phillips2014communication}.

It emerges from this discussion that the DFT concept of strong correlation is related on the one hand to its mono-determinant nature and to symmetry breaking on the other. Measures have been proposed along these lines. Thus, Yang and co-workers relate static correlation to fractional spins, and present an extension to DFT covering fractional spins including a measure based on this extension\cite{cohen2008fractional,cohen2008insights}. Grimme and Hansen offer a measure as well as a visualisation tool for static correlation effects based on fractional occupation numbers obtained from finite temperature DFT calculations\cite{grimme2015practicable}. Head-Gordon and co-workers use spin (and sometimes spatial) symmetry breaking as a criterion and suggest a measure based on spin contamination\cite{shee2021revealing}, which may be related to natural orbital occupation numbers, a criterion also used by others\cite{cremer2001density}. Still other authors use checks based on the idempotency of the one-body density, which is essentially used here as a check on the mono-determinant nature of the wavefunction under consideration\cite{lowdin1955quantum,muechler2022quantum}.

\section{Quantum Information Theory}

Quantum information theory offers a different perspective on the same phenomenon, albeit one that can be related to more traditional quantum chemical concepts\cite{benavides2017relating}. The starting point is a two-level quantum system, a qubit, that serves as an analogue of classical bits. We take spin-up and spin-down states to form the qubit in the following. A spin-up state can be labelled $\lvert 0\rangle$ and a spin-down state as $\lvert 1\rangle$, and a qubit can be any linear combination of these two basis states\cite{nielsen2010quantum}, $|m\rangle=\alpha\lvert 0\rangle + \beta\lvert 1\rangle$. If the spin of an electron prepared in such a state is measured in a field oriented along an axis parallel to the spin-up and spin-down basis states, then the probability of finding the electrons in state $\lvert 0\rangle$ or $\lvert 1\rangle$ is $|\alpha|^2$ and $|\beta|^2$, respectively. Along the same lines, one may prepare two electrons. Assuming at first that the two electrons do not interact before the measurement, the possible two-electron states may be described in the basis of the product states $\lvert 0\rangle\otimes\lvert 0\rangle$, $\lvert 0\rangle\otimes\lvert 1\rangle$, $\lvert 1\rangle\otimes\lvert 0\rangle$ and $\lvert 1\rangle\otimes\lvert 1\rangle$. They remain distinguishable and they are also independent in the sense that measuring the first qubit yields results that only depend on how the first electron was prepared and is not influenced by the second electron. Imagine now bringing spin-1/2 particles close enough together so that their spins align in an antiparallel fashion corresponding to the ground state and then separating them again far enough so that one is located at site $A$ the other at site $B$. The wavefunction that describes this state is 
\begin{equation}
|\Psi_0\rangle = \frac{1}{\sqrt{2}}(\lvert0\rangle_A\otimes\lvert1\rangle_B - \lvert1\rangle_A\otimes\lvert0\rangle_B).
\label{2Q}
\end{equation}
In classical computing, the value of the first bit in a two-bit string does not yield information about the value of the second bit, in the same way as the independently prepared quantum systems do not. The two-qubit quantum state in Eq.~\eqref{2Q} is different because measuring the spin only at site $A$ also reveals what the result of the measurement will be at site $B$. Thus, the two qubits at site $A$ and $B$ are entangled\cite{almeida2007introduction} in the sense that the result of a measurement at $A$ determines that at $B$. It is also often said that they are correlated, although entanglement is only a special kind of quantum correlation\cite{ding2022quantum}. This is different from the definition of strong correlation in terms of perturbation strength, but has the advantage that it is more easily quantifiable\cite{chan2012low}. Note that Eq.~\eqref{2Q} describes a system of distinguishable spin-1/2 particles, such as an electron and a positron, or two electrons at a large distance. Although electrons are identical particles, in cases like the dissociating H$_2$ molecule, each electron can be identified as the one localized around a specific nucleus if the two nuclei are separated enough so that the electronic wavefunctions no longer overlap\cite{omar2005particle}. Comparing Eq.~\eqref{2Q} to the VB wavefunction of H$_2$ at large bond distances in Eq.~\eqref{SState} and assuming that $\mu$ is at site $A$ and $\nu$ at site $B$ reveals that  
\begin{equation}
|\mu\bar{\nu}| = \frac{1}{\sqrt{2}}(
|0\rangle_A\otimes |1\rangle_B -
|1\rangle_B\otimes |0\rangle_A), \quad
|\bar{\mu}\nu| = \frac{1}{\sqrt{2}}(
|1\rangle_A\otimes |0\rangle_B -
|0\rangle_B\otimes |1\rangle_A),
\end{equation}
since the tensor product here is ordered w.r.t. the particle labels, e.g., $\mu(2)\alpha(2)\nu(1)\beta(1)\to |1\rangle_B\otimes |0\rangle_A$. Thus, Eq.~\eqref{SState} is the antisymmetrized version of Eq.~\eqref{2Q}. In a sense, this indicates another kind of symmetry breaking. In spin-unrestricted approaches discussed in the DFT context, only one of the determinants $|\mu\bar{\nu}|$ or $|\bar{\mu}\nu|$ in Eq.~\eqref{SState} are present at large bond distances, i.e., it is known which spin is at site $A$ (or $B$) but not which particle carries that spin. Eq.~\eqref{2Q} on the other hand breaks antisymmetry as a natural consequence of the assumption of distinguishability: in this case it is known which particle is at site $A$ (or $B$), but not the spin of that particle, which is the question a subsequent measurement can help resolve. As this discussion suggests, the established notion of entanglement\cite{nielsen2010quantum} assumes distinguishable subsystems, although extensions to indistinguishable particle systems exist\cite{benatti2020entanglement}. This should perhaps be kept in mind when comparing with quantum chemical notions, where indistinguishability of electrons in molecules is assumed (cf. also the discussion on product functions and generalized product functions\cite{mcweeny1959density} earlier). Still, the resonance structures of VB/GVB theories, for example, assume a degree of distinguishability and entanglement may be a useful tool in dealing with them\cite{chan2012low}. It is also worth mentioning that when measuring the spin of a single electron as discussed above, it is always possible to rotate the measurement field so that, say, the spin-up state is measured with probability 1. This is not possible for the singlet described in Eq.~\eqref{2Q}, since the expectation value of one of the spins along \emph{any} axis is zero. Due to this rotational invariance, the invariance of entanglement measures with respect to the choice of basis is often emphasized\cite{almeida2007introduction}, but this should not be confused with orbital rotations to be discussed below.

It may be said that the VB wavefunction is created by identifying the entangled local states (orbitals) associated with chemical bonds. In fact, just as traditional wavefunction approaches start from the independent particle model and build more accurate Ans\"atze by incorporating the most important excitations, one might start a similar hierarchy around independent subsystems as a starting point which explores the Hilbert space by encoding more entanglement\cite{chan2012low}. This leads to matrix product states (MPS) in general and the density matrix renormalization group (DMRG) approach in particular\cite{chan2012low}. The seniority-based CC theory of Scuseria and co-workers\cite{henderson2014seniority} can also be interpreted on a quantum information theory basis\cite{boguslawski2015orbital,ding2022quantifying}. This approach is built on orbital pairs rather than single orbitals with seniority being the number of unpaired electrons in a determinant that can be used to select the determinants for inclusion in a wavefunction expansion. The claim is that strong correlation can be treated by low-seniority contributions in a suitable orbital basis\cite{henderson2014seniority}, and it is possible to set up an entanglement-based measure to be used as a cost function\cite{ding2022quantifying}.

One measure of entanglement is based on the von Neumann entropy (the ``quantum Shannon entropy'') of a state and the quantum relative entropy\cite{nielsen2010quantum} which provides a measure of the distinguishability of two states\cite{ding2022quantum}. The relative entropy of entanglement can be given as\cite{chan2012low}
\begin{equation}
S=-\sum_i \sigma_i^2\log_2\sigma_i^2,
\end{equation}
in terms of the singular values $\sigma_i$ corresponding to a Schmidt decomposition\cite{nielsen2010quantum}: if $A$ and $B$ are distinguishable subsystems of the whole system, then a pure state $|\Psi\rangle$ can be expanded in the basis of functions of the type $|\Psi_A\rangle\otimes|\Psi_B\rangle$, 
\begin{equation}
|\Psi\rangle = \sum_{AB} C_{AB}|\Psi_A\rangle\otimes|\Psi_B\rangle,
\end{equation}
where $|\Psi_A\rangle$, $|\Psi_B\rangle$ are basis states in $A$ and $B$, and the singular values $\sigma_i$ are those of the expansion coefficient matrix $C_{AB}$. If only one singular value is nonzero, the entropy is zero and $|\Psi\rangle$ is separable, otherwise it is entangled, and it is maximally entangled if all singular values are the same\cite{chan2012low,almeida2007introduction}. Observe that entanglement has to do with the characterization of subsystems and depends on the partitioning of the system. For example, in Eq.~\eqref{2Q}, it is the two subsystems at sites $A$ and $B$ that are entangled because the total state cannot be written in the simple form $|m\rangle_A\otimes|n\rangle_B$. For further considerations on superselection rules and classical and quantum correlation types, see\cite{ding2022quantifying,ding2022quantum}. If the local states are chosen to be single orbitals, then we may speak about (single) orbital entropy\cite{legeza2003optimizing,boguslawski2015orbital} and use the eigenvalues of the orbital reduced density matrix instead of $\sigma_i^2$ in the definition of the entropy above, with the summation running over unoccupied, singly occupied spin-up and spin-down, and doubly occupied states. Orbital entropy has been used to measure multiconfigurational character\cite{stein2017measuring} and as a basis for automatic active space selection\cite{stein2016automated}. It has also been applied to the H$_2$ model system\cite{stein2017measuring} and to a number of other bond formation processes\cite{stein2017measuring,boguslawski2013orbital}. See also another entanglement-based study on 2-electron systems by Huang and Kais\cite{huang2005entanglement}. The dependence of entanglement-based measures on the orbital basis has been noted in the literature\cite{ding2022quantum,stein2017measuring} and they have also been used to set up a qualitative distinction between non-dynamic, static, dynamic and dispersion correlation effects\cite{boguslawski2012entanglement,boguslawski2015orbital} in the sense of Bartlett and Stanton\cite{bartlett1994applications}. In earlier approaches, natural orbital occupation numbers have also been substituted for $\sigma_i^2$ to calculate the von Neumann entropy as a measure of correlation strength\cite{ziesche1995correlation}. Recently, entropy-based measures have even served as a basis of a geometric interpretation and an upper bound to the correlation energy\cite{ghosh2022geometrical}. Finally, we note that the two-orbital entropy can be defined in an analogous manner, and serves as the basis for orbital interaction (mutual information) measures\cite{legeza2003optimizing,rissler2006measuring,boguslawski2015orbital} that can be used to find an orbital ordering that ensures good convergence in DMRG by placing strongly interacting ones near each other\cite{rissler2006measuring}.

So far we have discussed the application of quantum information theoretical concepts in quantum chemistry. However, these concepts also underlie the theory of quantum computing in which chemistry has also been identified as one of the potential areas of breakthrough\cite{cao2019quantum}, as exemplified by resource estimates in the study of Reiher et al. on FeMoco\cite{reiher2017elucidating}. The latter compound served as an example of a strongly correlated system that would especially benefit from the fact that quantum computers are quantum systems themselves that need not explicitly generate the FCI coefficients in Eq.~\eqref{FCI}. Thus, we will also briefly mention recent developments in this field. In particular, the need that the initial states should have a good overlap with the target state for certain quantum algorithms to succeed is well-known\cite{tubman2018postponing}. The fact that the preparation of spin eigenfunctions on quantum computers has been investigated recently\cite{carbone2022quantum,lacroix2022symmetry} can be seen as a step in this direction and it also underpins our efforts in the upcoming sections to provide a more nuanced classification of reference functions than the usual single vs. multi-determinant one. As far as the potential gains in scaling are concerned, a recent study suggests that the exponential nature of quantum advantage in quantum chemistry\cite{lee2022there} cannot be taken for granted in most cases, although the authors are careful to point out that quantum computers may still be very useful in quantum chemistry. Thus, the search is on for specific chemical systems where quantum computers could be most beneficially applied\cite{tazhigulov2022simulating,amsler2023quantum,rubin2023fault}.

\section{Symmetry Considerations}

We have already distinguished between DETs, CSFs and CFGs as various choices of basis states. It also follows from the previous discussion that symmetry plays a significant role in the classification of correlation strength. To get a better grip on these notions, let us remind ourselves about how an $N$-electron wavefunction is built from one electron functions (orbitals) and how various symmetries may be taken into account in the construction. The simplest approach is to approximate the $N$-electron wavefunction as a Hartree product (HPD) of $N$ orbitals and enforce various symmetries. In general, the operator $\hat{Q}$ encodes a symmetry of the Hamiltonian $\hat{H}$ if $[\hat{H},\hat{Q}]=0$, and thus the exact wavefunction is a simultaneous eigenstate of $\hat{H}$ and $\hat{Q}$. If an approximate wavefunction does not have a particular symmetry, an appropriate projection operator $\hat{O}_Q$ can be constructed that would produce an eigenfunction of $\hat{Q}$ with eigenvalue $Q$. We are not concerned here with all possible symmetries (e.g., the particle number operator, time reversal), but may limit ourselves to the following scheme,
\begin{equation}
\text{HPD}\quad\xrightarrow{\hat{\mathcal{A}}}\quad
\text{DET}\quad\xrightarrow{\hat{\mathcal{O}}_S}\quad
\text{CSF(CFG)}\quad\xrightarrow{\hat{\mathcal{O}}_R}\quad
\text{SA-CSF},
\end{equation}
where antisymmetrization ($\hat{\mathcal{A}}$) produces a Slater determinant (DET), spin projection ($\hat{O}_S$) a configuration state function (CSF), or, more generally a number of CSFs belonging to the same configuration (CFG, i.e., a sequence of spatial occupation numbers), while spatial symmetry can also be enforced via $\hat{O}_R$ of some symmetry operation $R$, producing a spatial symmetry-adapted CSF (SA-CSF). We will briefly examine these in the following. It only remains to note here that symmetry adaptation in general leads to higher energies in a variational procedure since it introduces constraints in the variational manifold, a phenomenon sometimes referred to as L\"owdin's dilemma\cite{lykos1963discussion}.

The original purpose of introducing the Slater determinant\cite{slater1929theory} was to find an $N$-electron basis that automatically satisfies the exclusion principle, or, rather, its generalization requiring a fermionic wavefunction to be antisymmetric. For this purpose, the HPD is constructed from the product of spin orbitals $\phi_i(\mathbf{x}_i)$, which itself is the product of a spatial orbital $\varphi_i(\mathbf{r}_i)$ and a spin function $\sigma_i(s_i)$, and the antisymmetrization $\hat{\mathcal{A}}$ is carried out by permuting the coordinates $\mathbf{x}_i=(\mathbf{r}_i,s_i)$ and summing over the permutations multiplied by their parity,
\begin{equation}
\Phi = \hat{\mathcal{A}}\prod_i^N \phi_i(\mathbf{x}_i) = 
\hat{\mathcal{A}}\prod_i^N \varphi_i(\mathbf{r}_i)\prod_i^N \sigma_i(s_i).
\label{DET}
\end{equation}
Since $\hat{\mathcal{A}}$ is a projector multiplied by a constant, i.e., $\hat{\mathcal{A}}^2=\sqrt{N!}\hat{\mathcal{A}}$, $\Phi$ is an eigenfunction of $\hat{\mathcal{A}}$. The correlation introduced by this formulation is usually illustrated\cite{tew2007electron,giner2016fermi} using the case of two parallel electrons, where the spatial part $\varphi_1(\mathbf{r}_1)\varphi_2(\mathbf{r}_2)-\varphi_2(\mathbf{r}_1)\varphi_1(\mathbf{r}_2)$ is obviously zero at $\mathbf{r}_1 = \mathbf{r}_2$. It has been observed that this Fermi hole is in fact non-local, because the spatial part also vanishes if $\varphi_1(\mathbf{r}_{1})=\varphi_1(\mathbf{r}_{2})=\pm\varphi_2(\mathbf{r}_{1})=\pm\varphi_2(\mathbf{r}_{2})$\cite{giner2016fermi}. Moreover, antisymmetrization also introduces anti-correlation in the opposite-spin case, which has been used to explain the fact that same-spin double excitations contribute less to the energy in post-Hartree--Fock calculations than opposite-spin ones\cite{giner2016fermi}.

Given that the Slater determinant is a much simpler entity than a general group theoretical construction of the wavefunction, it could be claimed with some plausibility that ``Slater had slain the `Gruppenpest',''\cite{slater1975solid} i.e., the pestilence of groups that threatened the conceptual clarity of quantum mechanics in his view. Were it only about the antisymmetry requirement, Slater's approach would be preferable to general group theoretical presentations given that it is much more transparent, a fact also responsible to a large extent for the establishment of the Slater determinant as the ``default choice'' as reference state. However, the question of spin adaptation still remains and is in some sense a natural extension of antisymmetrization. A spin-adapted CSF may be written as
\begin{equation}
\Theta = 
\hat{\mathcal{A}}\prod_i^N \varphi_i(\mathbf{r}_i)X_S(s_1,\ldots,s_N),
\label{CSF}
\end{equation}
where we have retained the simplest Hartree product in the spatial part, but assumed a generalized spin part $X_S$ which is an eigenfunction of $\hat{S}^2$ with total spin $S$. Since $X_S$ can be written as a linear combination of spin function products of the type appearing in Eq.~\eqref{DET}, it also follows that $\Theta$ can be written as a linear combination of determinants, although CSFs can be constructed directly without determinants as intermediates\cite{pauncz2012spin,chilkuri2021comparison}. The general construction of $X_S$ remains an elaborate group theoretical exercise. Matsen's spin-free approach\cite{matsen1964spin} developed in the 1960s relied on the observation that enforcing the $\mathrm{SU}(2)$ spin symmetry in the spin part of the wavefunction is equivalent to representing the $N$-particle spatial part in terms of the irreducible representations of the symmetric group S$_N$ of permutations\cite{pauncz2012spin}. On the other hand, the spin-free representation of the Hamiltonian is given in terms of the generators of the unitary group $\mathrm{U}(n)$ in the $n$-orbital basis. The latter observation is the basis of the unitary group approach\cite{paldus1981unitary} (UGA) and its graphical application\cite{shavitt1981graphical} (GUGA), which was recently discussed in detail by Dobrautz et al.\cite{dobrautz2019efficient} and the advantages of which have also been exploited in a number of publications\cite{li2020compression,manni2021modeling,li2021resolution}. We will return to some aspects of spin-adaptation and its consequences for the correlation problem in the next section, and similar considerations will also occur in the section on the role of orbitals.

Finally, we shall briefly consider the problem of spatial symmetry since it does appear in discussions on strong correlation\cite{shee2021revealing}. The prototypical example is the He$_2^+$ system\cite{mclean1985symmetry,ayala1998nonorthogonal} which dissociates into a He atom and a He$^+$ ion. At large distances, two electrons are localized on one of the He nuclei and one electron around the other. In the minimal basis, this means that the Hartree--Fock orbital associated with the two electrons has a larger spatial extent than the other one, and they do not reflect the point group symmetry of He$_2^+$. Moreover, there are two symmetry broken solutions depending on whether the two electrons are localized around nucleus $A$ or $B$, corresponding in essence to two resonance structures. It is possible to enforce spatial symmetry by constructing orbitals of the type $\varphi_A\pm\varphi_B$, where $\varphi_A$ and $\varphi_B$ are orbitals localized around $A$ and $B$. A determinant built from such orbitals could be expanded in terms of the resonance structures mentioned, but would also include high-energy ionic configurations, giving rise to an artificially large energy at long bond distances. Thus, this is the spatial equivalent of the dissociation problem of the H$_2$ molecule in which breaking spin (and spatial) symmetry leads to a lower energy. Usually, these problems are regarded multiconfigurational\cite{ayala1998nonorthogonal}, because more than one configuration is needed to give the correct energy without breaking symmetry. Unfortunately, at the level of approximate wavefunctions (often used as a reference), enforcing spatial symmetry may easily lead to additional complications. Since infinitesimal changes may disrupt the symmetric arrangement of the nuclei, this may lead to discontinuities in the energy or the need to use sub-optimal symmetry-adapted orbitals\cite{manne1972brillouin}. Thus, using SA-CSFs may introduce more conceptual uncertainty about symmetry-related correlation effects than simply using DETs or even CSFs (CFGs). We will not consider SA-CSFs here any further, but the review of Davidson and Borden\cite{davidson1983symmetry} discusses the issue of spatial symmetry breaking in all its aspects, be it ``real or artifactual’’, as they put it, and a series of papers by \v{C}\'{i}\v{z}ek and Paldus on the HF stability conditions\cite{cizek1967stability,civek1970stability,paldus1970stability} also devotes considerable space to this problem.

It has been noted before that the various categories for correlation are operational, i.e., what counts as static or non-dynamical correlation is practically whatever is covered by choice at the level of the reference calculation. Symmetry on the other hand introduces exact degeneracies that are known \emph{a priori} and reference functions can be constructed to account for them. Thus, in line with the general thrust of many discussions on the subject, it seems worthwhile to distinguish between correlation effects due to various kinds of symmetry and the rest. We may call these symmetry effects static correlation\cite{mcweeny1967nature,bartlett1994applications}. In the next section, we will discuss accounting for spin symmetry in the basis states. The remaining correlation effects may be weak or strong depending on the weight of the symmetry-adapted reference state in the final calculation. In this regard, near degeneracies are of interest since they do not arise from the symmetry effects discussed so far. The case of the Be atom is instructive\cite{tew2007electron}: due to the near degeneracy of the  2s and 2p orbitals, the 1s$^2$2s$^2$ configuration does not completely dominate the wavefunction expansion, the 1s$^2$2p$^2$ configurations must also be considered. Such non-dynamical effects are not captured by symmetry restrictions but are usually still included in the reference calculation, leaving any residual correlation effects, i.e., dynamic correlation, to be recovered by the wavefunction Ansatz built on that reference. Handy and co-workers\cite{mok1996dynamical} even go so far as to suggest that even this ``angular'' correlation in Be, so called because s and p orbitals are of different angular types, might be regarded as dynamical based on a model in which explicit dependence on the interelectronic distance appears. The Be example also indicates that the weight of the reference state is of crucial importance and what is regarded as non-dynamical or strong depends on a threshold value set on this weight. The dependence of weight-based measures on the orbital space will also be addressed in a subsequent section using a simple bond breaking process as an example, another situation in which strong correlation effects are typically expected.

\section{Representing the Wavefunction}

It has already been mentioned that DETs are only eigenfunctions of $\hat{S}_z$, while CSFs are also eigenfunctions of the total spin. The total number of DETs\cite{helgaker2014molecular} associated with a CFG and with a given spin-projection $M_S$ and number of unpaired electrons $N$ is
\begin{equation}
f_N^M = \binom{N}{\frac{1}{2}N-M_S},
\end{equation}
from which the CSFs with different possible total spin values $M_S\leq S$ can also be built. The number of CSFs\cite{helgaker2014molecular} with $S=M_S$ is
\begin{equation}
f_N^S = \binom{N}{\frac{1}{2}N-S}-\binom{N}{\frac{1}{2}N-S-1},
\end{equation}
which is obviously less than the number of determinants. In fact, summing over all possible $S$ values yields the number of determinants\cite{helgaker2014molecular}.

\begin{table}[h!]
\caption{Simple examples of $N$-electron basis states, including configurations (CFG), configuration state functions (CSF) and determinants (DET). Spatial orbitals are labelled simply as 1, 2 or 3 in DETs and CSFs, while the corresponding occupation numbers (0, 1 or 2) are indicated in CFGs.}
  \center{
  \begin{tabular}{ccccc}
    \hline\hline
      $S$ & $M_S$ & CFG & CSF & DET \\
    \hline
     0 & 0 & $20$ & $|1\bar{1}|$ & $|1\bar{1}|$ \\
     0 & 0 & $02$ & $|2\bar{2}|$ & $|2\bar{2}|$ \\
     0 & 0 & $11$ & $\frac{1}{\sqrt{2}}(|1\bar{2}|-|\bar{1}2|)$ & $|1\bar{2}|$, $|\bar{1}2|$ \\
     1 & 1 & $11$ & $|12|$ & $|12|$ \\
     1 & 0 & $11$ & $\frac{1}{\sqrt{2}}(|1\bar{2}|+|\bar{1}2|)$ & $|1\bar{2}|$, $|\bar{1}2|$ \\
     1 & $-$1 & $11$ & $|\bar{1}\bar{2}|$ & $|\bar{1}\bar{2}|$ \\
     $\frac{3}{2}$ & $\frac{3}{2}$ & $111$ & $|123|$ & $|123|$ \\
     $\frac{3}{2}$ & $\frac{1}{2}$ & $111$ & $\frac{1}{\sqrt{3}}(|\bar{1}23|+|1\bar{2}3|+|12\bar{3}|)$ & $|\bar{1}23|$, $|1\bar{2}3|$, $|12\bar{3}|$ \\
     $\frac{1}{2}$ & $\frac{1}{2}$ & $111$ & $\frac{1}{\sqrt{6}}(2|12\bar{3}|-|1\bar{2}3|-|\bar{1}23|)$, $\frac{1}{\sqrt{2}}(|1\bar{2}3|-|\bar{1}23|)$ & $|\bar{1}23|$, $|1\bar{2}3|$, $|12\bar{3}|$ \\
    \hline\hline
  \end{tabular}
}
\label{table1}
\end{table}

To help ground these notions, consider some simple examples collected in Table~\ref{table1}, and discussed in detail elsewhere\cite{helgaker2014molecular}. Taking the example of two electrons in two orbitals, the electron pair may be placed on the same orbital as in the configurations 20 and 02, or on different orbitals as in the configuration 11. In the former two cases, a single CSF consisting of a single determinant is enough to describe a closed-shell singlet, hence the success of determinants in many situations in chemistry. In the 11 case, a singlet and a triplet CSF can be built from two determinants, though the other two triplet states need only a single determinant. A slightly more complicated example involves distributing three electrons among three orbitals. For the configuration 111, the $S=3/2$ cases yield nothing conceptually new, they are described by CSFs consisting of a single or multiple determinants. On the other hand, the $S=1/2$, $M_S=1/2$ subspace can only be spanned by two degenerate CSFs, themselves consisting of multiple determinants. Since the reference function is often obtained at the Hartree--Fock level, it is worth recalling how this discussion affects such calculations. In more commonly occurring cases, the spin-adapted HF wavefunction can be constructed from either a single determinant, as in the typical closed-shell case, or from a single CSF, as in a low-spin open-shell case\cite{davidson1973spin}. In those open-shell cases, such as $S=1/2$, $M_S=1/2$ for 111, where a configuration cannot be uniquely described by a single CSF\cite{edwards1987generalized}, Edwards and Zerner recommend in a footnote that a HF calculation using a chosen CSF should be followed by a configuration interaction (CI) calculation\cite{edwards1987generalized}. This might minimally include only the CSFs belonging to the CFG in question, e.g., the two degenerate ones in the 111 case,  possibly with or without orbital optimization, though the question remains whether such a small CI calculation would accurately reflect the exact solution. Indeed, Edwards and Zerner recommend CI in the active space, i.e., presumably including multiple configurations that might produce the same total and projected spin (e.g., 201 in the three-electron case). Nevertheless, the expansion coefficients of the CSFs of a single CFG will be of interest here for analytic purposes.

It is worth noting that already at the HF level spin-restricted open-shell CSFs are considerably harder to optimize than the unrestricted determinants. This has led to several ways of circumventing the direct use of CSFs. A simplified version\cite{amos1961single} of L\"owdin’s projection operator\cite{lowdin1955quantum3} may be used to remove the spin-components that lie nearest to the desired state from the spin-unrestricted determinant, but in any case, spin-projection leads to artefacts if applied after optimization and falls short of size-consistency if made part of it\cite{mayer1980spin}. More recent projection approaches constrain the eigenvalues of the spin-density to obtain results that are identical to the more cumbersome conventional spin-restricted approach\cite{tsuchimochi2010communication}. Even more pragmatically, quasi-restricted approaches of various kinds have been proposed to obtain orbitals resembling restricted ones from some simple recipe\cite{rittby1988open,neese2006importance}. More complicated Ans\"atze are usually obtained from reference states by (ideally spin-adapted) excitation (or ionization) operators\cite{paldus1981unitary}, and in a similar vein, it is also possible to define spin-flip operators that change the spin state of the reference\cite{casanova2020spin}. In simple cases, such spin-flip operators may even be spin-adapted themselves\cite{roemelt2013excited}. Sometimes symmetry-broken solutions may be used creatively, e.g., to obtain some property of spin-states as in PO/BS-DFT discussed above, or in combining unrestricted determinants for quasi-spin-adaptation\cite{izsak2022second}, sometimes considerations about the CSF expansion are employed innovatively, e.g., to obtain singlet-triplet gaps from bi-determinantal $M_S=0$ states (see Table~\ref{table1}) in cases in which methods with a single determinant bias encounter multideterminant situations\cite{veis2018intricate}.

At this stage, there is some ambiguity whether we associate multireference with the requirement that the exact solution is dominated by more than a single determinant, CSF or configuration. In a normalized FCI expansion of the type in Eq.~\eqref{FCI}, a set of basis states  dominates the expansion if their cumulative weight, $\sum_P |C_P|^2$ is larger than the rest of the weights combined, i.e., $1-\sum_P |C_P|^2$. Assuming knowledge of the exact solution, the following categories are proposed (cf.\cite{cremer2001density}): 
\begin{itemize}
\item{
\textbf{Single Determinant (SD) Expansion:} The weight of a single determinant dominates and hence a single-determinant HF calculation is enough for a reference.
}
\item{
\textbf{Single Spin-Coupling (SS) Expansion:} It consists of a linear combination of determinants whose relative weights are fixed but the weight of the resulting CSF dominates the expansion. An restricted open-shell calculation should be a good reference.  
}
\item{
\textbf{Single Configurational (SC) Expansion:} Such an expansion has multiple CSFs belonging to the same configuration. An expansion would fall into this category if the cumulative weight of CSFs of a single configuration dominates. A restricted open-shell calculation followed by a minimal CI calculation to determine the relative weight of the CSFs involved might be a good reference. 
}
\item{
\textbf{Multiconfiguration (MC) Expansion:} No configuration alone dominates and hence this is the true multireference case since no single reference function of the kinds discussed before can be found.
}
\end{itemize}
Note that the single reference (SR) categories above may overlap, since e.g., the closed-shell ground state of a two-electron system is at the same time SD, SS and SC. An open-shell singlet is SS and SC, but not SD. The three-electron doublet that has two CSFs would be SC but not SS or SD, if the cumulative weight of those CSFs dominate the expansion. All these are examples of various SR expansions. The simplest example of an MR expansion would be a combination of the type $02+20$, as in the case of the dissociating H$_2$ molecule. In the next section, we shall discuss how the choice of orbital space impacts this classification. Once spin-coupling is taken care of, an important source of correlation effects is addressed and the question remains how strong the remaining correlation effects are. This reflects the distinction between static and non-dynamic correlation as used by Bartlett and Stanton\cite{bartlett1994applications}, with static correlation being eliminated by spin-adaptation. In this regard, this generalization of the excitation schemes can be compared with entanglement\cite{chan2012low} and seniority\cite{henderson2014seniority} approaches which also aim at finding a good reference based on different criteria. 

\section{The Role of the Orbital Space}

The idea that the weight of the reference function in an approximate CI calculation could be used as an indication of multireference character is an old one, but it was also observed long ago that due to the bias of canonical orbitals towards the reference (HF) state, it is of limited use\cite{lee1989theoretical}. Furthermore, it was also suggested that such CI calculations should be complete within at least an active space for this metric to be of use\cite{jiang2012multireference}. In that case, the leading CAS coefficient in the basis of natural orbitals using determinants or CSFs was found to be a sensible metric\cite{sears2008assessing1,sears2008assessing2}. The role of the orbital space has been discussed in connection to most approaches discussed in this perspective\cite{jiang2012multireference,cremer2001density,henderson2014seniority,ding2022quantum,stein2017measuring}, and orbital localization techniques form a part of practical DMRG\cite{olivares2015ab} calculations as well as reducing the complexity of CAS\cite{pandharkar2022localized} calculations. Within the DMRG context, orbital optimization techniques have been recently suggested as a means of reducing the multireference character of the wavefunction\cite{mate2022compressing}. A similar result has been observed in SCI methods, where orbital optimization leads to a more rapid convergence to the FCI solution\cite{smith2017}. The recent work of Li Manni and co-workers on GUGA based FCIQMC\cite{li2020compression,manni2021modeling,li2021resolution} and its application as a stochastic CASCI solver\cite{dobrautz2021spin} shed even more light on the role of the orbital space. This methodology has the following features relevant for the current discussion: a) using GUGA to calculate spin-adapted basis states efficiently; b) using a unitary orbital transformation to compress the wavefunction expansion; c) the combined use of unitary and symmetric group approaches means that the ordering of orbitals also affects the complexity of the wavefunction. The technique has been used to reduce the number of CSFs expected from the exponentially scaling combinatorial formula to a small number of CSFs, sometimes only a single one\cite{li2021resolution}.  Moreover, not only is the wavefunction more compact, but the Hamiltonian assumes a convenient block-diagonal form that allows for the selective targeting of excited states\cite{li2021resolution}. Interestingly, it was also found that the best orbital ordering for a spin-adapted CSF is not identical to the best site-ordering in the DMRG sense\cite{dobrautz2022combined}. More recent work has also attempted to place the orbital ordering process on a more systematic basis exploiting the commutation properties of cumulative spin operators and the Hamiltonian\cite{li2022resolution}. Here, the role of the orbital space will be illustrated via a simple example that will remain within a more traditional CAS/FCI context.

\begin{figure}
\includegraphics[scale=1]{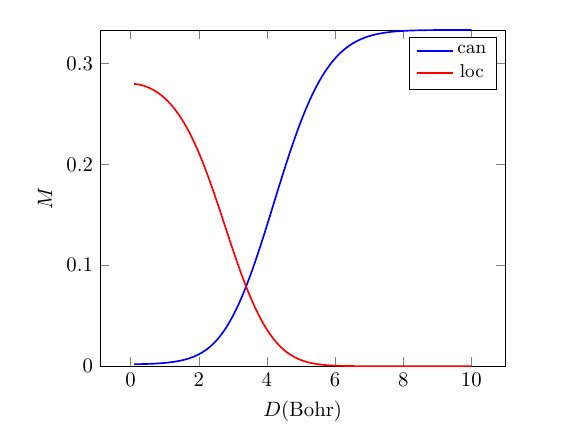}
\caption{The multireference character $M$ as a function of internuclear distance $D$ in the H$_2$ dissociation problem for the FCI/STO-1G ground state.  The blue curve corresponds to canonical orbitals (can), the red one to localized orbitals (loc).}
\label{fig1}
\end{figure}

In this perspective, we are concerned with the \emph{a posteriori} characterization of states rather than finding an \emph{a priori} indicator of multireference character, and hence we will assume that an FCI solution is available. We will illustrate the role of orbital spaces by considering the H$_2$ bond dissociation problem using canonical and localized orbitals\cite{pipek1989fast}. We will define the multireference character $M$ as
\begin{equation}
M=\frac{1- \sum_{P}|C_{P}|^2}{1+ \sum_{P}|C_{P}|^2},
\end{equation}
where the summation runs over the DETs or CSFs belonging to the dominant CFG. This measure will converge to 1 for infinitely many coefficients assuming they are all equal, i.e., the worst MR case. In SR cases, $M$ should be close to 0, and in the H$_2$ case $M=1/3$ corresponds to two configurations of equal weight. In practice, one might choose a different threshold on the cumulative weight of the configuration.  For example, requiring the dominant cumulative weights to be above 80\% leads to $M\leq 1/9$. If such a dominant configuration exists (and if it were known in advance), it would be the ideal reference function on which a correlated Ansatz could be built.

Fig.~\ref{fig1} shows the value of $M$ at different internuclear separations $D$. In the canonical basis, a single configuration $20$ is dominant around the equilibrium distance, while at infinite separation the CFGs $20$ and $02$ contribute equally. Thus, in the latter case, the expansion is multiconfigurational as well as multi-determinantal.  Evaluating $M$ in the basis of local orbitals yields to the opposite results. The local orbitals are each essentially localized on one atom, and at long bond distance, a single CSF of the 11 configurations can describe the dissociation. Thus, the expansion is SC and SS, but not SD. At shorter bond lengths, the localized expansion is much less appropriate as seen from the rising multireference character. Changing from the canonical to the local orbital basis has a similar effect on entanglement-based metrics\cite{stein2017measuring}: in the canonical basis entanglement changes between its minimal and maximal values as $D$ grows and the other way around in the local basis since in the latter all occupation patterns are equally likely at small $D$. This ties in also with our earlier remarks on the relationship between VB and MO approaches.

It has been argued earlier that an important source of correlation effects may be eliminated by spin adaptation. Yet, because of L\"owdin's dilemma, the constrained reference energy may increase and this could in turn lead to an increase in the correlation energy. In the H$_2$ example, the spin-restricted closed-shell reference function has been found qualitatively adequate around the equilibrium bond length, but at large distances it leads to a catastrophic increase in the magnitude of the reference and correlation energies. Using canonical orbitals, the remedy involves combining the $20$ and $02$ configurations. Thus, it seems that enforcing spin symmetry makes the problem more complicated by requiring a multiconfigurational treatment whereas the spin-unrestricted single determinant also converges to the qualitatively correct energy at large bond lengths. This is because it reduces to one of the symmetry broken determinants $\lvert\mu\bar{\nu}\rvert$ or $\lvert\bar{\mu}\nu\rvert$ in this limit. In general, these broken symmetry determinants produce an energy that lies between the open shell singlet in Eq.~\eqref{SState} and triplet states that can also be constructed from them, and as such, they are not a good approximation to either. At large bond distances however, the singlet-triplet gap closes and the energy converges to the correct limit while the wavefunction is not the exact one\cite{knowles2000ab,fulde1995electron}. While symmetry breaking may be desirable under some circumstances, for finite molecular systems symmetric solutions are in general preferred. In the open-shell singlet case, for example, spin-adaptation is indispensable for a qualitatively accurate description. It is here that orbital rotations may play some role: by localizing the orbitals at large distances, it is possible to convert the multiconfigurational representation of the dissociating H$_2$ molecule in the canonical basis into a single configurational open-shell singlet in the local basis.

The fact that the multireference character depends on the orbital representation warns against assigning the multiconfigurational nature of the wavefunction expansion as a physical property of a state of the system. At the very least, one should also demand that there be no unitary transformation originating in the orbital space that removes the multireference character of an expansion completely. Of course, the analysis as presented here is not a practical recipe for computation, since it assumes knowledge of the FCI solution, although it may be used fruitfully in interpreting less costly wavefunctions of the CASCI or CASSCF type. Furthermore, it does underline the potential of orbital rotations in reducing the multireference character. This important subject will be analysed in more detail in a forthcoming publication.

\section{Outlook}

The distinctions we introduced between single-determinantal, single spin-coupling, single configurational and multiconfigurational expansions should have a relevance in any chemical system, as also pointed out by other authors\cite{bartlett1994applications,tew2007electron}. These categories form a clear hierarchy of SR measures in the sense that they are increasingly less restrictive: SD implies SS and both SD and SS imply SC.  Unlike in extended systems, where symmetry broken solutions are often preferred, symmetry-adapted wavefunctions are often needed to reliably predict certain chemical properties of molecules. Again, unlike in periodic solids, strongly interacting regimes are usually easier to localize, and hence the complexity of the wavefunction can often be reduced. If spin coupling effects can be incorporated into a single reference function, then the remaining correlation effects might be weak, and in cases when they are not, such as in certain regions of bond dissociation curves, they can be typically taken care of by small active space calculations. As observed by Malrieu et al.\cite{malrieu2007bond}, the fact that one can always build a small complete active space representation by placing two electrons in a pair of localized bonding and antibonding orbitals that accounts for such correlation effects is perhaps the best unbiased confirmation of Lewis' idea of the chemical bond consisting of an electron pair\cite{lewis1916atom}. We have discussed the effect of orbital rotations on the multireference character in a simple bond dissociation process and pointed out that such orbital rotations have the potential of reducing it. If by strong correlation we also mean the large size of the active space necessary to remedy these non-dynamical effects, then it seems likely that most molecular systems are not strongly correlated, with the possible exception of multi-site antiferromagnetically coupled systems\cite{shee2021revealing,chen2011multireference,li2019electronic,khedkar2021modern,tarrago2021experimental,han2023magnetic}. Thus, the complexity of most finite chemical systems should be well below exponential and should in most cases involve only a few configurations. Devising an appropriate metric that can help find these \emph{a priori} is a challenge, but measures based on leading coefficients, density matrices, natural orbitals and orbital entanglement have been used for this purpose in the literature. 

\section*{Acknowledgements}

The authors would like to thank Earl T. Cambell, Giovanni Li Manni and Pavel Pokhilko for useful discussions on the manuscript.

\providecommand{\latin}[1]{#1}
\makeatletter
\providecommand{\doi}
  {\begingroup\let\do\@makeother\dospecials
  \catcode`\{=1 \catcode`\}=2 \doi@aux}
\providecommand{\doi@aux}[1]{\endgroup\texttt{#1}}
\makeatother
\providecommand*\mcitethebibliography{\thebibliography}
\csname @ifundefined\endcsname{endmcitethebibliography}
  {\let\endmcitethebibliography\endthebibliography}{}


\begin{mcitethebibliography}{252}
\providecommand*\natexlab[1]{#1}
\providecommand*\mciteSetBstSublistMode[1]{}
\providecommand*\mciteSetBstMaxWidthForm[2]{}
\providecommand*\mciteBstWouldAddEndPuncttrue
  {\def\EndOfBibitem{\unskip.}}
\providecommand*\mciteBstWouldAddEndPunctfalse
  {\let\EndOfBibitem\relax}
\providecommand*\mciteSetBstMidEndSepPunct[3]{}
\providecommand*\mciteSetBstSublistLabelBeginEnd[3]{}
\providecommand*\EndOfBibitem{}
\mciteSetBstSublistMode{f}
\mciteSetBstMaxWidthForm{subitem}{(\alph{mcitesubitemcount})}
\mciteSetBstSublistLabelBeginEnd
  {\mcitemaxwidthsubitemform\space}
  {\relax}
  {\relax}

\bibitem[Kurth and Perdew(2000)Kurth, and Perdew]{kurth2000role}
Kurth,~S.; Perdew,~J.~P. Role of the exchange–correlation energy: Nature's
  glue. \emph{Int. J. Quantum Chem.} \textbf{2000}, \emph{77}, 814--818\relax
\mciteBstWouldAddEndPuncttrue
\mciteSetBstMidEndSepPunct{\mcitedefaultmidpunct}
{\mcitedefaultendpunct}{\mcitedefaultseppunct}\relax
\EndOfBibitem
\bibitem[Martin(2022)]{martin2022electron}
Martin,~J.~M. Electron Correlation: Nature's Weird and Wonderful Chemical Glue.
  \emph{Isr. J. Chem.} \textbf{2022}, \emph{62}, e202100111\relax
\mciteBstWouldAddEndPuncttrue
\mciteSetBstMidEndSepPunct{\mcitedefaultmidpunct}
{\mcitedefaultendpunct}{\mcitedefaultseppunct}\relax
\EndOfBibitem
\bibitem[L{\"o}wdin(1959)]{lowdin1958correlation}
L{\"o}wdin,~P.-O. Correlation Problem in Many-Electron Quantum Mechanics I.
  Review of Different Approaches and Discussion of Some Current Ideas. In
  \emph{Advances in Chemical Physics}; Interscience: New York, 1959; Vol.~2;
  Chapter 7, pp 207--322\relax
\mciteBstWouldAddEndPuncttrue
\mciteSetBstMidEndSepPunct{\mcitedefaultmidpunct}
{\mcitedefaultendpunct}{\mcitedefaultseppunct}\relax
\EndOfBibitem
\bibitem[Kotani \latin{et~al.}(1961)Kotani, Ohno, and
  Kayama]{kotani1961quantum}
Kotani,~M.; Ohno,~K.; Kayama,~K. Quantum Mechanics of Electronic Structure of
  Simple Molecules. In \emph{Encyclopedia of Physics/Handbuch der Physik};
  Springer: Berlin, 1961; Vol. 37.2; Chapter 1, pp 1--172\relax
\mciteBstWouldAddEndPuncttrue
\mciteSetBstMidEndSepPunct{\mcitedefaultmidpunct}
{\mcitedefaultendpunct}{\mcitedefaultseppunct}\relax
\EndOfBibitem
\bibitem[Slater(1963)]{slater1963quantum}
Slater,~J.~C. \emph{Quantum Theory of Molecules and Solids. Vol. 1: Electronic
  Structure of Molecules}; McGraw-Hill: New York, 1963\relax
\mciteBstWouldAddEndPuncttrue
\mciteSetBstMidEndSepPunct{\mcitedefaultmidpunct}
{\mcitedefaultendpunct}{\mcitedefaultseppunct}\relax
\EndOfBibitem
\bibitem[Kutzelnigg(2002)]{kutzelnigg2002einfuhrung}
Kutzelnigg,~W. \emph{Einf{\"u}hrung in die theoretische Chemie}; Wiley-VCH:
  Weinheim, 2002; Vol.~2\relax
\mciteBstWouldAddEndPuncttrue
\mciteSetBstMidEndSepPunct{\mcitedefaultmidpunct}
{\mcitedefaultendpunct}{\mcitedefaultseppunct}\relax
\EndOfBibitem
\bibitem[Wigner(1934)]{wigner1934interaction}
Wigner,~E. On the interaction of electrons in metals. \emph{Phys. Rev.}
  \textbf{1934}, \emph{46}, 1002\relax
\mciteBstWouldAddEndPuncttrue
\mciteSetBstMidEndSepPunct{\mcitedefaultmidpunct}
{\mcitedefaultendpunct}{\mcitedefaultseppunct}\relax
\EndOfBibitem
\bibitem[Wigner(1938)]{wigner1938effects}
Wigner,~E. Effects of the electron interaction on the energy levels of
  electrons in metals. \emph{Trans. Faraday Soc.} \textbf{1938}, \emph{34},
  678--685\relax
\mciteBstWouldAddEndPuncttrue
\mciteSetBstMidEndSepPunct{\mcitedefaultmidpunct}
{\mcitedefaultendpunct}{\mcitedefaultseppunct}\relax
\EndOfBibitem
\bibitem[L{\"o}wdin(1995)]{lowdin1995historical}
L{\"o}wdin,~P.-O. The historical development of the electron correlation
  problem. \emph{Int. J. Quantum Chem.} \textbf{1995}, \emph{55}, 77--102\relax
\mciteBstWouldAddEndPuncttrue
\mciteSetBstMidEndSepPunct{\mcitedefaultmidpunct}
{\mcitedefaultendpunct}{\mcitedefaultseppunct}\relax
\EndOfBibitem
\bibitem[Sinano\u{g}lu(1961)]{sinanoglu1961many}
Sinano\u{g}lu,~O. Many-Electron Theory of Atoms and Molecules. \emph{Proc. Nat.
  Acad. Sci.} \textbf{1961}, \emph{47}, 1217--1226\relax
\mciteBstWouldAddEndPuncttrue
\mciteSetBstMidEndSepPunct{\mcitedefaultmidpunct}
{\mcitedefaultendpunct}{\mcitedefaultseppunct}\relax
\EndOfBibitem
\bibitem[Sinano\u{g}lu(1964)]{sinanoglu1964many}
Sinano\u{g}lu,~O. Many-Electron Theory of Atoms, Molecules and Their
  Interactions. In \emph{Advances in Chemical Physics}; John Wiley \& Sons,
  Ltd: London, 1964; Vol.~6; Chapter 7, pp 315--412\relax
\mciteBstWouldAddEndPuncttrue
\mciteSetBstMidEndSepPunct{\mcitedefaultmidpunct}
{\mcitedefaultendpunct}{\mcitedefaultseppunct}\relax
\EndOfBibitem
\bibitem[McWeeny(1967)]{mcweeny1967nature}
McWeeny,~R. The nature of electron correlation in molecules. \emph{Int. J.
  Quantum Chem.} \textbf{1967}, \emph{1}, 351--359\relax
\mciteBstWouldAddEndPuncttrue
\mciteSetBstMidEndSepPunct{\mcitedefaultmidpunct}
{\mcitedefaultendpunct}{\mcitedefaultseppunct}\relax
\EndOfBibitem
\bibitem[Kutzelnigg \latin{et~al.}(1968)Kutzelnigg, Del~Re, and
  Berthier]{kutzelnigg1968correlation}
Kutzelnigg,~W.; Del~Re,~G.; Berthier,~G. Correlation coefficients for
  electronic wave functions. \emph{Phys. Rev.} \textbf{1968}, \emph{172},
  49--59\relax
\mciteBstWouldAddEndPuncttrue
\mciteSetBstMidEndSepPunct{\mcitedefaultmidpunct}
{\mcitedefaultendpunct}{\mcitedefaultseppunct}\relax
\EndOfBibitem
\bibitem[von Herigonte(1972)]{herigonte1972electron}
von Herigonte,~P. Electron correlation in the seventies. In \emph{Structure and
  Bonding}; Springer: New York, 1972; Vol.~12; Chapter 1, pp 1--47\relax
\mciteBstWouldAddEndPuncttrue
\mciteSetBstMidEndSepPunct{\mcitedefaultmidpunct}
{\mcitedefaultendpunct}{\mcitedefaultseppunct}\relax
\EndOfBibitem
\bibitem[Kutzelnigg and von Herigonte(2000)Kutzelnigg, and von
  Herigonte]{kutzelnigg2000electron}
Kutzelnigg,~W.; von Herigonte,~P. Electron correlation at the dawn of the 21st
  century. In \emph{Advances in Quantum Chemistry}; Academic Press: San Diego,
  2000; Vol.~36; pp 185--229\relax
\mciteBstWouldAddEndPuncttrue
\mciteSetBstMidEndSepPunct{\mcitedefaultmidpunct}
{\mcitedefaultendpunct}{\mcitedefaultseppunct}\relax
\EndOfBibitem
\bibitem[Kutzelnigg(2003)]{kutzelnigg2003theory}
Kutzelnigg,~W. Theory of electron correlation. In \emph{Explicitly Correlated
  Wave Functions in Chemistry and Physics}; Springer Science \& Business Media:
  Dordrecht, 2003; Chapter 1, pp 3--90\relax
\mciteBstWouldAddEndPuncttrue
\mciteSetBstMidEndSepPunct{\mcitedefaultmidpunct}
{\mcitedefaultendpunct}{\mcitedefaultseppunct}\relax
\EndOfBibitem
\bibitem[Bartlett and Stanton(1994)Bartlett, and
  Stanton]{bartlett1994applications}
Bartlett,~R.~J.; Stanton,~J.~F. Applications of Post-Hartree—Fock Methods: A
  Tutorial. In \emph{Reviews in Computational Chemistry}; VCH Publishers: New
  York, 1994; Chapter 2, pp 65--169\relax
\mciteBstWouldAddEndPuncttrue
\mciteSetBstMidEndSepPunct{\mcitedefaultmidpunct}
{\mcitedefaultendpunct}{\mcitedefaultseppunct}\relax
\EndOfBibitem
\bibitem[Knowles \latin{et~al.}(2000)Knowles, Sch\"{u}tz, and
  Werner]{knowles2000ab}
Knowles,~P.~J.; Sch\"{u}tz,~M.; Werner,~H.-J. {Ab Initio Methods for Electron
  Correlation in Molecules}. In \emph{Modern Methods and Algorithms of Quantum
  Chemistry}; Grotendorst,~J., Ed.; NIC: J\"{u}lich, 2000; Vol.~1; pp
  69--151\relax
\mciteBstWouldAddEndPuncttrue
\mciteSetBstMidEndSepPunct{\mcitedefaultmidpunct}
{\mcitedefaultendpunct}{\mcitedefaultseppunct}\relax
\EndOfBibitem
\bibitem[Tew \latin{et~al.}(2007)Tew, Klopper, and Helgaker]{tew2007electron}
Tew,~D.~P.; Klopper,~W.; Helgaker,~T. Electron correlation: The many-body
  problem at the heart of chemistry. \emph{J. Comput. Chem.} \textbf{2007},
  \emph{28}, 1307--1320\relax
\mciteBstWouldAddEndPuncttrue
\mciteSetBstMidEndSepPunct{\mcitedefaultmidpunct}
{\mcitedefaultendpunct}{\mcitedefaultseppunct}\relax
\EndOfBibitem
\bibitem[Senatore and March(1994)Senatore, and March]{senatore1994recent}
Senatore,~G.; March,~N. Recent progress in the field of electron correlation.
  \emph{Rev. Mod. Phys.} \textbf{1994}, \emph{66}, 445--479\relax
\mciteBstWouldAddEndPuncttrue
\mciteSetBstMidEndSepPunct{\mcitedefaultmidpunct}
{\mcitedefaultendpunct}{\mcitedefaultseppunct}\relax
\EndOfBibitem
\bibitem[Loos and Gill(2016)Loos, and Gill]{loos2016uniform}
Loos,~P.-F.; Gill,~P.~M. The uniform electron gas. \emph{Wiley Interdiscip.
  Rev. Comput. Mol. Sci.} \textbf{2016}, \emph{6}, 410--429\relax
\mciteBstWouldAddEndPuncttrue
\mciteSetBstMidEndSepPunct{\mcitedefaultmidpunct}
{\mcitedefaultendpunct}{\mcitedefaultseppunct}\relax
\EndOfBibitem
\bibitem[Chan(2012)]{chan2012low}
Chan,~G. K.-L. Low entanglement wavefunctions. \emph{Wiley Interdiscip. Rev.
  Comput. Mol. Sci.} \textbf{2012}, \emph{2}, 907--920\relax
\mciteBstWouldAddEndPuncttrue
\mciteSetBstMidEndSepPunct{\mcitedefaultmidpunct}
{\mcitedefaultendpunct}{\mcitedefaultseppunct}\relax
\EndOfBibitem
\bibitem[Fulde(1995)]{fulde1995electron}
Fulde,~P. \emph{Electron correlations in molecules and solids};
  Springer-Verlag: Berlin, 1995\relax
\mciteBstWouldAddEndPuncttrue
\mciteSetBstMidEndSepPunct{\mcitedefaultmidpunct}
{\mcitedefaultendpunct}{\mcitedefaultseppunct}\relax
\EndOfBibitem
\bibitem[March(1996)]{march1996electron}
March,~N.~H. \emph{Electron correlation in molecules and condensed phases};
  Springer Science \& Business Media: New York, 1996\relax
\mciteBstWouldAddEndPuncttrue
\mciteSetBstMidEndSepPunct{\mcitedefaultmidpunct}
{\mcitedefaultendpunct}{\mcitedefaultseppunct}\relax
\EndOfBibitem
\bibitem[Wilson(2007)]{wilson2014electron}
Wilson,~S. \emph{Electron Correlation in Molecules}; Dover: Mineola, 2007\relax
\mciteBstWouldAddEndPuncttrue
\mciteSetBstMidEndSepPunct{\mcitedefaultmidpunct}
{\mcitedefaultendpunct}{\mcitedefaultseppunct}\relax
\EndOfBibitem
\bibitem[Lefebvre and Moser(1969)Lefebvre, and Moser]{advchemphys014}
Lefebvre,~R., Moser,~C., Eds. \emph{Advances in Chemical Physics}; John Wiley
  \& Sons, Ltd: London, 1969; Vol. 14: Correlation Effects in Atoms and
  Molecules\relax
\mciteBstWouldAddEndPuncttrue
\mciteSetBstMidEndSepPunct{\mcitedefaultmidpunct}
{\mcitedefaultendpunct}{\mcitedefaultseppunct}\relax
\EndOfBibitem
\bibitem[March(1999)]{march1999electron}
March,~N.~H., Ed. \emph{Electron Correlation in the Solid State}; Imperial
  Collage Press: London, 1999\relax
\mciteBstWouldAddEndPuncttrue
\mciteSetBstMidEndSepPunct{\mcitedefaultmidpunct}
{\mcitedefaultendpunct}{\mcitedefaultseppunct}\relax
\EndOfBibitem
\bibitem[Raimes(1972)]{raimes1972many}
Raimes,~S. \emph{Many-electron Theory}; North-Holland: Amsterdam, 1972\relax
\mciteBstWouldAddEndPuncttrue
\mciteSetBstMidEndSepPunct{\mcitedefaultmidpunct}
{\mcitedefaultendpunct}{\mcitedefaultseppunct}\relax
\EndOfBibitem
\bibitem[Gell-Mann and Brueckner(1957)Gell-Mann, and
  Brueckner]{gell-mann1957correlation}
Gell-Mann,~M.; Brueckner,~K.~A. Correlation energy of an electron gas at high
  density. \emph{Phys. Rev.} \textbf{1957}, \emph{106}, 364--368\relax
\mciteBstWouldAddEndPuncttrue
\mciteSetBstMidEndSepPunct{\mcitedefaultmidpunct}
{\mcitedefaultendpunct}{\mcitedefaultseppunct}\relax
\EndOfBibitem
\bibitem[Hund(1927)]{hund1927deutung}
Hund,~F. Zur Deutung der Molekelspektren. I. \emph{Z. Phys.} \textbf{1927},
  \emph{40}, 742--764\relax
\mciteBstWouldAddEndPuncttrue
\mciteSetBstMidEndSepPunct{\mcitedefaultmidpunct}
{\mcitedefaultendpunct}{\mcitedefaultseppunct}\relax
\EndOfBibitem
\bibitem[Mulliken(1928)]{mulliken1928assignment}
Mulliken,~R.~S. The assignment of quantum numbers for electrons in molecules.
  I. \emph{Phys. Rev.} \textbf{1928}, \emph{32}, 186--222\relax
\mciteBstWouldAddEndPuncttrue
\mciteSetBstMidEndSepPunct{\mcitedefaultmidpunct}
{\mcitedefaultendpunct}{\mcitedefaultseppunct}\relax
\EndOfBibitem
\bibitem[Lennard-Jones(1929)]{lennard1929electronic}
Lennard-Jones,~J.~E. The electronic structure of some diatomic molecules.
  \emph{Trans. Faraday Soc.} \textbf{1929}, \emph{25}, 668--686\relax
\mciteBstWouldAddEndPuncttrue
\mciteSetBstMidEndSepPunct{\mcitedefaultmidpunct}
{\mcitedefaultendpunct}{\mcitedefaultseppunct}\relax
\EndOfBibitem
\bibitem[Heitler and London(1927)Heitler, and
  London]{heitler1927wechselwirkung}
Heitler,~W.; London,~F. Wechselwirkung neutraler Atome und hom{\"o}opolare
  Bindung nach der Quantenmechanik. \emph{Z. Phys.} \textbf{1927}, \emph{44},
  455--472\relax
\mciteBstWouldAddEndPuncttrue
\mciteSetBstMidEndSepPunct{\mcitedefaultmidpunct}
{\mcitedefaultendpunct}{\mcitedefaultseppunct}\relax
\EndOfBibitem
\bibitem[Pauling(1931)]{pauling1931nature}
Pauling,~L. The nature of the chemical bond. Application of results obtained
  from the quantum mechanics and from a theory of paramagnetic susceptibility
  to the structure of molecules. \emph{J. Am. Chem. Soc.} \textbf{1931},
  \emph{53}, 1367--1400\relax
\mciteBstWouldAddEndPuncttrue
\mciteSetBstMidEndSepPunct{\mcitedefaultmidpunct}
{\mcitedefaultendpunct}{\mcitedefaultseppunct}\relax
\EndOfBibitem
\bibitem[Coulson and Fischer(1949)Coulson, and Fischer]{coulson1949notes}
Coulson,~C.~A.; Fischer,~I. Notes on the molecular orbital treatment of the
  hydrogen molecule. \emph{Phil. Mag.} \textbf{1949}, \emph{40}, 386--393\relax
\mciteBstWouldAddEndPuncttrue
\mciteSetBstMidEndSepPunct{\mcitedefaultmidpunct}
{\mcitedefaultendpunct}{\mcitedefaultseppunct}\relax
\EndOfBibitem
\bibitem[Kutzelnigg and Mukherjee(1999)Kutzelnigg, and
  Mukherjee]{kutzelnigg1999cumulant}
Kutzelnigg,~W.; Mukherjee,~D. Cumulant expansion of the reduced density
  matrices. \emph{J. Chem. Phys.} \textbf{1999}, \emph{110}, 2800--2809\relax
\mciteBstWouldAddEndPuncttrue
\mciteSetBstMidEndSepPunct{\mcitedefaultmidpunct}
{\mcitedefaultendpunct}{\mcitedefaultseppunct}\relax
\EndOfBibitem
\bibitem[Wigner and Seitz(1933)Wigner, and Seitz]{wigner1933constitution}
Wigner,~E.; Seitz,~F. On the Constitution of Metallic Sodium. \emph{Phys. Rev.}
  \textbf{1933}, \emph{43}, 804--810\relax
\mciteBstWouldAddEndPuncttrue
\mciteSetBstMidEndSepPunct{\mcitedefaultmidpunct}
{\mcitedefaultendpunct}{\mcitedefaultseppunct}\relax
\EndOfBibitem
\bibitem[Wigner and Seitz(1934)Wigner, and Seitz]{wigner1934constitution}
Wigner,~E.; Seitz,~F. On the Constitution of Metallic Sodium. II. \emph{Phys.
  Rev.} \textbf{1934}, \emph{46}, 509--524\relax
\mciteBstWouldAddEndPuncttrue
\mciteSetBstMidEndSepPunct{\mcitedefaultmidpunct}
{\mcitedefaultendpunct}{\mcitedefaultseppunct}\relax
\EndOfBibitem
\bibitem[McWeeny(1959)]{mcweeny1959density}
McWeeny,~R. The density matrix in many-electron quantum mechanics I.
  Generalized product functions. Factorization and physical interpretation of
  the density matrices. \emph{Proc. Roy. Soc. A} \textbf{1959}, \emph{253},
  242--259\relax
\mciteBstWouldAddEndPuncttrue
\mciteSetBstMidEndSepPunct{\mcitedefaultmidpunct}
{\mcitedefaultendpunct}{\mcitedefaultseppunct}\relax
\EndOfBibitem
\bibitem[Giner \latin{et~al.}(2016)Giner, Tenti, Angeli, and
  Malrieu]{giner2016fermi}
Giner,~E.; Tenti,~L.; Angeli,~C.; Malrieu,~J.-P. The “Fermi hole” and the
  correlation introduced by the symmetrization or the anti-symmetrization of
  the wave function. \emph{J. Chem. Phys.} \textbf{2016}, \emph{145},
  124114\relax
\mciteBstWouldAddEndPuncttrue
\mciteSetBstMidEndSepPunct{\mcitedefaultmidpunct}
{\mcitedefaultendpunct}{\mcitedefaultseppunct}\relax
\EndOfBibitem
\bibitem[Kutzelnigg and Mukherjee(1997)Kutzelnigg, and
  Mukherjee]{kutzelnigg1997normal}
Kutzelnigg,~W.; Mukherjee,~D. Normal order and extended Wick theorem for a
  multiconfiguration reference wave function. \emph{J. Chem. Phys.}
  \textbf{1997}, \emph{107}, 432--449\relax
\mciteBstWouldAddEndPuncttrue
\mciteSetBstMidEndSepPunct{\mcitedefaultmidpunct}
{\mcitedefaultendpunct}{\mcitedefaultseppunct}\relax
\EndOfBibitem
\bibitem[Evangelista(2022)]{evangelista2022automatic}
Evangelista,~F.~A. Automatic derivation of many-body theories based on general
  Fermi vacua. \emph{J. Chem. Phys.} \textbf{2022}, \emph{157}, 064111\relax
\mciteBstWouldAddEndPuncttrue
\mciteSetBstMidEndSepPunct{\mcitedefaultmidpunct}
{\mcitedefaultendpunct}{\mcitedefaultseppunct}\relax
\EndOfBibitem
\bibitem[Tsuchimochi and Scuseria(2009)Tsuchimochi, and
  Scuseria]{tsuchimochi2009strong}
Tsuchimochi,~T.; Scuseria,~G.~E. Strong correlations via constrained-pairing
  mean-field theory. \emph{J. Chem. Phys.} \textbf{2009}, \emph{131},
  121102\relax
\mciteBstWouldAddEndPuncttrue
\mciteSetBstMidEndSepPunct{\mcitedefaultmidpunct}
{\mcitedefaultendpunct}{\mcitedefaultseppunct}\relax
\EndOfBibitem
\bibitem[Kutzelnigg(2012)]{kutzelnigg2012separation}
Kutzelnigg,~W. Separation of strong (bond-breaking) from weak (dynamical)
  correlation. \emph{Chem. Phys.} \textbf{2012}, \emph{401}, 119--124\relax
\mciteBstWouldAddEndPuncttrue
\mciteSetBstMidEndSepPunct{\mcitedefaultmidpunct}
{\mcitedefaultendpunct}{\mcitedefaultseppunct}\relax
\EndOfBibitem
\bibitem[Georges and Kotliar(1992)Georges, and Kotliar]{georges_1992}
Georges,~A.; Kotliar,~G. Hubbard model in infinite dimensions. \emph{Phys. Rev.
  B} \textbf{1992}, \emph{45}, 6479--6483\relax
\mciteBstWouldAddEndPuncttrue
\mciteSetBstMidEndSepPunct{\mcitedefaultmidpunct}
{\mcitedefaultendpunct}{\mcitedefaultseppunct}\relax
\EndOfBibitem
\bibitem[Georges \latin{et~al.}(1996)Georges, Kotliar, Krauth, and
  Rozenberg]{georges_1996}
Georges,~A.; Kotliar,~G.; Krauth,~W.; Rozenberg,~M.~J. Dynamical mean-field
  theory of strongly correlated fermion systems and the limit of infinite
  dimensions. \emph{Rev. Mod. Phys.} \textbf{1996}, \emph{68}, 13--125\relax
\mciteBstWouldAddEndPuncttrue
\mciteSetBstMidEndSepPunct{\mcitedefaultmidpunct}
{\mcitedefaultendpunct}{\mcitedefaultseppunct}\relax
\EndOfBibitem
\bibitem[Anisimov and Izyumov(2010)Anisimov, and
  Izyumov]{anisimov2010electronic}
Anisimov,~V.; Izyumov,~Y. \emph{Electronic Structure of Strongly Correlated
  Materials}; Springer: Berlin, 2010\relax
\mciteBstWouldAddEndPuncttrue
\mciteSetBstMidEndSepPunct{\mcitedefaultmidpunct}
{\mcitedefaultendpunct}{\mcitedefaultseppunct}\relax
\EndOfBibitem
\bibitem[Kuramoto(2020)]{kuramoto2020quantum}
Kuramoto,~Y. \emph{Quantum Many-Body Physics}; Springer: Tokyo, 2020\relax
\mciteBstWouldAddEndPuncttrue
\mciteSetBstMidEndSepPunct{\mcitedefaultmidpunct}
{\mcitedefaultendpunct}{\mcitedefaultseppunct}\relax
\EndOfBibitem
\bibitem[Foulkes \latin{et~al.}(2001)Foulkes, Mitas, Needs, and
  Rajagopal]{foulkes2001quantum}
Foulkes,~W. M.~C.; Mitas,~L.; Needs,~R.~J.; Rajagopal,~G. Quantum Monte Carlo
  simulations of solids. \emph{Rev. Mod. Phys.} \textbf{2001}, \emph{73},
  33--83\relax
\mciteBstWouldAddEndPuncttrue
\mciteSetBstMidEndSepPunct{\mcitedefaultmidpunct}
{\mcitedefaultendpunct}{\mcitedefaultseppunct}\relax
\EndOfBibitem
\bibitem[Zhang \latin{et~al.}(2018)Zhang, Malone, and Morales]{shuai_2018}
Zhang,~S.; Malone,~F.~D.; Morales,~M.~A. Auxiliary-field quantum Monte Carlo
  calculations of the structural properties of nickel oxide. \emph{J. Chem.
  Phys.} \textbf{2018}, \emph{149}, 164102\relax
\mciteBstWouldAddEndPuncttrue
\mciteSetBstMidEndSepPunct{\mcitedefaultmidpunct}
{\mcitedefaultendpunct}{\mcitedefaultseppunct}\relax
\EndOfBibitem
\bibitem[Korepin and Essler(1994)Korepin, and Essler]{korepin1994exactly}
Korepin,~V.~E., Essler,~F.~H., Eds. \emph{Exactly solvable models of strongly
  correlated electrons}; World Scientific: Singapore, 1994\relax
\mciteBstWouldAddEndPuncttrue
\mciteSetBstMidEndSepPunct{\mcitedefaultmidpunct}
{\mcitedefaultendpunct}{\mcitedefaultseppunct}\relax
\EndOfBibitem
\bibitem[Gross(1996)]{gross1996role}
Gross,~D.~J. The role of symmetry in fundamental physics. \emph{Proc. Nat.
  Acad. Sci.} \textbf{1996}, \emph{93}, 14256--14259\relax
\mciteBstWouldAddEndPuncttrue
\mciteSetBstMidEndSepPunct{\mcitedefaultmidpunct}
{\mcitedefaultendpunct}{\mcitedefaultseppunct}\relax
\EndOfBibitem
\bibitem[Raimes(1963)]{raimes1963wave}
Raimes,~S. \emph{The wave mechanics of electrons in metals}; North-Holland:
  Amsterdam, 1963\relax
\mciteBstWouldAddEndPuncttrue
\mciteSetBstMidEndSepPunct{\mcitedefaultmidpunct}
{\mcitedefaultendpunct}{\mcitedefaultseppunct}\relax
\EndOfBibitem
\bibitem[Fulde(1999)]{fulde1999solids}
Fulde,~P. Solids with weak and strong electron correlations. In \emph{Electron
  Correlation in the Solid State}; Imperial Collage Press: London, 1999;
  Chapter 2, pp 47--102\relax
\mciteBstWouldAddEndPuncttrue
\mciteSetBstMidEndSepPunct{\mcitedefaultmidpunct}
{\mcitedefaultendpunct}{\mcitedefaultseppunct}\relax
\EndOfBibitem
\bibitem[Coldwell‐Horsfall and Maradudin(1960)Coldwell‐Horsfall, and
  Maradudin]{coldwell-horsfall1960zero}
Coldwell‐Horsfall,~R.~A.; Maradudin,~A.~A. Zero‐Point Energy of an Electron
  Lattice. \emph{J. Math. Phys.} \textbf{1960}, \emph{1}, 395--404\relax
\mciteBstWouldAddEndPuncttrue
\mciteSetBstMidEndSepPunct{\mcitedefaultmidpunct}
{\mcitedefaultendpunct}{\mcitedefaultseppunct}\relax
\EndOfBibitem
\bibitem[Bohm and Pines(1951)Bohm, and Pines]{bohm1951collective}
Bohm,~D.; Pines,~D. A Collective Description of Electron Interactions. I.
  Magnetic Interactions. \emph{Phys. Rev.} \textbf{1951}, \emph{82},
  625--634\relax
\mciteBstWouldAddEndPuncttrue
\mciteSetBstMidEndSepPunct{\mcitedefaultmidpunct}
{\mcitedefaultendpunct}{\mcitedefaultseppunct}\relax
\EndOfBibitem
\bibitem[Pines(1955)]{pines1956electron}
Pines,~D. Electron Interaction in Metals. In \emph{Solid State Physics};
  Academic Press: New York, 1955; Vol.~1; pp 367--450\relax
\mciteBstWouldAddEndPuncttrue
\mciteSetBstMidEndSepPunct{\mcitedefaultmidpunct}
{\mcitedefaultendpunct}{\mcitedefaultseppunct}\relax
\EndOfBibitem
\bibitem[Slater(1956)]{slater1956electronic}
Slater,~J.~C. The electronic structure of solids. In \emph{Encyclopedia of
  Physics/Handbuch der Physik}; Springer: Berlin, 1956; Vol.~19; pp
  1--136\relax
\mciteBstWouldAddEndPuncttrue
\mciteSetBstMidEndSepPunct{\mcitedefaultmidpunct}
{\mcitedefaultendpunct}{\mcitedefaultseppunct}\relax
\EndOfBibitem
\bibitem[Brueckner(1969)]{brueckner1969correlation}
Brueckner,~K.~A. The Correlation Energy of a Non-Uniform Electron Gas. In
  \emph{Advances in Chemical Physics}; John Wiley \& Sons, Ltd: London, 1969;
  Vol.~14; Chapter 7, pp 215--236\relax
\mciteBstWouldAddEndPuncttrue
\mciteSetBstMidEndSepPunct{\mcitedefaultmidpunct}
{\mcitedefaultendpunct}{\mcitedefaultseppunct}\relax
\EndOfBibitem
\bibitem[Slater(1933)]{slater1933virial}
Slater,~J.~C. The Virial and Molecular Structure. \emph{J. Chem. Phys.}
  \textbf{1933}, \emph{1}, 687--691\relax
\mciteBstWouldAddEndPuncttrue
\mciteSetBstMidEndSepPunct{\mcitedefaultmidpunct}
{\mcitedefaultendpunct}{\mcitedefaultseppunct}\relax
\EndOfBibitem
\bibitem[March(1958)]{march1958kinetic}
March,~N.~H. Kinetic and Potential Energies of an Electron Gas. \emph{Phys.
  Rev.} \textbf{1958}, \emph{110}, 604--605\relax
\mciteBstWouldAddEndPuncttrue
\mciteSetBstMidEndSepPunct{\mcitedefaultmidpunct}
{\mcitedefaultendpunct}{\mcitedefaultseppunct}\relax
\EndOfBibitem
\bibitem[Argyres(1967)]{argyres1967virial}
Argyres,~P.~N. Virial Theorem for the Homogeneous Electron Gas. \emph{Phys.
  Rev.} \textbf{1967}, \emph{154}, 410--413\relax
\mciteBstWouldAddEndPuncttrue
\mciteSetBstMidEndSepPunct{\mcitedefaultmidpunct}
{\mcitedefaultendpunct}{\mcitedefaultseppunct}\relax
\EndOfBibitem
\bibitem[Ruedenberg(1962)]{ruedenberg1962physical}
Ruedenberg,~K. The Physical Nature of the Chemical Bond. \emph{Rev. Mod. Phys.}
  \textbf{1962}, \emph{34}, 326--376\relax
\mciteBstWouldAddEndPuncttrue
\mciteSetBstMidEndSepPunct{\mcitedefaultmidpunct}
{\mcitedefaultendpunct}{\mcitedefaultseppunct}\relax
\EndOfBibitem
\bibitem[Hubbard(1963)]{hubbard1963electron}
Hubbard,~J. Electron correlations in narrow energy bands. \emph{Proc. Roy. Soc.
  A} \textbf{1963}, \emph{276}, 238--257\relax
\mciteBstWouldAddEndPuncttrue
\mciteSetBstMidEndSepPunct{\mcitedefaultmidpunct}
{\mcitedefaultendpunct}{\mcitedefaultseppunct}\relax
\EndOfBibitem
\bibitem[Janesko(2017)]{janesko2017strong}
Janesko,~B.~G. Strong correlation in surface chemistry. \emph{Mol. Simul.}
  \textbf{2017}, \emph{43}, 394--405\relax
\mciteBstWouldAddEndPuncttrue
\mciteSetBstMidEndSepPunct{\mcitedefaultmidpunct}
{\mcitedefaultendpunct}{\mcitedefaultseppunct}\relax
\EndOfBibitem
\bibitem[Motta \latin{et~al.}(2017)Motta, Ceperley, Chan, Gomez, Gull, Guo,
  Jim\'enez-Hoyos, Lan, Li, Ma, Millis, Prokof'ev, Ray, Scuseria, Sorella,
  Stoudenmire, Sun, Tupitsyn, White, Zgid, and Zhang]{motta2017towards}
Motta,~M.; Ceperley,~D.~M.; Chan,~G. K.-L.; Gomez,~J.~A.; Gull,~E.; Guo,~S.;
  Jim\'enez-Hoyos,~C.~A.; Lan,~T.~N.; Li,~J.; Ma,~F.; Millis,~A.~J.;
  Prokof'ev,~N.~V.; Ray,~U.; Scuseria,~G.~E.; Sorella,~S.; Stoudenmire,~E.~M.;
  Sun,~Q.; Tupitsyn,~I.~S.; White,~S.~R.; Zgid,~D.; Zhang,~S. Towards the
  Solution of the Many-Electron Problem in Real Materials: Equation of State of
  the Hydrogen Chain with State-of-the-Art Many-Body Methods. \emph{Phys. Rev.
  X} \textbf{2017}, \emph{7}, 031059\relax
\mciteBstWouldAddEndPuncttrue
\mciteSetBstMidEndSepPunct{\mcitedefaultmidpunct}
{\mcitedefaultendpunct}{\mcitedefaultseppunct}\relax
\EndOfBibitem
\bibitem[Motta \latin{et~al.}(2020)Motta, Genovese, Ma, Cui, Sawaya, Chan,
  Chepiga, Helms, Jim\'enez-Hoyos, Millis, Ray, Ronca, Shi, Sorella,
  Stoudenmire, White, and Zhang]{motta2020ground}
Motta,~M.; Genovese,~C.; Ma,~F.; Cui,~Z.-H.; Sawaya,~R.; Chan,~G. K.-L.;
  Chepiga,~N.; Helms,~P.; Jim\'enez-Hoyos,~C.; Millis,~A.~J.; Ray,~U.;
  Ronca,~E.; Shi,~H.; Sorella,~S.; Stoudenmire,~E.~M.; White,~S.~R.; Zhang,~S.
  Ground-State Properties of the Hydrogen Chain: Dimerization,
  Insulator-to-Metal Transition, and Magnetic Phases. \emph{Phys. Rev. X}
  \textbf{2020}, \emph{10}, 031058\relax
\mciteBstWouldAddEndPuncttrue
\mciteSetBstMidEndSepPunct{\mcitedefaultmidpunct}
{\mcitedefaultendpunct}{\mcitedefaultseppunct}\relax
\EndOfBibitem
\bibitem[Sinano{\u{g}}lu and Tuan(1963)Sinano{\u{g}}lu, and
  Tuan]{sinanouglu1963many}
Sinano{\u{g}}lu,~O.; Tuan,~D. F.-t. Many-Electron Theory of Atoms and
  Molecules. III. Effect of Correlation on Orbitals. \emph{J. Chem. Phys.}
  \textbf{1963}, \emph{38}, 1740--1748\relax
\mciteBstWouldAddEndPuncttrue
\mciteSetBstMidEndSepPunct{\mcitedefaultmidpunct}
{\mcitedefaultendpunct}{\mcitedefaultseppunct}\relax
\EndOfBibitem
\bibitem[Prendergast \latin{et~al.}(2001)Prendergast, Nolan, Filippi, Fahy, and
  Greer]{prendergast2001impact}
Prendergast,~D.; Nolan,~M.; Filippi,~C.; Fahy,~S.; Greer,~J. Impact of
  electron--electron cusp on configuration interaction energies. \emph{J. Chem.
  Phys.} \textbf{2001}, \emph{115}, 1626--1634\relax
\mciteBstWouldAddEndPuncttrue
\mciteSetBstMidEndSepPunct{\mcitedefaultmidpunct}
{\mcitedefaultendpunct}{\mcitedefaultseppunct}\relax
\EndOfBibitem
\bibitem[Mok \latin{et~al.}(1996)Mok, Neumann, and Handy]{mok1996dynamical}
Mok,~D.~K.; Neumann,~R.; Handy,~N.~C. Dynamical and nondynamical correlation.
  \emph{J. Phys. Chem.} \textbf{1996}, \emph{100}, 6225--6230\relax
\mciteBstWouldAddEndPuncttrue
\mciteSetBstMidEndSepPunct{\mcitedefaultmidpunct}
{\mcitedefaultendpunct}{\mcitedefaultseppunct}\relax
\EndOfBibitem
\bibitem[Hollett and Gill(2011)Hollett, and Gill]{hollett2011two}
Hollett,~J.~W.; Gill,~P.~M. The two faces of static correlation. \emph{J. Chem.
  Phys.} \textbf{2011}, \emph{134}, 114111\relax
\mciteBstWouldAddEndPuncttrue
\mciteSetBstMidEndSepPunct{\mcitedefaultmidpunct}
{\mcitedefaultendpunct}{\mcitedefaultseppunct}\relax
\EndOfBibitem
\bibitem[Benavides-Riveros \latin{et~al.}(2017)Benavides-Riveros, Lathiotakis,
  and Marques]{benavides2017towards}
Benavides-Riveros,~C.~L.; Lathiotakis,~N.~N.; Marques,~M.~A. Towards a formal
  definition of static and dynamic electronic correlations. \emph{Phys. Chem.
  Chem. Phys.} \textbf{2017}, \emph{19}, 12655--12664\relax
\mciteBstWouldAddEndPuncttrue
\mciteSetBstMidEndSepPunct{\mcitedefaultmidpunct}
{\mcitedefaultendpunct}{\mcitedefaultseppunct}\relax
\EndOfBibitem
\bibitem[Bulik \latin{et~al.}(2015)Bulik, Henderson, and
  Scuseria]{bulik2015can}
Bulik,~I.~W.; Henderson,~T.~M.; Scuseria,~G.~E. Can single-reference coupled
  cluster theory describe static correlation? \emph{J. Chem. Theory Comput.}
  \textbf{2015}, \emph{11}, 3171--3179\relax
\mciteBstWouldAddEndPuncttrue
\mciteSetBstMidEndSepPunct{\mcitedefaultmidpunct}
{\mcitedefaultendpunct}{\mcitedefaultseppunct}\relax
\EndOfBibitem
\bibitem[Karton \latin{et~al.}(2006)Karton, Rabinovich, Martin, and
  Ruscic]{karton2006w4}
Karton,~A.; Rabinovich,~E.; Martin,~J.~M.; Ruscic,~B. W4 theory for
  computational thermochemistry: In pursuit of confident sub-kJ/mol
  predictions. \emph{J. Chem. Phys.} \textbf{2006}, \emph{125}, 144108\relax
\mciteBstWouldAddEndPuncttrue
\mciteSetBstMidEndSepPunct{\mcitedefaultmidpunct}
{\mcitedefaultendpunct}{\mcitedefaultseppunct}\relax
\EndOfBibitem
\bibitem[Hait \latin{et~al.}(2019)Hait, Tubman, Levine, Whaley, and
  Head-Gordon]{hait2019levels}
Hait,~D.; Tubman,~N.~M.; Levine,~D.~S.; Whaley,~K.~B.; Head-Gordon,~M. What
  levels of coupled cluster theory are appropriate for transition metal
  systems? A study using near-exact quantum chemical values for 3d transition
  metal binary compounds. \emph{J. Chem. Theory Comput.} \textbf{2019},
  \emph{15}, 5370--5385\relax
\mciteBstWouldAddEndPuncttrue
\mciteSetBstMidEndSepPunct{\mcitedefaultmidpunct}
{\mcitedefaultendpunct}{\mcitedefaultseppunct}\relax
\EndOfBibitem
\bibitem[Roos \latin{et~al.}(1980)Roos, Taylor, and Sigbahn]{roos1980complete}
Roos,~B.~O.; Taylor,~P.~R.; Sigbahn,~P.~E. A complete active space SCF method
  (CASSCF) using a density matrix formulated super-CI approach. \emph{Chem.
  Phys.} \textbf{1980}, \emph{48}, 157--173\relax
\mciteBstWouldAddEndPuncttrue
\mciteSetBstMidEndSepPunct{\mcitedefaultmidpunct}
{\mcitedefaultendpunct}{\mcitedefaultseppunct}\relax
\EndOfBibitem
\bibitem[Chan and Sharma(2011)Chan, and Sharma]{chan2011density}
Chan,~G. K.-L.; Sharma,~S. The density matrix renormalization group in quantum
  chemistry. \emph{Annu. Rev. Phys. Chem.} \textbf{2011}, \emph{62},
  465--481\relax
\mciteBstWouldAddEndPuncttrue
\mciteSetBstMidEndSepPunct{\mcitedefaultmidpunct}
{\mcitedefaultendpunct}{\mcitedefaultseppunct}\relax
\EndOfBibitem
\bibitem[Mouesca(2014)]{mouesca2014density}
Mouesca,~J.-M. Density functional theory--broken symmetry (DFT--BS) methodology
  applied to electronic and magnetic properties of bioinorganic prosthetic
  groups. In \emph{Metalloproteins}; Springer: New York, 2014; pp
  269--296\relax
\mciteBstWouldAddEndPuncttrue
\mciteSetBstMidEndSepPunct{\mcitedefaultmidpunct}
{\mcitedefaultendpunct}{\mcitedefaultseppunct}\relax
\EndOfBibitem
\bibitem[Scuseria \latin{et~al.}(2011)Scuseria, Jim{\'e}nez-Hoyos, Henderson,
  Samanta, and Ellis]{scuseria2011projected}
Scuseria,~G.~E.; Jim{\'e}nez-Hoyos,~C.~A.; Henderson,~T.~M.; Samanta,~K.;
  Ellis,~J.~K. Projected quasiparticle theory for molecular electronic
  structure. \emph{J. Chem. Phys.} \textbf{2011}, \emph{135}, 124108\relax
\mciteBstWouldAddEndPuncttrue
\mciteSetBstMidEndSepPunct{\mcitedefaultmidpunct}
{\mcitedefaultendpunct}{\mcitedefaultseppunct}\relax
\EndOfBibitem
\bibitem[Helgaker \latin{et~al.}(2000)Helgaker, Jorgensen, and
  Olsen]{helgaker2014molecular}
Helgaker,~T.; Jorgensen,~P.; Olsen,~J. \emph{Molecular electronic-structure
  theory}; John Wiley \& Sons: Chichester, 2000\relax
\mciteBstWouldAddEndPuncttrue
\mciteSetBstMidEndSepPunct{\mcitedefaultmidpunct}
{\mcitedefaultendpunct}{\mcitedefaultseppunct}\relax
\EndOfBibitem
\bibitem[Shavitt and Bartlett(2009)Shavitt, and Bartlett]{shavitt2009many}
Shavitt,~I.; Bartlett,~R.~J. \emph{Many-body methods in chemistry and physics:
  MBPT and coupled-cluster theory}; Cambridge University Press: Cambridge,
  2009\relax
\mciteBstWouldAddEndPuncttrue
\mciteSetBstMidEndSepPunct{\mcitedefaultmidpunct}
{\mcitedefaultendpunct}{\mcitedefaultseppunct}\relax
\EndOfBibitem
\bibitem[{\v{C}}{\'\i}{\v{z}}ek(1966)]{cizek1966correlation}
{\v{C}}{\'\i}{\v{z}}ek,~J. On the {{Correlation Problem}} in {{Atomic}} and
  {{Molecular Systems}}. {{Calculation}} of {{Wavefunction Components}} in
  {{Ursell-Type Expansion Using Quantum-Field Theoretical Methods}}. \emph{J.
  Chem. Phys.} \textbf{1966}, \emph{45}, 4256--4266\relax
\mciteBstWouldAddEndPuncttrue
\mciteSetBstMidEndSepPunct{\mcitedefaultmidpunct}
{\mcitedefaultendpunct}{\mcitedefaultseppunct}\relax
\EndOfBibitem
\bibitem[Paldus \latin{et~al.}()Paldus, {\v{C}}{\'\i}{\v{z}}ek, and
  Shavitt]{paldus1972correlation}
Paldus,~J.; {\v{C}}{\'\i}{\v{z}}ek,~J.; Shavitt,~I. Correlation {{Problems}} in
  {{Atomic}} and {{Molecular Systems}}. {{IV}}. {{Extended Coupled-Pair
  Many-Electron Theory}} and {{Its Application}} to the
  {{BH}}{\textsubscript{3}} {{Molecule}}. \emph{Phys. Rev. A} \emph{5},
  50--67\relax
\mciteBstWouldAddEndPuncttrue
\mciteSetBstMidEndSepPunct{\mcitedefaultmidpunct}
{\mcitedefaultendpunct}{\mcitedefaultseppunct}\relax
\EndOfBibitem
\bibitem[Arponen(1983)]{arponen1983variational}
Arponen,~J. Variational principles and linked-cluster exp S expansions for
  static and dynamic many-body problems. \emph{Ann. Phys.} \textbf{1983},
  \emph{151}, 311--382\relax
\mciteBstWouldAddEndPuncttrue
\mciteSetBstMidEndSepPunct{\mcitedefaultmidpunct}
{\mcitedefaultendpunct}{\mcitedefaultseppunct}\relax
\EndOfBibitem
\bibitem[Faulstich \latin{et~al.}(2019)Faulstich, Laestadius, Legeza,
  Schneider, and Kvaal]{faulstich2019analysis}
Faulstich,~F.~M.; Laestadius,~A.; Legeza,~O.; Schneider,~R.; Kvaal,~S. Analysis
  of the Tailored Coupled-Cluster Method in Quantum Chemistry. \emph{SIAM J.
  Numer. Anal.} \textbf{2019}, \emph{57}, 2579--2607\relax
\mciteBstWouldAddEndPuncttrue
\mciteSetBstMidEndSepPunct{\mcitedefaultmidpunct}
{\mcitedefaultendpunct}{\mcitedefaultseppunct}\relax
\EndOfBibitem
\bibitem[Csirik and Laestadius(2021)Csirik, and Laestadius]{csirik2021coupled}
Csirik,~M.~A.; Laestadius,~A. Coupled-cluster theory revisited. \emph{arXiv
  preprint arXiv:2105.13134} \textbf{2021}, \relax
\mciteBstWouldAddEndPunctfalse
\mciteSetBstMidEndSepPunct{\mcitedefaultmidpunct}
{}{\mcitedefaultseppunct}\relax
\EndOfBibitem
\bibitem[Sch{\"u}tz and Werner(2001)Sch{\"u}tz, and Werner]{schutz2001low}
Sch{\"u}tz,~M.; Werner,~H.-J. Low-order scaling local electron correlation
  methods. IV. Linear scaling local coupled-cluster (LCCSD). \emph{J. Chem.
  Phys.} \textbf{2001}, \emph{114}, 661--681\relax
\mciteBstWouldAddEndPuncttrue
\mciteSetBstMidEndSepPunct{\mcitedefaultmidpunct}
{\mcitedefaultendpunct}{\mcitedefaultseppunct}\relax
\EndOfBibitem
\bibitem[Riplinger and Neese(2013)Riplinger, and Neese]{riplinger2013efficient}
Riplinger,~C.; Neese,~F. An efficient and near linear scaling pair natural
  orbital based local coupled cluster method. \emph{J. Chem. Phys.}
  \textbf{2013}, \emph{138}, 034106\relax
\mciteBstWouldAddEndPuncttrue
\mciteSetBstMidEndSepPunct{\mcitedefaultmidpunct}
{\mcitedefaultendpunct}{\mcitedefaultseppunct}\relax
\EndOfBibitem
\bibitem[Shiozaki \latin{et~al.}(2009)Shiozaki, Valeev, and
  Hirata]{shiozaki2009explicitly}
Shiozaki,~T.; Valeev,~E.~F.; Hirata,~S. Explicitly correlated coupled-cluster
  methods. In \emph{Annual Reports in Computational Chemistry}; Elsevier:
  Amsterdam, 2009; Vol.~5; Chapter 6, pp 131--148\relax
\mciteBstWouldAddEndPuncttrue
\mciteSetBstMidEndSepPunct{\mcitedefaultmidpunct}
{\mcitedefaultendpunct}{\mcitedefaultseppunct}\relax
\EndOfBibitem
\bibitem[Izs{\'a}k(2020)]{izsak2020single}
Izs{\'a}k,~R. Single-reference coupled cluster methods for computing excitation
  energies in large molecules: The efficiency and accuracy of approximations.
  \emph{Wiley Interdiscip. Rev. Comput. Mol. Sci.} \textbf{2020}, \emph{10},
  e1445\relax
\mciteBstWouldAddEndPuncttrue
\mciteSetBstMidEndSepPunct{\mcitedefaultmidpunct}
{\mcitedefaultendpunct}{\mcitedefaultseppunct}\relax
\EndOfBibitem
\bibitem[Lyakh \latin{et~al.}(2012)Lyakh, Musia{\l}, Lotrich, and
  Bartlett]{lyakh2012multireference}
Lyakh,~D.~I.; Musia{\l},~M.; Lotrich,~V.~F.; Bartlett,~R.~J. Multireference
  nature of chemistry: The coupled-cluster view. \emph{Chem. Rev.}
  \textbf{2012}, \emph{112}, 182--243\relax
\mciteBstWouldAddEndPuncttrue
\mciteSetBstMidEndSepPunct{\mcitedefaultmidpunct}
{\mcitedefaultendpunct}{\mcitedefaultseppunct}\relax
\EndOfBibitem
\bibitem[Huron \latin{et~al.}(1973)Huron, Malrieu, and
  Rancurel]{huron1973iterative}
Huron,~B.; Malrieu,~J.~P.; Rancurel,~P. Iterative perturbation calculations of
  ground and excited state energies from multiconfigurational zeroth‐order
  wavefunctions. \emph{J. Chem. Phys.} \textbf{1973}, \emph{58},
  5745--5759\relax
\mciteBstWouldAddEndPuncttrue
\mciteSetBstMidEndSepPunct{\mcitedefaultmidpunct}
{\mcitedefaultendpunct}{\mcitedefaultseppunct}\relax
\EndOfBibitem
\bibitem[Austin \latin{et~al.}(2012)Austin, Zubarev, and
  Lester~Jr]{austin2012quantum}
Austin,~B.~M.; Zubarev,~D.~Y.; Lester~Jr,~W.~A. Quantum Monte Carlo and related
  approaches. \emph{Chem. Rev.} \textbf{2012}, \emph{112}, 263--288\relax
\mciteBstWouldAddEndPuncttrue
\mciteSetBstMidEndSepPunct{\mcitedefaultmidpunct}
{\mcitedefaultendpunct}{\mcitedefaultseppunct}\relax
\EndOfBibitem
\bibitem[Booth \latin{et~al.}(2009)Booth, Thom, and Alavi]{booth2009fermion}
Booth,~G.~H.; Thom,~A.~J.; Alavi,~A. Fermion Monte Carlo without fixed nodes: A
  game of life, death, and annihilation in Slater determinant space. \emph{J.
  Chem. Phys.} \textbf{2009}, \emph{131}, 054106\relax
\mciteBstWouldAddEndPuncttrue
\mciteSetBstMidEndSepPunct{\mcitedefaultmidpunct}
{\mcitedefaultendpunct}{\mcitedefaultseppunct}\relax
\EndOfBibitem
\bibitem[Petruzielo \latin{et~al.}(2012)Petruzielo, Holmes, Changlani,
  Nightingale, and Umrigar]{petruzielo2012}
Petruzielo,~F.~R.; Holmes,~A.~A.; Changlani,~H.~J.; Nightingale,~M.~P.;
  Umrigar,~C.~J. Semistochastic Projector Monte Carlo Method. \emph{Phys. Rev.
  Lett.} \textbf{2012}, \emph{109}, 230201\relax
\mciteBstWouldAddEndPuncttrue
\mciteSetBstMidEndSepPunct{\mcitedefaultmidpunct}
{\mcitedefaultendpunct}{\mcitedefaultseppunct}\relax
\EndOfBibitem
\bibitem[Spencer \latin{et~al.}(2012)Spencer, Blunt, and Foulkes]{spencer2012}
Spencer,~J.~S.; Blunt,~N.~S.; Foulkes,~W.~M. The sign problem and population
  dynamics in the full configuration interaction quantum Monte Carlo method.
  \emph{J. Chem. Phys.} \textbf{2012}, \emph{136}, 054110\relax
\mciteBstWouldAddEndPuncttrue
\mciteSetBstMidEndSepPunct{\mcitedefaultmidpunct}
{\mcitedefaultendpunct}{\mcitedefaultseppunct}\relax
\EndOfBibitem
\bibitem[Cleland \latin{et~al.}(2010)Cleland, Booth, and Alavi]{cleland2010}
Cleland,~D.; Booth,~G.~H.; Alavi,~A. Communications: Survival of the fittest:
  Accelerating convergence in full configuration-interaction quantum Monte
  Carlo. \emph{J. Chem. Phys.} \textbf{2010}, \emph{132}, 041103\relax
\mciteBstWouldAddEndPuncttrue
\mciteSetBstMidEndSepPunct{\mcitedefaultmidpunct}
{\mcitedefaultendpunct}{\mcitedefaultseppunct}\relax
\EndOfBibitem
\bibitem[Cleland \latin{et~al.}(2011)Cleland, Booth, and Alavi]{cleland2011}
Cleland,~D.~M.; Booth,~G.~H.; Alavi,~A. A study of electron affinities using
  the initiator approach to full configuration interaction quantum Monte Carlo.
  \emph{J. Chem. Phys.} \textbf{2011}, \emph{134}, 024112\relax
\mciteBstWouldAddEndPuncttrue
\mciteSetBstMidEndSepPunct{\mcitedefaultmidpunct}
{\mcitedefaultendpunct}{\mcitedefaultseppunct}\relax
\EndOfBibitem
\bibitem[Eriksen(2020)]{eriksen2020shape}
Eriksen,~J.~J. The shape of full configuration interaction to come. \emph{J.
  Phys. Chem. Lett.} \textbf{2020}, \emph{12}, 418--432\relax
\mciteBstWouldAddEndPuncttrue
\mciteSetBstMidEndSepPunct{\mcitedefaultmidpunct}
{\mcitedefaultendpunct}{\mcitedefaultseppunct}\relax
\EndOfBibitem
\bibitem[Eriksen and Gauss(2018)Eriksen, and Gauss]{eriksen2018many}
Eriksen,~J.~J.; Gauss,~J. Many-body expanded full configuration interaction. I.
  Weakly correlated regime. \emph{J. Chem. Theory Comput.} \textbf{2018},
  \emph{14}, 5180--5191\relax
\mciteBstWouldAddEndPuncttrue
\mciteSetBstMidEndSepPunct{\mcitedefaultmidpunct}
{\mcitedefaultendpunct}{\mcitedefaultseppunct}\relax
\EndOfBibitem
\bibitem[Eriksen and Gauss(2019)Eriksen, and Gauss]{eriksen2019many}
Eriksen,~J.~J.; Gauss,~J. Many-body expanded full configuration interaction.
  II. Strongly correlated regime. \emph{J. Chem. Theory Comput.} \textbf{2019},
  \emph{15}, 4873--4884\relax
\mciteBstWouldAddEndPuncttrue
\mciteSetBstMidEndSepPunct{\mcitedefaultmidpunct}
{\mcitedefaultendpunct}{\mcitedefaultseppunct}\relax
\EndOfBibitem
\bibitem[White(1992)]{white1992density}
White,~S.~R. Density matrix formulation for quantum renormalization groups.
  \emph{Phys. Rev. Lett.} \textbf{1992}, \emph{69}, 2863--2866\relax
\mciteBstWouldAddEndPuncttrue
\mciteSetBstMidEndSepPunct{\mcitedefaultmidpunct}
{\mcitedefaultendpunct}{\mcitedefaultseppunct}\relax
\EndOfBibitem
\bibitem[Chilkuri and Neese(2021)Chilkuri, and Neese]{chilkuri2021comparison}
Chilkuri,~V.~G.; Neese,~F. Comparison of many-particle representations for
  selected-CI I: A tree based approach. \emph{J. Comput. Chem.} \textbf{2021},
  \emph{42}, 982--1005\relax
\mciteBstWouldAddEndPuncttrue
\mciteSetBstMidEndSepPunct{\mcitedefaultmidpunct}
{\mcitedefaultendpunct}{\mcitedefaultseppunct}\relax
\EndOfBibitem
\bibitem[{Li~Manni} \latin{et~al.}(2020){Li~Manni}, Dobrautz, and
  Alavi]{li2020compression}
{Li~Manni},~G.; Dobrautz,~W.; Alavi,~A. Compression of spin-adapted
  multiconfigurational wave functions in exchange-coupled polynuclear spin
  systems. \emph{J. Chem. Theory Comput.} \textbf{2020}, \emph{16},
  2202--2215\relax
\mciteBstWouldAddEndPuncttrue
\mciteSetBstMidEndSepPunct{\mcitedefaultmidpunct}
{\mcitedefaultendpunct}{\mcitedefaultseppunct}\relax
\EndOfBibitem
\bibitem[{Li~Manni}(2021)]{manni2021modeling}
{Li~Manni},~G. Modeling magnetic interactions in high-valent trinuclear
  [Mn$_3^{(IV)}$O$_4$]$^{4+}$ complexes through highly compressed
  multi-configurational wave functions. \emph{Phys. Chem. Chem. Phys.}
  \textbf{2021}, \emph{23}, 19766--19780\relax
\mciteBstWouldAddEndPuncttrue
\mciteSetBstMidEndSepPunct{\mcitedefaultmidpunct}
{\mcitedefaultendpunct}{\mcitedefaultseppunct}\relax
\EndOfBibitem
\bibitem[{Li~Manni} \latin{et~al.}(2021){Li~Manni}, Dobrautz, Bogdanov, Guther,
  and Alavi]{li2021resolution}
{Li~Manni},~G.; Dobrautz,~W.; Bogdanov,~N.~A.; Guther,~K.; Alavi,~A. Resolution
  of low-energy states in spin-exchange transition-metal clusters: Case study
  of singlet states in [Fe(III)$_4$S$_4$] cubanes. \emph{J. Phys. Chem. A}
  \textbf{2021}, \emph{125}, 4727--4740\relax
\mciteBstWouldAddEndPuncttrue
\mciteSetBstMidEndSepPunct{\mcitedefaultmidpunct}
{\mcitedefaultendpunct}{\mcitedefaultseppunct}\relax
\EndOfBibitem
\bibitem[Shee \latin{et~al.}(2021)Shee, Loipersberger, Hait, Lee, and
  Head-Gordon]{shee2021revealing}
Shee,~J.; Loipersberger,~M.; Hait,~D.; Lee,~J.; Head-Gordon,~M. Revealing the
  nature of electron correlation in transition metal complexes with symmetry
  breaking and chemical intuition. \emph{J. Chem. Phys.} \textbf{2021},
  \emph{154}, 194109\relax
\mciteBstWouldAddEndPuncttrue
\mciteSetBstMidEndSepPunct{\mcitedefaultmidpunct}
{\mcitedefaultendpunct}{\mcitedefaultseppunct}\relax
\EndOfBibitem
\bibitem[Benavides-Riveros \latin{et~al.}(2017)Benavides-Riveros, Lathiotakis,
  Schilling, and Marques]{benavides2017relating}
Benavides-Riveros,~C.~L.; Lathiotakis,~N.~N.; Schilling,~C.; Marques,~M.~A.
  Relating correlation measures: The importance of the energy gap. \emph{Phys.
  Rev. A} \textbf{2017}, \emph{95}, 032507\relax
\mciteBstWouldAddEndPuncttrue
\mciteSetBstMidEndSepPunct{\mcitedefaultmidpunct}
{\mcitedefaultendpunct}{\mcitedefaultseppunct}\relax
\EndOfBibitem
\bibitem[Jiang \latin{et~al.}(2012)Jiang, DeYonker, and
  Wilson]{jiang2012multireference}
Jiang,~W.; DeYonker,~N.~J.; Wilson,~A.~K. Multireference character for 3d
  transition-metal-containing molecules. \emph{J. Chem. Theory Comput.}
  \textbf{2012}, \emph{8}, 460--468\relax
\mciteBstWouldAddEndPuncttrue
\mciteSetBstMidEndSepPunct{\mcitedefaultmidpunct}
{\mcitedefaultendpunct}{\mcitedefaultseppunct}\relax
\EndOfBibitem
\bibitem[Lee \latin{et~al.}(1989)Lee, Rice, Scuseria, and
  Schaefer]{lee1989theoretical}
Lee,~T.~J.; Rice,~J.~E.; Scuseria,~G.~E.; Schaefer,~H.~F. Theoretical
  investigations of molecules composed only of fluorine, oxygen and nitrogen:
  determination of the equilibrium structures of FOOF, (NO)$_2$ and FNNF and
  the transition state structure for FNNF cis-trans isomerization. \emph{Theor.
  Chim. Acta} \textbf{1989}, \emph{75}, 81--98\relax
\mciteBstWouldAddEndPuncttrue
\mciteSetBstMidEndSepPunct{\mcitedefaultmidpunct}
{\mcitedefaultendpunct}{\mcitedefaultseppunct}\relax
\EndOfBibitem
\bibitem[Liakos and Neese(2011)Liakos, and Neese]{liakos2011interplay}
Liakos,~D.~G.; Neese,~F. Interplay of Correlation and Relativistic Effects in
  Correlated Calculations on Transition-Metal Complexes: The
  (Cu$_2$O$_2$)$^{2+}$ Core Revisited. \emph{J. Chem. Theory Comput.}
  \textbf{2011}, \emph{7}, 1511--1523\relax
\mciteBstWouldAddEndPuncttrue
\mciteSetBstMidEndSepPunct{\mcitedefaultmidpunct}
{\mcitedefaultendpunct}{\mcitedefaultseppunct}\relax
\EndOfBibitem
\bibitem[Head-Gordon(2003)]{head2003characterizing}
Head-Gordon,~M. Characterizing unpaired electrons from the one-particle density
  matrix. \emph{Chem. Phys. Lett.} \textbf{2003}, \emph{372}, 508--511\relax
\mciteBstWouldAddEndPuncttrue
\mciteSetBstMidEndSepPunct{\mcitedefaultmidpunct}
{\mcitedefaultendpunct}{\mcitedefaultseppunct}\relax
\EndOfBibitem
\bibitem[Ramos-Cordoba and Matito(2017)Ramos-Cordoba, and
  Matito]{ramos-cordoba2017local}
Ramos-Cordoba,~E.; Matito,~E. Local Descriptors of Dynamic and Nondynamic
  Correlation. \emph{J. Chem. Theory Comput.} \textbf{2017}, \emph{13},
  2705--2711\relax
\mciteBstWouldAddEndPuncttrue
\mciteSetBstMidEndSepPunct{\mcitedefaultmidpunct}
{\mcitedefaultendpunct}{\mcitedefaultseppunct}\relax
\EndOfBibitem
\bibitem[L{\"o}wdin(1955)]{lowdin1955quantum}
L{\"o}wdin,~P.-O. Quantum theory of many-particle systems. I. Physical
  interpretations by means of density matrices, natural spin-orbitals, and
  convergence problems in the method of configurational interaction.
  \emph{Phys. Rev.} \textbf{1955}, \emph{97}, 1474--1489\relax
\mciteBstWouldAddEndPuncttrue
\mciteSetBstMidEndSepPunct{\mcitedefaultmidpunct}
{\mcitedefaultendpunct}{\mcitedefaultseppunct}\relax
\EndOfBibitem
\bibitem[Ramos-Cordoba \latin{et~al.}(2016)Ramos-Cordoba, Salvador, and
  Matito]{ramos2016separation}
Ramos-Cordoba,~E.; Salvador,~P.; Matito,~E. Separation of dynamic and
  nondynamic correlation. \emph{Phys. Chem. Chem. Phys.} \textbf{2016},
  \emph{18}, 24015--24023\relax
\mciteBstWouldAddEndPuncttrue
\mciteSetBstMidEndSepPunct{\mcitedefaultmidpunct}
{\mcitedefaultendpunct}{\mcitedefaultseppunct}\relax
\EndOfBibitem
\bibitem[Juh{\'a}sz and Mazziotti(2006)Juh{\'a}sz, and
  Mazziotti]{juhasz2006cumulant}
Juh{\'a}sz,~T.; Mazziotti,~D.~A. The cumulant two-particle reduced density
  matrix as a measure of electron correlation and entanglement. \emph{J. Chem.
  Phys.} \textbf{2006}, \emph{125}, 174105\relax
\mciteBstWouldAddEndPuncttrue
\mciteSetBstMidEndSepPunct{\mcitedefaultmidpunct}
{\mcitedefaultendpunct}{\mcitedefaultseppunct}\relax
\EndOfBibitem
\bibitem[Crittenden(2013)]{crittenden2013hierarchy}
Crittenden,~D.~L. A Hierarchy of Static Correlation Models. \emph{J. Phys.
  Chem. A} \textbf{2013}, \emph{117}, 3852--3860\relax
\mciteBstWouldAddEndPuncttrue
\mciteSetBstMidEndSepPunct{\mcitedefaultmidpunct}
{\mcitedefaultendpunct}{\mcitedefaultseppunct}\relax
\EndOfBibitem
\bibitem[Pauncz(1979)]{pauncz2012spin}
Pauncz,~R. \emph{Spin Eigenfunctions: Construction and Use}; Plenum Press: New
  York, 1979\relax
\mciteBstWouldAddEndPuncttrue
\mciteSetBstMidEndSepPunct{\mcitedefaultmidpunct}
{\mcitedefaultendpunct}{\mcitedefaultseppunct}\relax
\EndOfBibitem
\bibitem[Perdew \latin{et~al.}(2021)Perdew, Ruzsinszky, Sun, Nepal, and
  Kaplan]{perdew2021interpretations}
Perdew,~J.~P.; Ruzsinszky,~A.; Sun,~J.; Nepal,~N.~K.; Kaplan,~A.~D.
  Interpretations of ground-state symmetry breaking and strong correlation in
  wavefunction and density functional theories. \emph{Proc. Nat. Acad. Sci.}
  \textbf{2021}, \emph{118}, e2017850118\relax
\mciteBstWouldAddEndPuncttrue
\mciteSetBstMidEndSepPunct{\mcitedefaultmidpunct}
{\mcitedefaultendpunct}{\mcitedefaultseppunct}\relax
\EndOfBibitem
\bibitem[Perdew \latin{et~al.}(2009)Perdew, Ruzsinszky, Constantin, Sun, and
  Csonka]{perdew2009some}
Perdew,~J.~P.; Ruzsinszky,~A.; Constantin,~L.~A.; Sun,~J.; Csonka,~G.~I. Some
  fundamental issues in ground-state density functional theory: A guide for the
  perplexed. \emph{J. Chem. Theory Comput.} \textbf{2009}, \emph{5},
  902--908\relax
\mciteBstWouldAddEndPuncttrue
\mciteSetBstMidEndSepPunct{\mcitedefaultmidpunct}
{\mcitedefaultendpunct}{\mcitedefaultseppunct}\relax
\EndOfBibitem
\bibitem[Filatov and Shaik(1998)Filatov, and Shaik]{filatov1998spin}
Filatov,~M.; Shaik,~S. Spin-restricted density functional approach to the
  open-shell problem. \emph{Chem. Phys. Lett.} \textbf{1998}, \emph{288},
  689--697\relax
\mciteBstWouldAddEndPuncttrue
\mciteSetBstMidEndSepPunct{\mcitedefaultmidpunct}
{\mcitedefaultendpunct}{\mcitedefaultseppunct}\relax
\EndOfBibitem
\bibitem[Gagliardi \latin{et~al.}(2017)Gagliardi, Truhlar, Li~Manni, Carlson,
  Hoyer, and Bao]{gagliardi2017multiconfiguration}
Gagliardi,~L.; Truhlar,~D.~G.; Li~Manni,~G.; Carlson,~R.~K.; Hoyer,~C.~E.;
  Bao,~J.~L. Multiconfiguration pair-density functional theory: A new way to
  treat strongly correlated systems. \emph{Acc. Chem. Res.} \textbf{2017},
  \emph{50}, 66--73\relax
\mciteBstWouldAddEndPuncttrue
\mciteSetBstMidEndSepPunct{\mcitedefaultmidpunct}
{\mcitedefaultendpunct}{\mcitedefaultseppunct}\relax
\EndOfBibitem
\bibitem[Cremer(2001)]{cremer2001density}
Cremer,~D. Density functional theory: coverage of dynamic and non-dynamic
  electron correlation effects. \emph{Mol. Phys.} \textbf{2001}, \emph{99},
  1899--1940\relax
\mciteBstWouldAddEndPuncttrue
\mciteSetBstMidEndSepPunct{\mcitedefaultmidpunct}
{\mcitedefaultendpunct}{\mcitedefaultseppunct}\relax
\EndOfBibitem
\bibitem[Ziegler \latin{et~al.}(1977)Ziegler, Rauk, and
  Baerends]{ziegler1977calculation}
Ziegler,~T.; Rauk,~A.; Baerends,~E.~J. On the calculation of multiplet energies
  by the Hartree-Fock-Slater method. \emph{Theor. Chim. Acta} \textbf{1977},
  \emph{43}, 261--271\relax
\mciteBstWouldAddEndPuncttrue
\mciteSetBstMidEndSepPunct{\mcitedefaultmidpunct}
{\mcitedefaultendpunct}{\mcitedefaultseppunct}\relax
\EndOfBibitem
\bibitem[Noodleman(1981)]{noodleman1981valence}
Noodleman,~L. Valence bond description of antiferromagnetic coupling in
  transition metal dimers. \emph{J. Chem. Phys.} \textbf{1981}, \emph{74},
  5737--5743\relax
\mciteBstWouldAddEndPuncttrue
\mciteSetBstMidEndSepPunct{\mcitedefaultmidpunct}
{\mcitedefaultendpunct}{\mcitedefaultseppunct}\relax
\EndOfBibitem
\bibitem[Daul(1994)]{daul1994density}
Daul,~C. Density functional theory applied to the excited states of
  coordination compounds. \emph{Int. J. Quantum Chem.} \textbf{1994},
  \emph{52}, 867--877\relax
\mciteBstWouldAddEndPuncttrue
\mciteSetBstMidEndSepPunct{\mcitedefaultmidpunct}
{\mcitedefaultendpunct}{\mcitedefaultseppunct}\relax
\EndOfBibitem
\bibitem[Neese(2004)]{neese2004definition}
Neese,~F. Definition of corresponding orbitals and the diradical character in
  broken symmetry DFT calculations on spin coupled systems. \emph{J. Phys.
  Chem. Solids} \textbf{2004}, \emph{65}, 781--785\relax
\mciteBstWouldAddEndPuncttrue
\mciteSetBstMidEndSepPunct{\mcitedefaultmidpunct}
{\mcitedefaultendpunct}{\mcitedefaultseppunct}\relax
\EndOfBibitem
\bibitem[Gunnarsson and Lundqvist(1976)Gunnarsson, and
  Lundqvist]{gunnarsson1976exchange}
Gunnarsson,~O.; Lundqvist,~B.~I. Exchange and correlation in atoms, molecules,
  and solids by the spin-density-functional formalism. \emph{Phys. Rev. B}
  \textbf{1976}, \emph{13}, 4274--4298\relax
\mciteBstWouldAddEndPuncttrue
\mciteSetBstMidEndSepPunct{\mcitedefaultmidpunct}
{\mcitedefaultendpunct}{\mcitedefaultseppunct}\relax
\EndOfBibitem
\bibitem[Anderson(1972)]{anderson1972more}
Anderson,~P.~W. More is different: broken symmetry and the nature of the
  hierarchical structure of science. \emph{Science} \textbf{1972}, \emph{177},
  393--396\relax
\mciteBstWouldAddEndPuncttrue
\mciteSetBstMidEndSepPunct{\mcitedefaultmidpunct}
{\mcitedefaultendpunct}{\mcitedefaultseppunct}\relax
\EndOfBibitem
\bibitem[Amos and Hall(1961)Amos, and Hall]{amos1961single}
Amos,~A.; Hall,~G. Single determinant wave functions. \emph{Proc. Roy. Soc. A}
  \textbf{1961}, \emph{263}, 483--493\relax
\mciteBstWouldAddEndPuncttrue
\mciteSetBstMidEndSepPunct{\mcitedefaultmidpunct}
{\mcitedefaultendpunct}{\mcitedefaultseppunct}\relax
\EndOfBibitem
\bibitem[Ladner and Goddard~III(1969)Ladner, and
  Goddard~III]{ladner1969improved}
Ladner,~R.~C.; Goddard~III,~W.~A. Improved Quantum Theory of Many-Electron
  Systems. V. The Spin-Coupling Optimized GI Method. \emph{J. Chem. Phys.}
  \textbf{1969}, \emph{51}, 1073--1087\relax
\mciteBstWouldAddEndPuncttrue
\mciteSetBstMidEndSepPunct{\mcitedefaultmidpunct}
{\mcitedefaultendpunct}{\mcitedefaultseppunct}\relax
\EndOfBibitem
\bibitem[Bobrowicz and Goddard(1977)Bobrowicz, and Goddard]{bobrowicz1977self}
Bobrowicz,~F.~W.; Goddard,~W.~A. The Self-Consistent Field Equations for
  Generalized Valence Bond and Open-Shell Hartree—Fock Wave Functions. In
  \emph{Methods of Electronic Structure Theory}; Springer: New York, 1977;
  Chapter 4, pp 79--127\relax
\mciteBstWouldAddEndPuncttrue
\mciteSetBstMidEndSepPunct{\mcitedefaultmidpunct}
{\mcitedefaultendpunct}{\mcitedefaultseppunct}\relax
\EndOfBibitem
\bibitem[Ghosh \latin{et~al.}(2003)Ghosh, Bill, Weyherm{\"u}ller, Neese, and
  Wieghardt]{ghosh2003noninnocence}
Ghosh,~P.; Bill,~E.; Weyherm{\"u}ller,~T.; Neese,~F.; Wieghardt,~K.
  Noninnocence of the Ligand Glyoxal-bis (2-mercaptoanil). The Electronic
  Structures of
  [Fe(gma)]$_2$,[Fe(gma)(py)].py,[Fe(gma)(CN)]$^{1-/0}$,[Fe(gma)I], and
  [Fe(gma)(PR$_3$)$_n$] (n= 1, 2). Experimental and Theoretical Evidence for
  “Excited State” Coordination. \emph{J. Am. Chem. Soc.} \textbf{2003},
  \emph{125}, 1293--1308\relax
\mciteBstWouldAddEndPuncttrue
\mciteSetBstMidEndSepPunct{\mcitedefaultmidpunct}
{\mcitedefaultendpunct}{\mcitedefaultseppunct}\relax
\EndOfBibitem
\bibitem[Herebian \latin{et~al.}(2003)Herebian, Wieghardt, and
  Neese]{herebian2003analysis}
Herebian,~D.; Wieghardt,~K.~E.; Neese,~F. Analysis and interpretation of
  metal-radical coupling in a series of square planar nickel complexes:
  correlated ab initio and density functional investigation of
  [Ni(L$^{ISQ}$)$_2$](L$^{ISQ}$=
  3,5-di-tert-butyl-o-diiminobenzosemiquinonate(1-)). \emph{J. Am. Chem. Soc.}
  \textbf{2003}, \emph{125}, 10997--11005\relax
\mciteBstWouldAddEndPuncttrue
\mciteSetBstMidEndSepPunct{\mcitedefaultmidpunct}
{\mcitedefaultendpunct}{\mcitedefaultseppunct}\relax
\EndOfBibitem
\bibitem[Ch{\l}opek \latin{et~al.}(2007)Ch{\l}opek, Muresan, Neese, and
  Wieghardt]{chlopek2007electronic}
Ch{\l}opek,~K.; Muresan,~N.; Neese,~F.; Wieghardt,~K. Electronic Structures of
  Five-Coordinate Complexes of Iron Containing Zero, One, or Two $\pi$-Radical
  Ligands: A Broken-Symmetry Density Functional Theoretical Study. \emph{Chem.
  Eur. Journal} \textbf{2007}, \emph{13}, 8390--8403\relax
\mciteBstWouldAddEndPuncttrue
\mciteSetBstMidEndSepPunct{\mcitedefaultmidpunct}
{\mcitedefaultendpunct}{\mcitedefaultseppunct}\relax
\EndOfBibitem
\bibitem[Neese(2009)]{neese2009prediction}
Neese,~F. Prediction of molecular properties and molecular spectroscopy with
  density functional theory: From fundamental theory to exchange-coupling.
  \emph{Coord. Chem. Rev.} \textbf{2009}, \emph{253}, 526--563\relax
\mciteBstWouldAddEndPuncttrue
\mciteSetBstMidEndSepPunct{\mcitedefaultmidpunct}
{\mcitedefaultendpunct}{\mcitedefaultseppunct}\relax
\EndOfBibitem
\bibitem[Hohenberg and Kohn(1964)Hohenberg, and
  Kohn]{hohenberg1964inhomogeneous}
Hohenberg,~P.; Kohn,~W. Inhomogeneous electron gas. \emph{Phys. Rev.}
  \textbf{1964}, \emph{136}, B864--B871\relax
\mciteBstWouldAddEndPuncttrue
\mciteSetBstMidEndSepPunct{\mcitedefaultmidpunct}
{\mcitedefaultendpunct}{\mcitedefaultseppunct}\relax
\EndOfBibitem
\bibitem[Kutzelnigg(2006)]{kutzelnigg2006dft}
Kutzelnigg,~W. Density functional theory in terms of a Legendre transformation
  for beginners. \emph{J. Mol. Struct.: THEOCHEM} \textbf{2006}, \emph{768},
  163--173\relax
\mciteBstWouldAddEndPuncttrue
\mciteSetBstMidEndSepPunct{\mcitedefaultmidpunct}
{\mcitedefaultendpunct}{\mcitedefaultseppunct}\relax
\EndOfBibitem
\bibitem[Lieb(1983)]{lieb2002density}
Lieb,~E.~H. Density functionals for Coulomb systems. \emph{Int. J. Quantum
  Chem.} \textbf{1983}, \emph{24}, 243--277\relax
\mciteBstWouldAddEndPuncttrue
\mciteSetBstMidEndSepPunct{\mcitedefaultmidpunct}
{\mcitedefaultendpunct}{\mcitedefaultseppunct}\relax
\EndOfBibitem
\bibitem[Penz \latin{et~al.}(2022)Penz, Tellgren, Csirik, Ruggenthaler, and
  Laestadius]{penz2022structure}
Penz,~M.; Tellgren,~E.~I.; Csirik,~M.~A.; Ruggenthaler,~M.; Laestadius,~A. The
  structure of the density-potential mapping. Part I: Standard
  density-functional theory. \emph{arXiv preprint arXiv:2211.16627}
  \textbf{2022}, \relax
\mciteBstWouldAddEndPunctfalse
\mciteSetBstMidEndSepPunct{\mcitedefaultmidpunct}
{}{\mcitedefaultseppunct}\relax
\EndOfBibitem
\bibitem[Casida and Huix-Rotllant(2012)Casida, and
  Huix-Rotllant]{casida2012progress}
Casida,~M.~E.; Huix-Rotllant,~M. Progress in time-dependent density-functional
  theory. \emph{Annu. Rev. Phys. Chem.} \textbf{2012}, \emph{63},
  287--323\relax
\mciteBstWouldAddEndPuncttrue
\mciteSetBstMidEndSepPunct{\mcitedefaultmidpunct}
{\mcitedefaultendpunct}{\mcitedefaultseppunct}\relax
\EndOfBibitem
\bibitem[Kohn(1996)]{kohn1996density}
Kohn,~W. Density Functional and Density Matrix Method Scaling Linearly with the
  Number of Atoms. \emph{Phys. Rev. Lett.} \textbf{1996}, \emph{76},
  3168--3171\relax
\mciteBstWouldAddEndPuncttrue
\mciteSetBstMidEndSepPunct{\mcitedefaultmidpunct}
{\mcitedefaultendpunct}{\mcitedefaultseppunct}\relax
\EndOfBibitem
\bibitem[Prodan and Kohn(2005)Prodan, and Kohn]{prodan2005nearsightedness}
Prodan,~E.; Kohn,~W. Nearsightedness of electronic matter. \emph{Proc. Nat.
  Acad. Sci.} \textbf{2005}, \emph{102}, 11635--11638\relax
\mciteBstWouldAddEndPuncttrue
\mciteSetBstMidEndSepPunct{\mcitedefaultmidpunct}
{\mcitedefaultendpunct}{\mcitedefaultseppunct}\relax
\EndOfBibitem
\bibitem[Li and Burke(2020)Li, and Burke]{li2020recent}
Li,~L.; Burke,~K. Recent developments in density functional approximations. In
  \emph{Handbook of Materials Modeling: Methods: Theory and Modeling};
  Springer: Cham, 2020; pp 213--226\relax
\mciteBstWouldAddEndPuncttrue
\mciteSetBstMidEndSepPunct{\mcitedefaultmidpunct}
{\mcitedefaultendpunct}{\mcitedefaultseppunct}\relax
\EndOfBibitem
\bibitem[Mayer(1955)]{mayer1955electron}
Mayer,~J.~E. Electron Correlation. \emph{Phys. Rev.} \textbf{1955}, \emph{100},
  1579--1586\relax
\mciteBstWouldAddEndPuncttrue
\mciteSetBstMidEndSepPunct{\mcitedefaultmidpunct}
{\mcitedefaultendpunct}{\mcitedefaultseppunct}\relax
\EndOfBibitem
\bibitem[Bopp(1959)]{bopp1959ableitung}
Bopp,~F. Ableitung der Bindungsenergie vonN-Teilchen-Systemen aus
  2-Teilchen-Dichtematrizen. \emph{Z. Phys.} \textbf{1959}, \emph{156},
  348--359\relax
\mciteBstWouldAddEndPuncttrue
\mciteSetBstMidEndSepPunct{\mcitedefaultmidpunct}
{\mcitedefaultendpunct}{\mcitedefaultseppunct}\relax
\EndOfBibitem
\bibitem[Mazziotti(2012)]{mazziotti2012structure}
Mazziotti,~D.~A. Structure of Fermionic Density Matrices: Complete
  $N$-Representability Conditions. \emph{Phys. Rev. Lett.} \textbf{2012},
  \emph{108}, 263002\relax
\mciteBstWouldAddEndPuncttrue
\mciteSetBstMidEndSepPunct{\mcitedefaultmidpunct}
{\mcitedefaultendpunct}{\mcitedefaultseppunct}\relax
\EndOfBibitem
\bibitem[Gilbert(1975)]{gilbert1975hohenberg}
Gilbert,~T.~L. Hohenberg-Kohn theorem for nonlocal external potentials.
  \emph{Phys. Rev. B} \textbf{1975}, \emph{12}, 2111--2120\relax
\mciteBstWouldAddEndPuncttrue
\mciteSetBstMidEndSepPunct{\mcitedefaultmidpunct}
{\mcitedefaultendpunct}{\mcitedefaultseppunct}\relax
\EndOfBibitem
\bibitem[Müller(1984)]{muller1984explicit}
Müller,~A. Explicit approximate relation between reduced two- and one-particle
  density matrices. \emph{Phys. Lett. A} \textbf{1984}, \emph{105},
  446--452\relax
\mciteBstWouldAddEndPuncttrue
\mciteSetBstMidEndSepPunct{\mcitedefaultmidpunct}
{\mcitedefaultendpunct}{\mcitedefaultseppunct}\relax
\EndOfBibitem
\bibitem[Kutzelnigg(2006)]{kutzelnigg2006density}
Kutzelnigg,~W. Density-cumulant functional theory. \emph{J. Chem. Phys.}
  \textbf{2006}, \emph{125}, 171101\relax
\mciteBstWouldAddEndPuncttrue
\mciteSetBstMidEndSepPunct{\mcitedefaultmidpunct}
{\mcitedefaultendpunct}{\mcitedefaultseppunct}\relax
\EndOfBibitem
\bibitem[Piris and Ugalde(2014)Piris, and Ugalde]{piris2014perspective}
Piris,~M.; Ugalde,~J.~M. Perspective on natural orbital functional theory.
  \emph{Int. J. Quantum Chem.} \textbf{2014}, \emph{114}, 1169--1175\relax
\mciteBstWouldAddEndPuncttrue
\mciteSetBstMidEndSepPunct{\mcitedefaultmidpunct}
{\mcitedefaultendpunct}{\mcitedefaultseppunct}\relax
\EndOfBibitem
\bibitem[Pernal and Giesbertz(2016)Pernal, and Giesbertz]{pernal2016reduced}
Pernal,~K.; Giesbertz,~K. J.~H. Reduced Density Matrix Functional Theory
  (RDMFT) and Linear Response Time-Dependent RDMFT (TD-RDMFT). In
  \emph{Density-Functional Methods for Excited States}; Springer: Cham, 2016;
  pp 125--183\relax
\mciteBstWouldAddEndPuncttrue
\mciteSetBstMidEndSepPunct{\mcitedefaultmidpunct}
{\mcitedefaultendpunct}{\mcitedefaultseppunct}\relax
\EndOfBibitem
\bibitem[Coleman(1963)]{coleman1963structure}
Coleman,~A.~J. Structure of fermion density matrices. \emph{Rev. Mod. Phys.}
  \textbf{1963}, \emph{35}, 668--686\relax
\mciteBstWouldAddEndPuncttrue
\mciteSetBstMidEndSepPunct{\mcitedefaultmidpunct}
{\mcitedefaultendpunct}{\mcitedefaultseppunct}\relax
\EndOfBibitem
\bibitem[Cioslowski \latin{et~al.}(2003)Cioslowski, Pernal, and
  Buchowiecki]{cioslowski2003approximate}
Cioslowski,~J.; Pernal,~K.; Buchowiecki,~M. Approximate one-matrix functionals
  for the electron--electron repulsion energy from geminal theories. \emph{J.
  Chem. Phys.} \textbf{2003}, \emph{119}, 6443--6447\relax
\mciteBstWouldAddEndPuncttrue
\mciteSetBstMidEndSepPunct{\mcitedefaultmidpunct}
{\mcitedefaultendpunct}{\mcitedefaultseppunct}\relax
\EndOfBibitem
\bibitem[Piris(2013)]{piris2013natural}
Piris,~M. A natural orbital functional based on an explicit approach of the
  two-electron cumulant. \emph{Int. J. Quantum Chem.} \textbf{2013},
  \emph{113}, 620--630\relax
\mciteBstWouldAddEndPuncttrue
\mciteSetBstMidEndSepPunct{\mcitedefaultmidpunct}
{\mcitedefaultendpunct}{\mcitedefaultseppunct}\relax
\EndOfBibitem
\bibitem[Simmonett \latin{et~al.}(2010)Simmonett, Wilke, Schaefer~III, and
  Kutzelnigg]{simmonett2010density}
Simmonett,~A.~C.; Wilke,~J.~J.; Schaefer~III,~H.~F.; Kutzelnigg,~W. Density
  cumulant functional theory: First implementation and benchmark results for
  the DCFT-06 model. \emph{J. Chem. Phys.} \textbf{2010}, \emph{133},
  174122\relax
\mciteBstWouldAddEndPuncttrue
\mciteSetBstMidEndSepPunct{\mcitedefaultmidpunct}
{\mcitedefaultendpunct}{\mcitedefaultseppunct}\relax
\EndOfBibitem
\bibitem[Kutzelnigg(2006)]{kutzelnigg2006dft-aiqc}
Kutzelnigg,~W. Density Functional Theory (DFT) and ab-initio Quantum Chemistry
  (AIQC). Story of a difficult partnership. In \emph{Trends and Perspectives in
  Modern Computational Science}; Brill: Leiden, 2006; pp 23--62\relax
\mciteBstWouldAddEndPuncttrue
\mciteSetBstMidEndSepPunct{\mcitedefaultmidpunct}
{\mcitedefaultendpunct}{\mcitedefaultseppunct}\relax
\EndOfBibitem
\bibitem[Copan and Sokolov(2018)Copan, and Sokolov]{copan2018linear}
Copan,~A.~V.; Sokolov,~A.~Y. Linear-response density cumulant theory for
  excited electronic states. \emph{J. Chem. Theory Comput.} \textbf{2018},
  \emph{14}, 4097--4108\relax
\mciteBstWouldAddEndPuncttrue
\mciteSetBstMidEndSepPunct{\mcitedefaultmidpunct}
{\mcitedefaultendpunct}{\mcitedefaultseppunct}\relax
\EndOfBibitem
\bibitem[Cioslowski(2000)]{cioslowski2000many}
Cioslowski,~J., Ed. \emph{Many-electron densities and reduced density
  matrices}; Springer Science \& Business Media: New York, 2000\relax
\mciteBstWouldAddEndPuncttrue
\mciteSetBstMidEndSepPunct{\mcitedefaultmidpunct}
{\mcitedefaultendpunct}{\mcitedefaultseppunct}\relax
\EndOfBibitem
\bibitem[Gidofalvi and Mazziotti(2008)Gidofalvi, and
  Mazziotti]{gidofalvi2008active}
Gidofalvi,~G.; Mazziotti,~D.~A. Active-space two-electron
  reduced-density-matrix method: Complete active-space calculations without
  diagonalization of the N-electron Hamiltonian. \emph{J. Chem. Phys.}
  \textbf{2008}, \emph{129}, 134108\relax
\mciteBstWouldAddEndPuncttrue
\mciteSetBstMidEndSepPunct{\mcitedefaultmidpunct}
{\mcitedefaultendpunct}{\mcitedefaultseppunct}\relax
\EndOfBibitem
\bibitem[Mullinax \latin{et~al.}(2015)Mullinax, Sokolov, and
  Schaefer~III]{mullinax2015can}
Mullinax,~J.~W.; Sokolov,~A.~Y.; Schaefer~III,~H.~F. Can density cumulant
  functional theory describe static correlation effects? \emph{J. Chem. Theory
  Comput.} \textbf{2015}, \emph{11}, 2487--2495\relax
\mciteBstWouldAddEndPuncttrue
\mciteSetBstMidEndSepPunct{\mcitedefaultmidpunct}
{\mcitedefaultendpunct}{\mcitedefaultseppunct}\relax
\EndOfBibitem
\bibitem[Lathiotakis \latin{et~al.}(2014)Lathiotakis, Helbig, Rubio, and
  Gidopoulos]{lathiotakis2014local}
Lathiotakis,~N.~N.; Helbig,~N.; Rubio,~A.; Gidopoulos,~N.~I. Local
  reduced-density-matrix-functional theory: Incorporating static correlation
  effects in Kohn-Sham equations. \emph{Phys. Rev. A} \textbf{2014}, \emph{90},
  032511\relax
\mciteBstWouldAddEndPuncttrue
\mciteSetBstMidEndSepPunct{\mcitedefaultmidpunct}
{\mcitedefaultendpunct}{\mcitedefaultseppunct}\relax
\EndOfBibitem
\bibitem[Piris(2017)]{piris2017global}
Piris,~M. Global Method for Electron Correlation. \emph{Phys. Rev. Lett.}
  \textbf{2017}, \emph{119}, 063002\relax
\mciteBstWouldAddEndPuncttrue
\mciteSetBstMidEndSepPunct{\mcitedefaultmidpunct}
{\mcitedefaultendpunct}{\mcitedefaultseppunct}\relax
\EndOfBibitem
\bibitem[Gibney \latin{et~al.}(2022)Gibney, Boyn, and
  Mazziotti]{gibney2022density}
Gibney,~D.; Boyn,~J.-N.; Mazziotti,~D.~A. Density Functional Theory Transformed
  into a One-Electron Reduced-Density-Matrix Functional Theory for the Capture
  of Static Correlation. \emph{J. Phys. Chem. Lett.} \textbf{2022}, \emph{13},
  1382--1388\relax
\mciteBstWouldAddEndPuncttrue
\mciteSetBstMidEndSepPunct{\mcitedefaultmidpunct}
{\mcitedefaultendpunct}{\mcitedefaultseppunct}\relax
\EndOfBibitem
\bibitem[Martin and Schwinger(1959)Martin, and Schwinger]{martin1959theory}
Martin,~P.~C.; Schwinger,~J. Theory of many-particle systems. I. \emph{Phys.
  Rev.} \textbf{1959}, \emph{115}, 1342--1373\relax
\mciteBstWouldAddEndPuncttrue
\mciteSetBstMidEndSepPunct{\mcitedefaultmidpunct}
{\mcitedefaultendpunct}{\mcitedefaultseppunct}\relax
\EndOfBibitem
\bibitem[Luttinger and Ward(1960)Luttinger, and Ward]{luttinger1960ground}
Luttinger,~J.~M.; Ward,~J.~C. Ground-state energy of a many-fermion system. II.
  \emph{Phys. Rev.} \textbf{1960}, \emph{118}, 1417--1427\relax
\mciteBstWouldAddEndPuncttrue
\mciteSetBstMidEndSepPunct{\mcitedefaultmidpunct}
{\mcitedefaultendpunct}{\mcitedefaultseppunct}\relax
\EndOfBibitem
\bibitem[Baym and Kadanoff(1961)Baym, and Kadanoff]{baym1961conservation}
Baym,~G.; Kadanoff,~L.~P. Conservation laws and correlation functions.
  \emph{Phys. Rev.} \textbf{1961}, \emph{124}, 287--299\relax
\mciteBstWouldAddEndPuncttrue
\mciteSetBstMidEndSepPunct{\mcitedefaultmidpunct}
{\mcitedefaultendpunct}{\mcitedefaultseppunct}\relax
\EndOfBibitem
\bibitem[Ziesche(2000)]{ziesche2000cumulant}
Ziesche,~P. Cumulant expansions of reduced densities, reduced density matrices,
  and Green’s functions. In \emph{Many-Electron Densities and Reduced Density
  Matrices}; Springer: New York, 2000; Chapter 3, pp 33--56\relax
\mciteBstWouldAddEndPuncttrue
\mciteSetBstMidEndSepPunct{\mcitedefaultmidpunct}
{\mcitedefaultendpunct}{\mcitedefaultseppunct}\relax
\EndOfBibitem
\bibitem[Sun and Chan(2016)Sun, and Chan]{sun2016quantum}
Sun,~Q.; Chan,~G. K.-L. Quantum embedding theories. \emph{Acc. Chem. Res.}
  \textbf{2016}, \emph{49}, 2705--2712\relax
\mciteBstWouldAddEndPuncttrue
\mciteSetBstMidEndSepPunct{\mcitedefaultmidpunct}
{\mcitedefaultendpunct}{\mcitedefaultseppunct}\relax
\EndOfBibitem
\bibitem[Kotliar \latin{et~al.}(2006)Kotliar, Savrasov, Haule, Oudovenko,
  Parcollet, and Marianetti]{kotliar2006electronic}
Kotliar,~G.; Savrasov,~S.~Y.; Haule,~K.; Oudovenko,~V.~S.; Parcollet,~O.;
  Marianetti,~C. Electronic structure calculations with dynamical mean-field
  theory. \emph{Rev. Mod. Phys.} \textbf{2006}, \emph{78}, 865--951\relax
\mciteBstWouldAddEndPuncttrue
\mciteSetBstMidEndSepPunct{\mcitedefaultmidpunct}
{\mcitedefaultendpunct}{\mcitedefaultseppunct}\relax
\EndOfBibitem
\bibitem[Lin \latin{et~al.}(2011)Lin, Marianetti, Millis, and
  Reichman]{lin2011dynamical}
Lin,~N.; Marianetti,~C.; Millis,~A.~J.; Reichman,~D.~R. Dynamical mean-field
  theory for quantum chemistry. \emph{Phys. Rev. Lett.} \textbf{2011},
  \emph{106}, 096402\relax
\mciteBstWouldAddEndPuncttrue
\mciteSetBstMidEndSepPunct{\mcitedefaultmidpunct}
{\mcitedefaultendpunct}{\mcitedefaultseppunct}\relax
\EndOfBibitem
\bibitem[Zgid and Chan(2011)Zgid, and Chan]{zgid2011dynamical}
Zgid,~D.; Chan,~G. K.-L. Dynamical mean-field theory from a quantum chemical
  perspective. \emph{J. Chem. Phys.} \textbf{2011}, \emph{134}, 094115\relax
\mciteBstWouldAddEndPuncttrue
\mciteSetBstMidEndSepPunct{\mcitedefaultmidpunct}
{\mcitedefaultendpunct}{\mcitedefaultseppunct}\relax
\EndOfBibitem
\bibitem[Hedin(1965)]{hedin1965new}
Hedin,~L. New method for calculating the one-particle Green's function with
  application to the electron-gas problem. \emph{Phys. Rev.} \textbf{1965},
  \emph{139}, A796--A823\relax
\mciteBstWouldAddEndPuncttrue
\mciteSetBstMidEndSepPunct{\mcitedefaultmidpunct}
{\mcitedefaultendpunct}{\mcitedefaultseppunct}\relax
\EndOfBibitem
\bibitem[Aryasetiawan and Gunnarsson(1998)Aryasetiawan, and
  Gunnarsson]{aryasetiawan1998gw}
Aryasetiawan,~F.; Gunnarsson,~O. The GW method. \emph{Rep. Prog. Phys.}
  \textbf{1998}, \emph{61}, 237--312\relax
\mciteBstWouldAddEndPuncttrue
\mciteSetBstMidEndSepPunct{\mcitedefaultmidpunct}
{\mcitedefaultendpunct}{\mcitedefaultseppunct}\relax
\EndOfBibitem
\bibitem[Reining(2018)]{reining2018gw}
Reining,~L. The GW approximation: content, successes and limitations.
  \emph{Wiley Interdiscip. Rev. Comput. Mol. Sci.} \textbf{2018}, \emph{8},
  e1344\relax
\mciteBstWouldAddEndPuncttrue
\mciteSetBstMidEndSepPunct{\mcitedefaultmidpunct}
{\mcitedefaultendpunct}{\mcitedefaultseppunct}\relax
\EndOfBibitem
\bibitem[Phillips and Zgid(2014)Phillips, and Zgid]{phillips2014communication}
Phillips,~J.~J.; Zgid,~D. Communication: The description of strong correlation
  within self-consistent Green's function second-order perturbation theory.
  \emph{J. Chem. Phys.} \textbf{2014}, \emph{140}, 241101\relax
\mciteBstWouldAddEndPuncttrue
\mciteSetBstMidEndSepPunct{\mcitedefaultmidpunct}
{\mcitedefaultendpunct}{\mcitedefaultseppunct}\relax
\EndOfBibitem
\bibitem[Pokhilko and Zgid(2021)Pokhilko, and Zgid]{pokhilko2021interpretation}
Pokhilko,~P.; Zgid,~D. Interpretation of multiple solutions in fully iterative
  GF2 and GW schemes using local analysis of two-particle density matrices.
  \emph{J. Chem. Phys.} \textbf{2021}, \emph{155}, 024101\relax
\mciteBstWouldAddEndPuncttrue
\mciteSetBstMidEndSepPunct{\mcitedefaultmidpunct}
{\mcitedefaultendpunct}{\mcitedefaultseppunct}\relax
\EndOfBibitem
\bibitem[Blase \latin{et~al.}(2018)Blase, Duchemin, and
  Jacquemin]{blase2018bethe}
Blase,~X.; Duchemin,~I.; Jacquemin,~D. The Bethe--Salpeter equation in
  chemistry: relations with TD-DFT, applications and challenges. \emph{Chem.
  Soc. Rev.} \textbf{2018}, \emph{47}, 1022--1043\relax
\mciteBstWouldAddEndPuncttrue
\mciteSetBstMidEndSepPunct{\mcitedefaultmidpunct}
{\mcitedefaultendpunct}{\mcitedefaultseppunct}\relax
\EndOfBibitem
\bibitem[Pokhilko \latin{et~al.}(2021)Pokhilko, Iskakov, Yeh, and
  Zgid]{pokhilko2021evaluation}
Pokhilko,~P.; Iskakov,~S.; Yeh,~C.-N.; Zgid,~D. Evaluation of two-particle
  properties within finite-temperature self-consistent one-particle Green’s
  function methods: Theory and application to GW and GF2. \emph{J. Chem. Phys.}
  \textbf{2021}, \emph{155}, 024119\relax
\mciteBstWouldAddEndPuncttrue
\mciteSetBstMidEndSepPunct{\mcitedefaultmidpunct}
{\mcitedefaultendpunct}{\mcitedefaultseppunct}\relax
\EndOfBibitem
\bibitem[Cohen \latin{et~al.}(2008)Cohen, Mori-S{\'a}nchez, and
  Yang]{cohen2008fractional}
Cohen,~A.~J.; Mori-S{\'a}nchez,~P.; Yang,~W. Fractional spins and static
  correlation error in density functional theory. \emph{J. Chem. Phys.}
  \textbf{2008}, \emph{129}, 121104\relax
\mciteBstWouldAddEndPuncttrue
\mciteSetBstMidEndSepPunct{\mcitedefaultmidpunct}
{\mcitedefaultendpunct}{\mcitedefaultseppunct}\relax
\EndOfBibitem
\bibitem[Cohen \latin{et~al.}(2008)Cohen, Mori-S{\'a}nchez, and
  Yang]{cohen2008insights}
Cohen,~A.~J.; Mori-S{\'a}nchez,~P.; Yang,~W. Insights into current limitations
  of density functional theory. \emph{Science} \textbf{2008}, \emph{321},
  792--794\relax
\mciteBstWouldAddEndPuncttrue
\mciteSetBstMidEndSepPunct{\mcitedefaultmidpunct}
{\mcitedefaultendpunct}{\mcitedefaultseppunct}\relax
\EndOfBibitem
\bibitem[Grimme and Hansen(2015)Grimme, and Hansen]{grimme2015practicable}
Grimme,~S.; Hansen,~A. A practicable real-space measure and visualization of
  static electron-correlation effects. \emph{Angew. Chem. Int. Ed.}
  \textbf{2015}, \emph{54}, 12308--12313\relax
\mciteBstWouldAddEndPuncttrue
\mciteSetBstMidEndSepPunct{\mcitedefaultmidpunct}
{\mcitedefaultendpunct}{\mcitedefaultseppunct}\relax
\EndOfBibitem
\bibitem[Muechler \latin{et~al.}(2022)Muechler, Badrtdinov, Hampel, Cano,
  R{\"o}sner, and Dreyer]{muechler2022quantum}
Muechler,~L.; Badrtdinov,~D.~I.; Hampel,~A.; Cano,~J.; R{\"o}sner,~M.;
  Dreyer,~C.~E. Quantum embedding methods for correlated excited states of
  point defects: Case studies and challenges. \emph{Phys. Rev. B}
  \textbf{2022}, \emph{105}, 235104\relax
\mciteBstWouldAddEndPuncttrue
\mciteSetBstMidEndSepPunct{\mcitedefaultmidpunct}
{\mcitedefaultendpunct}{\mcitedefaultseppunct}\relax
\EndOfBibitem
\bibitem[Nielsen and Chuang(2010)Nielsen, and Chuang]{nielsen2010quantum}
Nielsen,~M.~A.; Chuang,~I.~L. \emph{Quantum Computation and Quantum
  Information}; Cambridge University Press: Cambridge, 2010\relax
\mciteBstWouldAddEndPuncttrue
\mciteSetBstMidEndSepPunct{\mcitedefaultmidpunct}
{\mcitedefaultendpunct}{\mcitedefaultseppunct}\relax
\EndOfBibitem
\bibitem[Almeida \latin{et~al.}(2007)Almeida, Omar, and
  Rocha~Vieira]{almeida2007introduction}
Almeida,~M.; Omar,~Y.; Rocha~Vieira,~V. Introduction to entanglement and
  applications to the simulation of many-body quantum systems. In
  \emph{Strongly Correlated Systems, Coherence And Entanglement}; World
  Scientific: Singapore, 2007; Chapter 19, pp 525--547\relax
\mciteBstWouldAddEndPuncttrue
\mciteSetBstMidEndSepPunct{\mcitedefaultmidpunct}
{\mcitedefaultendpunct}{\mcitedefaultseppunct}\relax
\EndOfBibitem
\bibitem[Ding \latin{et~al.}(2022)Ding, Knecht, Zimbor{\'a}s, and
  Schilling]{ding2022quantum}
Ding,~L.; Knecht,~S.; Zimbor{\'a}s,~Z.; Schilling,~C. Quantum Correlations in
  Molecules: from Quantum Resourcing to Chemical Bonding. \emph{Quantum Sci.
  Technol.} \textbf{2022}, \emph{8}, 015015\relax
\mciteBstWouldAddEndPuncttrue
\mciteSetBstMidEndSepPunct{\mcitedefaultmidpunct}
{\mcitedefaultendpunct}{\mcitedefaultseppunct}\relax
\EndOfBibitem
\bibitem[Omar(2005)]{omar2005particle}
Omar,~Y. Particle statistics in quantum information processing. \emph{Int. J.
  Quantum Inf.} \textbf{2005}, \emph{3}, 201--205\relax
\mciteBstWouldAddEndPuncttrue
\mciteSetBstMidEndSepPunct{\mcitedefaultmidpunct}
{\mcitedefaultendpunct}{\mcitedefaultseppunct}\relax
\EndOfBibitem
\bibitem[Benatti \latin{et~al.}(2020)Benatti, Floreanini, Franchini, and
  Marzolino]{benatti2020entanglement}
Benatti,~F.; Floreanini,~R.; Franchini,~F.; Marzolino,~U. Entanglement in
  indistinguishable particle systems. \emph{Phys. Rep.} \textbf{2020},
  \emph{878}, 1--27\relax
\mciteBstWouldAddEndPuncttrue
\mciteSetBstMidEndSepPunct{\mcitedefaultmidpunct}
{\mcitedefaultendpunct}{\mcitedefaultseppunct}\relax
\EndOfBibitem
\bibitem[Henderson \latin{et~al.}(2014)Henderson, Bulik, Stein, and
  Scuseria]{henderson2014seniority}
Henderson,~T.~M.; Bulik,~I.~W.; Stein,~T.; Scuseria,~G.~E. Seniority-based
  coupled cluster theory. \emph{J. Chem. Phys.} \textbf{2014}, \emph{141},
  244104\relax
\mciteBstWouldAddEndPuncttrue
\mciteSetBstMidEndSepPunct{\mcitedefaultmidpunct}
{\mcitedefaultendpunct}{\mcitedefaultseppunct}\relax
\EndOfBibitem
\bibitem[Boguslawski and Tecmer(2015)Boguslawski, and
  Tecmer]{boguslawski2015orbital}
Boguslawski,~K.; Tecmer,~P. Orbital entanglement in quantum chemistry.
  \emph{Int. J. Quantum Chem.} \textbf{2015}, \emph{115}, 1289--1295\relax
\mciteBstWouldAddEndPuncttrue
\mciteSetBstMidEndSepPunct{\mcitedefaultmidpunct}
{\mcitedefaultendpunct}{\mcitedefaultseppunct}\relax
\EndOfBibitem
\bibitem[Ding \latin{et~al.}(2022)Ding, Zimboras, and
  Schilling]{ding2022quantifying}
Ding,~L.; Zimboras,~Z.; Schilling,~C. Quantifying Electron Entanglement
  Faithfully. \emph{arXiv preprint arXiv:2207.03377} \textbf{2022}, \relax
\mciteBstWouldAddEndPunctfalse
\mciteSetBstMidEndSepPunct{\mcitedefaultmidpunct}
{}{\mcitedefaultseppunct}\relax
\EndOfBibitem
\bibitem[Legeza and S{\'o}lyom(2003)Legeza, and
  S{\'o}lyom]{legeza2003optimizing}
Legeza,~{\"O}.; S{\'o}lyom,~J. Optimizing the density-matrix renormalization
  group method using quantum information entropy. \emph{Phys. Rev. B}
  \textbf{2003}, \emph{68}, 195116\relax
\mciteBstWouldAddEndPuncttrue
\mciteSetBstMidEndSepPunct{\mcitedefaultmidpunct}
{\mcitedefaultendpunct}{\mcitedefaultseppunct}\relax
\EndOfBibitem
\bibitem[Stein and Reiher(2017)Stein, and Reiher]{stein2017measuring}
Stein,~C.~J.; Reiher,~M. Measuring multi-configurational character by orbital
  entanglement. \emph{Mol. Phys.} \textbf{2017}, \emph{115}, 2110--2119\relax
\mciteBstWouldAddEndPuncttrue
\mciteSetBstMidEndSepPunct{\mcitedefaultmidpunct}
{\mcitedefaultendpunct}{\mcitedefaultseppunct}\relax
\EndOfBibitem
\bibitem[Stein and Reiher(2016)Stein, and Reiher]{stein2016automated}
Stein,~C.~J.; Reiher,~M. Automated selection of active orbital spaces. \emph{J.
  Chem. Theory Comput.} \textbf{2016}, \emph{12}, 1760--1771\relax
\mciteBstWouldAddEndPuncttrue
\mciteSetBstMidEndSepPunct{\mcitedefaultmidpunct}
{\mcitedefaultendpunct}{\mcitedefaultseppunct}\relax
\EndOfBibitem
\bibitem[Boguslawski \latin{et~al.}(2013)Boguslawski, Tecmer, Barcza, Legeza,
  and Reiher]{boguslawski2013orbital}
Boguslawski,~K.; Tecmer,~P.; Barcza,~G.; Legeza,~O.; Reiher,~M. Orbital
  entanglement in bond-formation processes. \emph{J. Chem. Theory Comput.}
  \textbf{2013}, \emph{9}, 2959--2973\relax
\mciteBstWouldAddEndPuncttrue
\mciteSetBstMidEndSepPunct{\mcitedefaultmidpunct}
{\mcitedefaultendpunct}{\mcitedefaultseppunct}\relax
\EndOfBibitem
\bibitem[Huang and Kais(2005)Huang, and Kais]{huang2005entanglement}
Huang,~Z.; Kais,~S. Entanglement as measure of electron--electron correlation
  in quantum chemistry calculations. \emph{Chem. Phys. Lett.} \textbf{2005},
  \emph{413}, 1--5\relax
\mciteBstWouldAddEndPuncttrue
\mciteSetBstMidEndSepPunct{\mcitedefaultmidpunct}
{\mcitedefaultendpunct}{\mcitedefaultseppunct}\relax
\EndOfBibitem
\bibitem[Boguslawski \latin{et~al.}(2012)Boguslawski, Tecmer, Legeza, and
  Reiher]{boguslawski2012entanglement}
Boguslawski,~K.; Tecmer,~P.; Legeza,~O.; Reiher,~M. Entanglement measures for
  single-and multireference correlation effects. \emph{J. Phys. Chem. Lett.}
  \textbf{2012}, \emph{3}, 3129--3135\relax
\mciteBstWouldAddEndPuncttrue
\mciteSetBstMidEndSepPunct{\mcitedefaultmidpunct}
{\mcitedefaultendpunct}{\mcitedefaultseppunct}\relax
\EndOfBibitem
\bibitem[Ziesche(1995)]{ziesche1995correlation}
Ziesche,~P. Correlation strength and information entropy. \emph{Int. J. Quantum
  Chem.} \textbf{1995}, \emph{56}, 363--369\relax
\mciteBstWouldAddEndPuncttrue
\mciteSetBstMidEndSepPunct{\mcitedefaultmidpunct}
{\mcitedefaultendpunct}{\mcitedefaultseppunct}\relax
\EndOfBibitem
\bibitem[Ghosh \latin{et~al.}(2022)Ghosh, Kais, and
  Herschbach]{ghosh2022geometrical}
Ghosh,~K.~J.; Kais,~S.; Herschbach,~D.~R. Geometrical picture of the
  electron--electron correlation at the large-D limit. \emph{Phys. Chem. Chem.
  Phys.} \textbf{2022}, \emph{24}, 9298--9307\relax
\mciteBstWouldAddEndPuncttrue
\mciteSetBstMidEndSepPunct{\mcitedefaultmidpunct}
{\mcitedefaultendpunct}{\mcitedefaultseppunct}\relax
\EndOfBibitem
\bibitem[Rissler \latin{et~al.}(2006)Rissler, Noack, and
  White]{rissler2006measuring}
Rissler,~J.; Noack,~R.~M.; White,~S.~R. Measuring orbital interaction using
  quantum information theory. \emph{Chem. Phys.} \textbf{2006}, \emph{323},
  519--531\relax
\mciteBstWouldAddEndPuncttrue
\mciteSetBstMidEndSepPunct{\mcitedefaultmidpunct}
{\mcitedefaultendpunct}{\mcitedefaultseppunct}\relax
\EndOfBibitem
\bibitem[Cao \latin{et~al.}(2019)Cao, Romero, Olson, Degroote, Johnson,
  Kieferová, Kivlichan, Menke, Peropadre, Sawaya, Sim, Veis, and
  Aspuru-Guzik]{cao2019quantum}
Cao,~Y.; Romero,~J.; Olson,~J.~P.; Degroote,~M.; Johnson,~P.~D.;
  Kieferová,~M.; Kivlichan,~I.~D.; Menke,~T.; Peropadre,~B.; Sawaya,~N. P.~D.;
  Sim,~S.; Veis,~L.; Aspuru-Guzik,~A. Quantum Chemistry in the Age of Quantum
  Computing. \emph{Chem. Rev.} \textbf{2019}, \emph{119}, 10856--10915\relax
\mciteBstWouldAddEndPuncttrue
\mciteSetBstMidEndSepPunct{\mcitedefaultmidpunct}
{\mcitedefaultendpunct}{\mcitedefaultseppunct}\relax
\EndOfBibitem
\bibitem[Reiher \latin{et~al.}(2017)Reiher, Wiebe, Svore, Wecker, and
  Troyer]{reiher2017elucidating}
Reiher,~M.; Wiebe,~N.; Svore,~K.~M.; Wecker,~D.; Troyer,~M. Elucidating
  reaction mechanisms on quantum computers. \emph{Proc. Nat. Acad. Sci.}
  \textbf{2017}, \emph{114}, 7555--7560\relax
\mciteBstWouldAddEndPuncttrue
\mciteSetBstMidEndSepPunct{\mcitedefaultmidpunct}
{\mcitedefaultendpunct}{\mcitedefaultseppunct}\relax
\EndOfBibitem
\bibitem[Tubman \latin{et~al.}(2018)Tubman, Mejuto-Zaera, Epstein, Hait,
  Levine, Huggins, Jiang, McClean, Babbush, Head-Gordon, and
  Whaley]{tubman2018postponing}
Tubman,~N.~M.; Mejuto-Zaera,~C.; Epstein,~J.~M.; Hait,~D.; Levine,~D.~S.;
  Huggins,~W.; Jiang,~Z.; McClean,~J.~R.; Babbush,~R.; Head-Gordon,~M.;
  Whaley,~K.~B. Postponing the orthogonality catastrophe: efficient state
  preparation for electronic structure simulations on quantum devices.
  \emph{arXiv preprint arXiv:1809.05523} \textbf{2018}, \relax
\mciteBstWouldAddEndPunctfalse
\mciteSetBstMidEndSepPunct{\mcitedefaultmidpunct}
{}{\mcitedefaultseppunct}\relax
\EndOfBibitem
\bibitem[Carbone \latin{et~al.}(2022)Carbone, Galli, Motta, and
  Jones]{carbone2022quantum}
Carbone,~A.; Galli,~D.~E.; Motta,~M.; Jones,~B. Quantum circuits for the
  preparation of spin eigenfunctions on quantum computers. \emph{Symmetry}
  \textbf{2022}, \emph{14}, 624\relax
\mciteBstWouldAddEndPuncttrue
\mciteSetBstMidEndSepPunct{\mcitedefaultmidpunct}
{\mcitedefaultendpunct}{\mcitedefaultseppunct}\relax
\EndOfBibitem
\bibitem[Lacroix \latin{et~al.}(2023)Lacroix, Ruiz~Guzman, and
  Siwach]{lacroix2022symmetry}
Lacroix,~D.; Ruiz~Guzman,~E.~A.; Siwach,~P. Symmetry breaking/symmetry
  preserving circuits and symmetry restoration on quantum computers. \emph{Eur.
  Phys. J. A} \textbf{2023}, \emph{59}, 3\relax
\mciteBstWouldAddEndPuncttrue
\mciteSetBstMidEndSepPunct{\mcitedefaultmidpunct}
{\mcitedefaultendpunct}{\mcitedefaultseppunct}\relax
\EndOfBibitem
\bibitem[Lee \latin{et~al.}(2022)Lee, Lee, Zhai, Tong, Dalzell, Kumar, Helms,
  Gray, Cui, Liu, Kastoryano, Babbush, Preskill, Reichman, Campbell, Valeev,
  Lin, and Chan]{lee2022there}
Lee,~S.; Lee,~J.; Zhai,~H.; Tong,~Y.; Dalzell,~A.~M.; Kumar,~A.; Helms,~P.;
  Gray,~J.; Cui,~Z.-H.; Liu,~W.; Kastoryano,~M.; Babbush,~R.; Preskill,~J.;
  Reichman,~D.~R.; Campbell,~E.~T.; Valeev,~E.~F.; Lin,~L.; Chan,~G. K.-L. Is
  there evidence for exponential quantum advantage in quantum chemistry?
  \emph{arXiv preprint arXiv:2208.02199} \textbf{2022}, \relax
\mciteBstWouldAddEndPunctfalse
\mciteSetBstMidEndSepPunct{\mcitedefaultmidpunct}
{}{\mcitedefaultseppunct}\relax
\EndOfBibitem
\bibitem[Tazhigulov \latin{et~al.}(2022)Tazhigulov, Sun, Haghshenas, Zhai, Tan,
  Rubin, Babbush, Minnich, and Chan]{tazhigulov2022simulating}
Tazhigulov,~R.~N.; Sun,~S.-N.; Haghshenas,~R.; Zhai,~H.; Tan,~A.~T.;
  Rubin,~N.~C.; Babbush,~R.; Minnich,~A.~J.; Chan,~G. K.-L. Simulating models
  of challenging correlated molecules and materials on the Sycamore quantum
  processor. \emph{PRX Quantum} \textbf{2022}, \emph{3}, 040318\relax
\mciteBstWouldAddEndPuncttrue
\mciteSetBstMidEndSepPunct{\mcitedefaultmidpunct}
{\mcitedefaultendpunct}{\mcitedefaultseppunct}\relax
\EndOfBibitem
\bibitem[Amsler \latin{et~al.}(2023)Amsler, Deglmann, Degroote, Kaicher, Kiser,
  Kühn, Kumar, Maier, Samsonidze, Schroeder, Streif, Vodola, and
  Wever]{amsler2023quantum}
Amsler,~M.; Deglmann,~P.; Degroote,~M.; Kaicher,~M.~P.; Kiser,~M.; Kühn,~M.;
  Kumar,~C.; Maier,~A.; Samsonidze,~G.; Schroeder,~A.; Streif,~M.; Vodola,~D.;
  Wever,~C. Quantum-enhanced quantum Monte Carlo: an industrial view.
  \emph{arXiv preprint arXiv:2301.11838} \textbf{2023}, \relax
\mciteBstWouldAddEndPunctfalse
\mciteSetBstMidEndSepPunct{\mcitedefaultmidpunct}
{}{\mcitedefaultseppunct}\relax
\EndOfBibitem
\bibitem[Rubin \latin{et~al.}(2023)Rubin, Berry, Malone, White, Khattar, III,
  Sicolo, K{\"u}hn, Kaicher, Lee, and Babbush]{rubin2023fault}
Rubin,~N.~C.; Berry,~D.~W.; Malone,~F.~D.; White,~A.~F.; Khattar,~T.; III,~A.
  E.~D.; Sicolo,~S.; K{\"u}hn,~M.; Kaicher,~M.; Lee,~J.; Babbush,~R.
  Fault-tolerant quantum simulation of materials using Bloch orbitals.
  \emph{arXiv preprint arXiv:2302.05531} \textbf{2023}, \relax
\mciteBstWouldAddEndPunctfalse
\mciteSetBstMidEndSepPunct{\mcitedefaultmidpunct}
{}{\mcitedefaultseppunct}\relax
\EndOfBibitem
\bibitem[Lykos and Pratt(1963)Lykos, and Pratt]{lykos1963discussion}
Lykos,~P.; Pratt,~G.~W. Discussion on The Hartree-Fock Approximation.
  \emph{Rev. Mod. Phys.} \textbf{1963}, \emph{35}, 496--501\relax
\mciteBstWouldAddEndPuncttrue
\mciteSetBstMidEndSepPunct{\mcitedefaultmidpunct}
{\mcitedefaultendpunct}{\mcitedefaultseppunct}\relax
\EndOfBibitem
\bibitem[Slater(1929)]{slater1929theory}
Slater,~J.~C. The Theory of Complex Spectra. \emph{Phys. Rev.} \textbf{1929},
  \emph{34}, 1293--1322\relax
\mciteBstWouldAddEndPuncttrue
\mciteSetBstMidEndSepPunct{\mcitedefaultmidpunct}
{\mcitedefaultendpunct}{\mcitedefaultseppunct}\relax
\EndOfBibitem
\bibitem[Slater(1975)]{slater1975solid}
Slater,~J.~C. \emph{Solid-state and molecular theory: a scientific biography};
  Wiley-Interscience: New York, 1975\relax
\mciteBstWouldAddEndPuncttrue
\mciteSetBstMidEndSepPunct{\mcitedefaultmidpunct}
{\mcitedefaultendpunct}{\mcitedefaultseppunct}\relax
\EndOfBibitem
\bibitem[Matsen(1964)]{matsen1964spin}
Matsen,~F. Spin-free quantum chemistry. In \emph{Advances in Quantum
  Chemistry}; Interscience: New York, 1964; Vol.~1; pp 59--114\relax
\mciteBstWouldAddEndPuncttrue
\mciteSetBstMidEndSepPunct{\mcitedefaultmidpunct}
{\mcitedefaultendpunct}{\mcitedefaultseppunct}\relax
\EndOfBibitem
\bibitem[Paldus(1981)]{paldus1981unitary}
Paldus,~J. The Unitary Group for the Evaluation of Electronic Energy Matrix
  Elements. In \emph{The Unitary Group for the Evaluation of Electronic Energy
  Matrix Elements}; Springer: Berlin, 1981; pp 1--50\relax
\mciteBstWouldAddEndPuncttrue
\mciteSetBstMidEndSepPunct{\mcitedefaultmidpunct}
{\mcitedefaultendpunct}{\mcitedefaultseppunct}\relax
\EndOfBibitem
\bibitem[Shavitt(1981)]{shavitt1981graphical}
Shavitt,~I. The Graphical Unitary Group Approach and its Application to Direct
  Configuration Interaction Calculations. In \emph{The Unitary Group for the
  Evaluation of Electronic Energy Matrix Elements}; Springer: Berlin, 1981; pp
  51--99\relax
\mciteBstWouldAddEndPuncttrue
\mciteSetBstMidEndSepPunct{\mcitedefaultmidpunct}
{\mcitedefaultendpunct}{\mcitedefaultseppunct}\relax
\EndOfBibitem
\bibitem[Dobrautz \latin{et~al.}(2019)Dobrautz, Smart, and
  Alavi]{dobrautz2019efficient}
Dobrautz,~W.; Smart,~S.~D.; Alavi,~A. Efficient formulation of full
  configuration interaction quantum Monte Carlo in a spin eigenbasis via the
  graphical unitary group approach. \emph{J. Chem. Phys.} \textbf{2019},
  \emph{151}, 094104\relax
\mciteBstWouldAddEndPuncttrue
\mciteSetBstMidEndSepPunct{\mcitedefaultmidpunct}
{\mcitedefaultendpunct}{\mcitedefaultseppunct}\relax
\EndOfBibitem
\bibitem[McLean \latin{et~al.}(1985)McLean, Lengsfield, Pacansky, and
  Ellinger]{mclean1985symmetry}
McLean,~A.~D.; Lengsfield,~B.~H.; Pacansky,~J.; Ellinger,~Y. Symmetry breaking
  in molecular calculations and the reliable prediction of equilibrium
  geometries. The formyloxyl radical as an example. \emph{J. Chem. Phys.}
  \textbf{1985}, \emph{83}, 3567--3576\relax
\mciteBstWouldAddEndPuncttrue
\mciteSetBstMidEndSepPunct{\mcitedefaultmidpunct}
{\mcitedefaultendpunct}{\mcitedefaultseppunct}\relax
\EndOfBibitem
\bibitem[Ayala and Schlegel(1998)Ayala, and Schlegel]{ayala1998nonorthogonal}
Ayala,~P.~Y.; Schlegel,~H.~B. A nonorthogonal CI treatment of symmetry breaking
  in sigma formyloxyl radical. \emph{J. Chem. Phys.} \textbf{1998}, \emph{108},
  7560--7567\relax
\mciteBstWouldAddEndPuncttrue
\mciteSetBstMidEndSepPunct{\mcitedefaultmidpunct}
{\mcitedefaultendpunct}{\mcitedefaultseppunct}\relax
\EndOfBibitem
\bibitem[Manne(1972)]{manne1972brillouin}
Manne,~R. Brillouin's theorem in Roothaan's open-shell SCF method. \emph{Mol.
  Phys.} \textbf{1972}, \emph{24}, 935--944\relax
\mciteBstWouldAddEndPuncttrue
\mciteSetBstMidEndSepPunct{\mcitedefaultmidpunct}
{\mcitedefaultendpunct}{\mcitedefaultseppunct}\relax
\EndOfBibitem
\bibitem[Davidson and Borden(1983)Davidson, and Borden]{davidson1983symmetry}
Davidson,~E.~R.; Borden,~W.~T. Symmetry breaking in polyatomic molecules: real
  and artifactual. \emph{J. Phys. Chem.} \textbf{1983}, \emph{87},
  4783--4790\relax
\mciteBstWouldAddEndPuncttrue
\mciteSetBstMidEndSepPunct{\mcitedefaultmidpunct}
{\mcitedefaultendpunct}{\mcitedefaultseppunct}\relax
\EndOfBibitem
\bibitem[{\v{C}}{\'\i}{\v{z}}ek and Paldus(1967){\v{C}}{\'\i}{\v{z}}ek, and
  Paldus]{cizek1967stability}
{\v{C}}{\'\i}{\v{z}}ek,~J.; Paldus,~J. Stability Conditions for the Solutions
  of the Hartree—Fock Equations for Atomic and Molecular Systems. Application
  to the Pi-Electron Model of Cyclic Polyenes. \emph{J. Chem. Phys.}
  \textbf{1967}, \emph{47}, 3976--3985\relax
\mciteBstWouldAddEndPuncttrue
\mciteSetBstMidEndSepPunct{\mcitedefaultmidpunct}
{\mcitedefaultendpunct}{\mcitedefaultseppunct}\relax
\EndOfBibitem
\bibitem[{\v{C}}{\'\i}{\v{z}}ek and Paldus(1970){\v{C}}{\'\i}{\v{z}}ek, and
  Paldus]{civek1970stability}
{\v{C}}{\'\i}{\v{z}}ek,~J.; Paldus,~J. Stability Conditions for the Solutions
  of the Hartree--Fock Equations for Atomic and Molecular Systems. III. Rules
  for the Singlet Stability of Hartree--Fock Solutions of $\pi$-Electronic
  Systems. \emph{J. Chem. Phys.} \textbf{1970}, \emph{53}, 821--829\relax
\mciteBstWouldAddEndPuncttrue
\mciteSetBstMidEndSepPunct{\mcitedefaultmidpunct}
{\mcitedefaultendpunct}{\mcitedefaultseppunct}\relax
\EndOfBibitem
\bibitem[Paldus and {\v{C}}{\'\i}{\v{z}}ek(1970)Paldus, and
  {\v{C}}{\'\i}{\v{z}}ek]{paldus1970stability}
Paldus,~J.; {\v{C}}{\'\i}{\v{z}}ek,~J. Stability Conditions for the Solutions
  of the Hartree-Fock Equations for Atomic and Molecular Systems. VI.
  Singlet-Type Instabilities and Charge-Density-Wave Hartree-Fock Solutions for
  Cyclic Polyenes. \emph{Phys. Rev. A} \textbf{1970}, \emph{2},
  2268--2283\relax
\mciteBstWouldAddEndPuncttrue
\mciteSetBstMidEndSepPunct{\mcitedefaultmidpunct}
{\mcitedefaultendpunct}{\mcitedefaultseppunct}\relax
\EndOfBibitem
\bibitem[Davidson(1973)]{davidson1973spin}
Davidson,~E.~R. Spin-restricted open-shell self-consistent-field theory.
  \emph{Chem. Phys. Lett.} \textbf{1973}, \emph{21}, 565--567\relax
\mciteBstWouldAddEndPuncttrue
\mciteSetBstMidEndSepPunct{\mcitedefaultmidpunct}
{\mcitedefaultendpunct}{\mcitedefaultseppunct}\relax
\EndOfBibitem
\bibitem[Edwards and Zerner(1987)Edwards, and Zerner]{edwards1987generalized}
Edwards,~W.~D.; Zerner,~M.~C. A generalized restricted open-shell Fock
  operator. \emph{Theor. Chim. Acta} \textbf{1987}, \emph{72}, 347--361\relax
\mciteBstWouldAddEndPuncttrue
\mciteSetBstMidEndSepPunct{\mcitedefaultmidpunct}
{\mcitedefaultendpunct}{\mcitedefaultseppunct}\relax
\EndOfBibitem
\bibitem[L\"owdin(1955)]{lowdin1955quantum3}
L\"owdin,~P.-O. Quantum Theory of Many-Particle Systems. III. Extension of the
  Hartree-Fock Scheme to Include Degenerate Systems and Correlation Effects.
  \emph{Phys. Rev.} \textbf{1955}, \emph{97}, 1509--1520\relax
\mciteBstWouldAddEndPuncttrue
\mciteSetBstMidEndSepPunct{\mcitedefaultmidpunct}
{\mcitedefaultendpunct}{\mcitedefaultseppunct}\relax
\EndOfBibitem
\bibitem[Mayer(1980)]{mayer1980spin}
Mayer,~I. The spin-projected extended Hartree-Fock method. In \emph{Advances in
  Quantum Chemistry}; Academic Press: New York, 1980; Vol.~12; pp
  189--262\relax
\mciteBstWouldAddEndPuncttrue
\mciteSetBstMidEndSepPunct{\mcitedefaultmidpunct}
{\mcitedefaultendpunct}{\mcitedefaultseppunct}\relax
\EndOfBibitem
\bibitem[Tsuchimochi and Scuseria(2010)Tsuchimochi, and
  Scuseria]{tsuchimochi2010communication}
Tsuchimochi,~T.; Scuseria,~G.~E. Communication: ROHF theory made simple.
  \emph{J. Chem. Phys.} \textbf{2010}, \emph{133}, 141102\relax
\mciteBstWouldAddEndPuncttrue
\mciteSetBstMidEndSepPunct{\mcitedefaultmidpunct}
{\mcitedefaultendpunct}{\mcitedefaultseppunct}\relax
\EndOfBibitem
\bibitem[Rittby and Bartlett(1988)Rittby, and Bartlett]{rittby1988open}
Rittby,~M.; Bartlett,~R.~J. An open-shell spin-restricted coupled cluster
  method: application to ionization potentials in nitrogen. \emph{J. Phys.
  Chem.} \textbf{1988}, \emph{92}, 3033--3036\relax
\mciteBstWouldAddEndPuncttrue
\mciteSetBstMidEndSepPunct{\mcitedefaultmidpunct}
{\mcitedefaultendpunct}{\mcitedefaultseppunct}\relax
\EndOfBibitem
\bibitem[Neese(2006)]{neese2006importance}
Neese,~F. Importance of direct spin-spin coupling and spin-flip excitations for
  the zero-field splittings of transition metal complexes: A case study.
  \emph{J. Am. Chem. Soc.} \textbf{2006}, \emph{128}, 10213--10222\relax
\mciteBstWouldAddEndPuncttrue
\mciteSetBstMidEndSepPunct{\mcitedefaultmidpunct}
{\mcitedefaultendpunct}{\mcitedefaultseppunct}\relax
\EndOfBibitem
\bibitem[Casanova and Krylov(2020)Casanova, and Krylov]{casanova2020spin}
Casanova,~D.; Krylov,~A.~I. Spin-flip methods in quantum chemistry. \emph{Phys.
  Chem. Chem. Phys.} \textbf{2020}, \emph{22}, 4326--4342\relax
\mciteBstWouldAddEndPuncttrue
\mciteSetBstMidEndSepPunct{\mcitedefaultmidpunct}
{\mcitedefaultendpunct}{\mcitedefaultseppunct}\relax
\EndOfBibitem
\bibitem[Roemelt and Neese(2013)Roemelt, and Neese]{roemelt2013excited}
Roemelt,~M.; Neese,~F. Excited states of large open-shell molecules: an
  efficient, general, and spin-adapted approach based on a restricted
  open-shell ground state wave function. \emph{J. Phys. Chem. A} \textbf{2013},
  \emph{117}, 3069--3083\relax
\mciteBstWouldAddEndPuncttrue
\mciteSetBstMidEndSepPunct{\mcitedefaultmidpunct}
{\mcitedefaultendpunct}{\mcitedefaultseppunct}\relax
\EndOfBibitem
\bibitem[Izs{\'a}k(2022)]{izsak2022second}
Izs{\'a}k,~R. Second quantisation for unrestricted references: formalism and
  quasi-spin-adaptation of excitation and spin-flip operators. \emph{Mol. Phy.}
  \textbf{2022}, e2126802\relax
\mciteBstWouldAddEndPuncttrue
\mciteSetBstMidEndSepPunct{\mcitedefaultmidpunct}
{\mcitedefaultendpunct}{\mcitedefaultseppunct}\relax
\EndOfBibitem
\bibitem[Veis \latin{et~al.}(2018)Veis, Antalik, Legeza, Alavi, and
  Pittner]{veis2018intricate}
Veis,~L.; Antalik,~A.; Legeza,~O.; Alavi,~A.; Pittner,~J. The intricate case of
  tetramethyleneethane: A full configuration interaction quantum Monte Carlo
  benchmark and multireference coupled cluster studies. \emph{J. Chem. Theory
  Comput.} \textbf{2018}, \emph{14}, 2439--2445\relax
\mciteBstWouldAddEndPuncttrue
\mciteSetBstMidEndSepPunct{\mcitedefaultmidpunct}
{\mcitedefaultendpunct}{\mcitedefaultseppunct}\relax
\EndOfBibitem
\bibitem[Sears and Sherrill(2008)Sears, and Sherrill]{sears2008assessing1}
Sears,~J.~S.; Sherrill,~C.~D. Assessing the Performance of Density Functional
  Theory for the Electronic Structure of Metal-Salens: The 3d$^0$-Metals.
  \emph{J. Phys. Chem. A} \textbf{2008}, \emph{112}, 3466--3477\relax
\mciteBstWouldAddEndPuncttrue
\mciteSetBstMidEndSepPunct{\mcitedefaultmidpunct}
{\mcitedefaultendpunct}{\mcitedefaultseppunct}\relax
\EndOfBibitem
\bibitem[Sears and Sherrill(2008)Sears, and Sherrill]{sears2008assessing2}
Sears,~J.~S.; Sherrill,~C.~D. Assessing the performance of Density Functional
  Theory for the electronic structure of metal-salens: the d$^2$-metals.
  \emph{J. Phys. Chem. A} \textbf{2008}, \emph{112}, 6741--6752\relax
\mciteBstWouldAddEndPuncttrue
\mciteSetBstMidEndSepPunct{\mcitedefaultmidpunct}
{\mcitedefaultendpunct}{\mcitedefaultseppunct}\relax
\EndOfBibitem
\bibitem[Olivares-Amaya \latin{et~al.}(2015)Olivares-Amaya, Hu, Nakatani,
  Sharma, Yang, and Chan]{olivares2015ab}
Olivares-Amaya,~R.; Hu,~W.; Nakatani,~N.; Sharma,~S.; Yang,~J.; Chan,~G. K.-L.
  The ab-initio density matrix renormalization group in practice. \emph{J.
  Chem. Phys.} \textbf{2015}, \emph{142}, 034102\relax
\mciteBstWouldAddEndPuncttrue
\mciteSetBstMidEndSepPunct{\mcitedefaultmidpunct}
{\mcitedefaultendpunct}{\mcitedefaultseppunct}\relax
\EndOfBibitem
\bibitem[Pandharkar \latin{et~al.}(2022)Pandharkar, Hermes, Cramer, and
  Gagliardi]{pandharkar2022localized}
Pandharkar,~R.; Hermes,~M.~R.; Cramer,~C.~J.; Gagliardi,~L. Localized Active
  Space-State Interaction: a Multireference Method for Chemical Insight.
  \emph{J. Chem. Theory Comput.} \textbf{2022}, \emph{18}, 6557--6566\relax
\mciteBstWouldAddEndPuncttrue
\mciteSetBstMidEndSepPunct{\mcitedefaultmidpunct}
{\mcitedefaultendpunct}{\mcitedefaultseppunct}\relax
\EndOfBibitem
\bibitem[M{\'a}t{\'e} \latin{et~al.}(2023)M{\'a}t{\'e}, Petrov, Szalay, and
  Legeza]{mate2022compressing}
M{\'a}t{\'e},~M.; Petrov,~K.; Szalay,~S.; Legeza,~{\"O}. Compressing
  multireference character of wave functions via fermionic mode optimization.
  \emph{J. Math. Chem.} \textbf{2023}, \emph{61}, 362--375\relax
\mciteBstWouldAddEndPuncttrue
\mciteSetBstMidEndSepPunct{\mcitedefaultmidpunct}
{\mcitedefaultendpunct}{\mcitedefaultseppunct}\relax
\EndOfBibitem
\bibitem[Smith \latin{et~al.}(2017)Smith, Mussard, Holmes, and
  Sharma]{smith2017}
Smith,~J. E.~T.; Mussard,~B.; Holmes,~A.~A.; Sharma,~S. Cheap and Near Exact
  CASSCF with Large Active Spaces. \emph{J. Chem. Theory Comput.}
  \textbf{2017}, \emph{13}, 5468--5478\relax
\mciteBstWouldAddEndPuncttrue
\mciteSetBstMidEndSepPunct{\mcitedefaultmidpunct}
{\mcitedefaultendpunct}{\mcitedefaultseppunct}\relax
\EndOfBibitem
\bibitem[Dobrautz \latin{et~al.}(2021)Dobrautz, Weser, Bogdanov, Alavi, and
  {Li~Manni}]{dobrautz2021spin}
Dobrautz,~W.; Weser,~O.; Bogdanov,~N.~A.; Alavi,~A.; {Li~Manni},~G. Spin-Pure
  Stochastic-CASSCF via GUGA-FCIQMC Applied to Iron--Sulfur Clusters. \emph{J.
  Chem. Theory Comput.} \textbf{2021}, \emph{17}, 5684--5703\relax
\mciteBstWouldAddEndPuncttrue
\mciteSetBstMidEndSepPunct{\mcitedefaultmidpunct}
{\mcitedefaultendpunct}{\mcitedefaultseppunct}\relax
\EndOfBibitem
\bibitem[Dobrautz \latin{et~al.}(2022)Dobrautz, Katukuri, Bogdanov, Kats,
  {Li~Manni}, and Alavi]{dobrautz2022combined}
Dobrautz,~W.; Katukuri,~V.~M.; Bogdanov,~N.~A.; Kats,~D.; {Li~Manni},~G.;
  Alavi,~A. Combined unitary and symmetric group approach applied to
  low-dimensional Heisenberg spin systems. \emph{Phys. Rev. B} \textbf{2022},
  \emph{105}, 195123\relax
\mciteBstWouldAddEndPuncttrue
\mciteSetBstMidEndSepPunct{\mcitedefaultmidpunct}
{\mcitedefaultendpunct}{\mcitedefaultseppunct}\relax
\EndOfBibitem
\bibitem[{Li~Manni} \latin{et~al.}(2022){Li~Manni}, Kats, and
  Liebermann]{li2022resolution}
{Li~Manni},~G.; Kats,~D.; Liebermann,~N. Resolution of Electronic States in
  Heisenberg Cluster Models within the Unitary Group Approach. \emph{ChemRxiv
  10.26434/chemrxiv-2022-rfmhk-v2} \textbf{2022}, \relax
\mciteBstWouldAddEndPunctfalse
\mciteSetBstMidEndSepPunct{\mcitedefaultmidpunct}
{}{\mcitedefaultseppunct}\relax
\EndOfBibitem
\bibitem[Pipek and Mezey(1989)Pipek, and Mezey]{pipek1989fast}
Pipek,~J.; Mezey,~P.~G. A fast intrinsic localization procedure applicable for
  abinitio and semiempirical linear combination of atomic orbital wave
  functions. \emph{J. Chem. Phys.} \textbf{1989}, \emph{90}, 4916--4926\relax
\mciteBstWouldAddEndPuncttrue
\mciteSetBstMidEndSepPunct{\mcitedefaultmidpunct}
{\mcitedefaultendpunct}{\mcitedefaultseppunct}\relax
\EndOfBibitem
\bibitem[Malrieu \latin{et~al.}(2007)Malrieu, Guih{\'e}ry, Calzado, and
  Angeli]{malrieu2007bond}
Malrieu,~J.-P.; Guih{\'e}ry,~N.; Calzado,~C.~J.; Angeli,~C. Bond electron pair:
  Its relevance and analysis from the quantum chemistry point of view. \emph{J.
  Comput. Chem.} \textbf{2007}, \emph{28}, 35--50\relax
\mciteBstWouldAddEndPuncttrue
\mciteSetBstMidEndSepPunct{\mcitedefaultmidpunct}
{\mcitedefaultendpunct}{\mcitedefaultseppunct}\relax
\EndOfBibitem
\bibitem[Lewis(1916)]{lewis1916atom}
Lewis,~G.~N. The atom and the molecule. \emph{J. Am. Chem. Soc.} \textbf{1916},
  \emph{38}, 762--785\relax
\mciteBstWouldAddEndPuncttrue
\mciteSetBstMidEndSepPunct{\mcitedefaultmidpunct}
{\mcitedefaultendpunct}{\mcitedefaultseppunct}\relax
\EndOfBibitem
\bibitem[Chen \latin{et~al.}(2011)Chen, Lai, and Shaik]{chen2011multireference}
Chen,~H.; Lai,~W.; Shaik,~S. Multireference and Multiconfiguration Ab Initio
  Methods in Heme-Related Systems: What Have We Learned So Far? \emph{J. Phys.
  Chem. B} \textbf{2011}, \emph{115}, 1727--1742\relax
\mciteBstWouldAddEndPuncttrue
\mciteSetBstMidEndSepPunct{\mcitedefaultmidpunct}
{\mcitedefaultendpunct}{\mcitedefaultseppunct}\relax
\EndOfBibitem
\bibitem[Li \latin{et~al.}(2019)Li, Guo, Sun, and Chan]{li2019electronic}
Li,~Z.; Guo,~S.; Sun,~Q.; Chan,~G. K.-L. Electronic landscape of the P-cluster
  of nitrogenase as revealed through many-electron quantum wavefunction
  simulations. \emph{Nat. Chem.} \textbf{2019}, \emph{11}, 1026--1033\relax
\mciteBstWouldAddEndPuncttrue
\mciteSetBstMidEndSepPunct{\mcitedefaultmidpunct}
{\mcitedefaultendpunct}{\mcitedefaultseppunct}\relax
\EndOfBibitem
\bibitem[Khedkar and Roemelt(2021)Khedkar, and Roemelt]{khedkar2021modern}
Khedkar,~A.; Roemelt,~M. Modern multireference methods and their application in
  transition metal chemistry. \emph{Phys. Chem. Chem. Phys.} \textbf{2021},
  \emph{23}, 17097--17112\relax
\mciteBstWouldAddEndPuncttrue
\mciteSetBstMidEndSepPunct{\mcitedefaultmidpunct}
{\mcitedefaultendpunct}{\mcitedefaultseppunct}\relax
\EndOfBibitem
\bibitem[Tarrago \latin{et~al.}(2021)Tarrago, Römelt, Nehrkorn, Schnegg,
  Neese, Bill, and Ye]{tarrago2021experimental}
Tarrago,~M.; Römelt,~C.; Nehrkorn,~J.; Schnegg,~A.; Neese,~F.; Bill,~E.;
  Ye,~S. Experimental and Theoretical Evidence for an Unusual Almost Triply
  Degenerate Electronic Ground State of Ferrous Tetraphenylporphyrin.
  \emph{Inorg. Chem.} \textbf{2021}, \emph{60}, 4966--4985\relax
\mciteBstWouldAddEndPuncttrue
\mciteSetBstMidEndSepPunct{\mcitedefaultmidpunct}
{\mcitedefaultendpunct}{\mcitedefaultseppunct}\relax
\EndOfBibitem
\bibitem[Han \latin{et~al.}(2023)Han, Luber, and {Li~Manni}]{han2023magnetic}
Han,~R.; Luber,~S.; {Li~Manni},~G. Magnetic Interactions in a
  [Co(II)$_3$Er(III)(OR)$_4$] Model Cubane Through Forefront
  Multiconfigurational Methods. \emph{ChemRxiv 10.26434/chemrxiv-2023-xd0wv}
  \textbf{2023}, \relax
\mciteBstWouldAddEndPunctfalse
\mciteSetBstMidEndSepPunct{\mcitedefaultmidpunct}
{}{\mcitedefaultseppunct}\relax
\EndOfBibitem
\end{mcitethebibliography}
\end{document}